\documentclass[twocolumn]{aastex6}
\pdfoutput=1 
\usepackage{amsmath,amstext}
\usepackage{apjfonts} 
\usepackage{afterpage}
\usepackage{subfigure,color}

\newcommand{\NHsub}[1]{\mbox{$N_{\rm H,#1}$}}
\newcommand{\kms}{\mbox{km s$^{-1}$}}

\newcommand{\be}{\begin{equation}}
\newcommand{\ee}{\end{equation}}

\shorttitle{An Aspherical Cow with a Central Engine}
\shortauthors{Margutti et al.}

\begin{document}

\title{An embedded x-ray source shines through the aspherical AT2018cow: \\ revealing the inner workings of the most luminous fast-evolving optical transients}

\author{{R.~Margutti}\altaffilmark{1}}
\author{{B.~D. Metzger}\altaffilmark{2}}
\author{{R. Chornock}\altaffilmark{3}}
\author{{I. Vurm}\altaffilmark{4}}
\author{{N. Roth}\altaffilmark{5,6}}
\author{{B.~W. Grefenstette}\altaffilmark{7}}
\author{{V. Savchenko}\altaffilmark{8}}
\author{{R. Cartier}\altaffilmark{9,10}}
\author{{J.~F. Steiner}\altaffilmark{11,12}}
\author{{G. Terreran}\altaffilmark{1}}
\author{{G. Migliori}\altaffilmark{13,14}}
\author{{D. Milisavljevic}\altaffilmark{15}}
\author{{K.~D. Alexander}\altaffilmark{1,12}}
\author{{M. Bietenholz}\altaffilmark{16,17}}
\author{{P.~K. Blanchard}\altaffilmark{18}} 
\author{{E. Bozzo}\altaffilmark{8}}
\author{{D. Brethauer}\altaffilmark{1}}
\author{{I.~V. Chilingarian}\altaffilmark{18,19}}  
\author{{D.~L. Coppejans}\altaffilmark{1}}
\author{{L. Ducci}\altaffilmark{8,20}}
\author{{C. Ferrigno}\altaffilmark{8}}
\author{{W. Fong}\altaffilmark{1}}
\author{{D. G\"{o}tz}\altaffilmark{21}}
\author{{C. Guidorzi}\altaffilmark{22}}
\author{{A. Hajela}\altaffilmark{1}}
\author{{K. Hurley}\altaffilmark{23}}
\author{{E. Kuulkers}\altaffilmark{24}}
\author{{P. Laurent}\altaffilmark{20}}
\author{{S. Mereghetti}\altaffilmark{25}}
\author{{M. Nicholl}\altaffilmark{26,18}}
\author{{D. Patnaude}\altaffilmark{18}}
\author{{P. Ubertini}\altaffilmark{27}}
\author{{J. Banovetz}\altaffilmark{15}}
\author{{N. Bartel}\altaffilmark{17}}
\author{{E. Berger}\altaffilmark{18}}
\author{{E.~R. Coughlin}\altaffilmark{2,12}}
\author{{T. Eftekhari}\altaffilmark{18}}
\author{{D.~D. Frederiks}\altaffilmark{28}}
\author{{A.~V. Kozlova}\altaffilmark{28}}
\author{{T. Laskar}\altaffilmark{29,30}}
\author{{D.~S. Svinkin}\altaffilmark{28}}
\author{{M.~R. Drout}\altaffilmark{31,32}}
\author{{A. MacFadyen}\altaffilmark{33}}
\author{{K. Paterson}\altaffilmark{1}}

\altaffiltext{1}{Center for Interdisciplinary Exploration and Research in Astrophysics (CIERA) and Department of Physics and Astronomy, Northwestern University, Evanston, IL 60208.}
\altaffiltext{2}{Department of Physics and Columbia Astrophysics Laboratory, Columbia University, New York, NY 10027, USA.}
\altaffiltext{3}{Astrophysical Institute, Department of Physics and Astronomy, 251B Clippinger Lab, Ohio University, Athens, OH 45701, USA.}
\altaffiltext{4}{Tartu Observatory, University of Tartu, T\~{o}ravere 61602, Tartumaa, Estonia.}
\altaffiltext{5}{Department of Astronomy, University of Maryland, College Park, MD 20742, USA.}
\altaffiltext{6}{Joint Space-Science Institute, University of Maryland, College Park, MD 20742, USA.}
\altaffiltext{7}{Cahill Center for Astrophysics, 1216 E. California Boulevard, California Institute of Technology, Pasadena, CA 91125, USA.}
\altaffiltext{8}{ISDC, Department of Astronomy, University of Geneva, Chemin d'\'{E}cogia, 16 CH-1290 Versoix, Switzerland.}
\altaffiltext{9}{Millennium Institute of Astrophysics, Casilla 36-D, Santiago, Chile.}
\altaffiltext{10}{Departamento de Astronom\'{i}a, Universidad de Chile, Casilla 36-D, Santiago, Chile.}
\altaffiltext{11}{MIT Kavli Institute for Astrophysics and Space Research, MIT, 70 Vassar Street, Cambridge, MA 02139, USA.}
\altaffiltext{12}{Einstein Fellow.}
\altaffiltext{13}{Dipartimento di Fisica e Astronomia, Alma Mater Studiorum, Universit\`a degli Studi di Bologna, Via Gobetti 93/2, 40129 Bologna, Italy.} 
\altaffiltext{14}{INAF Istituto di Radioastronomia, via Gobetti 101, 40129 Bologna, Italy.} 
\altaffiltext{15}{Department of Physics and Astronomy, Purdue University, 525 Northwester Avenue, West Lafayette, IN 47907, USA.}
\altaffiltext{16}{Hartebeesthoek Radio Observatory, PO Box 443, Krugersdorp, 1740, South Africa.}
\altaffiltext{17}{Department of Physics and Astronomy, York University, Toronto, M3J~1P3, Ontario, Canada.}
\altaffiltext{18}{Harvard-Smithsonian Center for Astrophysics, 60 Garden St. Cambridge MA 02138.} 
\altaffiltext{19}{Sternberg Astronomical Institute, M.V.Lomonosov Moscow State University, Universitetsky prospect 13, Moscow, 119234, Russia.} 
\altaffiltext{20}{Institut f\"{u}r Astronomie und Astrophysik, Kepler Center for Astro and Particle Physics, Eberhard Karls Universitat, Sand 1, 72076, Tubingen, Germany.}
\altaffiltext{21}{CEA Saclay - Irfu/D'epartement d'Astrophysique, Orme des Merisiers, Bat. 709, F91191 Gif-sur-Yvette.}
\altaffiltext{22}{Department of Physics and Earth Science, University of Ferrara, via Saragat 1, I--44122, Ferrara, Italy.}
\altaffiltext{23}{University of California at Berkeley, Space Sciences Laboratory, 7 Gauss Way, Berkeley, CA 94720.}
\altaffiltext{24}{European Space Astronomy Centre (ESA/ESAC), Science Operations Department, 28691 Villanueva de la Canada, Madrid, Spain.}
\altaffiltext{25}{INAF - Istituto di Astrofisica Spaziale e Fisica Cosmica Milano, Via E. Bassini 15, 20133 Milano, Italy.}
\altaffiltext{26}{Institute for Astronomy, University of Edinburgh, Royal Observatory, Blackford Hill, Edinburgh EH9 3HJ, UK.}
\altaffiltext{27}{Istituto di Astrofisica e Planetologia Spaziali, INAF Via Fosso del Cavaliere 100, 00133 Rome, Italy.} 
\altaffiltext{28}{Ioffe Physical-Technical Institute, Politekhnicheskaya 26, St. Petersburg 194021, Russia.}
\altaffiltext{29}{National Radio Astronomy Observatory, 520 Edgemont Road, Charlottesville, VA 22903, USA.}
\altaffiltext{30}{Department of Astronomy, University of California, 501 Campbell Hall, Berkeley, CA 94720-3411, USA.}
\altaffiltext{31}{The Observatories of the Carnegie Institution for Science, 813 Santa Barbara St., Pasadena, CA 91101, USA.}
\altaffiltext{32}{Department of Astronomy and Astrophysics, University of Toronto, 50 St. George Street, Toronto, Ontario, M5S 3H4 Canada.}
\altaffiltext{33}{Center for Cosmology and Particle Physics, New York University, 726 Broadway, New York, NY 10003, USA.}

\begin{abstract}
 We present the first extensive radio to $\gamma$-ray observations of a fast-rising blue optical transient (FBOT), AT\,2018cow, over its first $\sim$100 days.  AT\,2018cow rose over a few days to a peak luminosity $L_{\rm{pk}}\sim4\times 10^{44}\,\rm{erg\,s^{-1}}$ exceeding those of superluminous supernovae (SNe), before declining as $L\propto t^{-2}$.  Initial spectra at $\delta t\lesssim 15$ days were mostly featureless and indicated large expansion velocities $v\sim0.1\,c$ and temperatures reaching $T\sim3\times 10^{4}$ K.  Later spectra revealed a persistent optically-thick photosphere and the emergence of H and He emission features with $v\sim4000$~\kms with no evidence for ejecta cooling.  Our broad-band monitoring revealed a hard X-ray spectral component at $E\ge 10$ keV, in addition to luminous and highly variable soft X-rays, with properties unprecedented among astronomical transients. An abrupt change in the X-ray decay rate and variability appears to accompany the change in optical spectral properties. AT\,2018cow showed bright radio emission consistent with the interaction of a blastwave with $v_{sh}\sim0.1\,c$ with a dense environment ($\dot M\sim10^{-3}-10^{-4}\,\rm{M_{\sun}yr^{-1}}$ for $v_w=1000$~\kms). While these properties exclude $^{56}$Ni-powered transients, our multi-wavelength analysis instead indicates that AT\,2018cow harbored a "central engine", either a compact object (magnetar or black hole) or an embedded internal shock produced by interaction with a compact, dense circumstellar medium. The engine released $\sim10^{50}-10^{51.5}$ erg over $\sim10^3-10^5$ s and resides within low-mass fast-moving material with equatorial-polar density asymmetry ($M_{\rm{ej,fast}}\lesssim0.3\,\rm{M_{\sun}}$).  Successful SNe from low-mass H-rich stars (like electron-capture SNe) or failed explosions from blue supergiants satisfy these constraints.  Intermediate-mass black-holes are disfavored by the large environmental density probed by the radio observations. 
\end{abstract}


\section{Introduction}
Recent high-cadence surveys have uncovered a plethora of rapidly-evolving transients with diverse observed properties that challenge our current notions of stellar death (e.g., \citealt{Drout14,Arcavi16,Tanaka16,Pursiainen18} for recent sample compilations).
Such rapid evolution is generally attributed to a small mass of ejecta $M\lesssim 1\,M_{\sun}$. However, the wide range of observed properties  (i.e., luminosities, energetics, chemical composition and environments), reveals them to be an extremely heterogeneous class and likely reflects a diverse range of intrinsic origins. 

Fast evolving transients can be either rich or poor in hydrogen, and span a wide range of peak luminosities. Some are less luminous than normal H-stripped core-collapse SNe (i.e., Ibc SNe, e.g., SN\,2005E, \citealt{Perets10}; SN\,2008ha, \citealt{Valenti09,Foley09}) or populate the low-end of the luminosity function of Ibc SNe (e.g., SNe 2005ek, 2010X; \citealt{Drout+2013,Kasliwal10}). The relatively old stellar environments of some of these transients and their low luminosities have inspired connections with models of He-shell detonations on white dwarf (WD) progenitors (``Ia'' SNe, \citealt{Shen10}). However, the oxygen-dominated ejecta of SN\,2005ek and the young stellar environments of other rapidly-evolving transients are instead more readily explained as the explosions of massive stars which have been efficiently stripped of their envelopes by binary interaction (\citealt{Drout+2013,Tauris13,Kleiser14,Tauris15,Suwa15,Moriya17}), or ``cooling envelope" emission from the explosion of radially-extended red supergiant stars (\citealt{Tanaka16}).

Some rapidly-evolving transients can  compete in luminosity with Ibc-SNe (e.g., SN\,2002bj ; \citealt{Poznanski10}) or even outshine normal core-collapse SNe (\citealt{Arcavi16}).  The short timescales, high peak luminosities and lack  of  UV  line  blanketing  observed  in  many  of  these transients are in tension with traditional SN models powered by the radioactive decay of $^{56}\rm{Ni}$ (e.g., \citealt{Poznanski10,Drout14,Pursiainen18,Rest18}). These objects typically show blue colors and have been referred to in the literature as ``Fast Evolving Luminous Transients'' (FELTs, \citealt{Rest18}) or ``Fast Blue Optical Transients'' (FBOTs, \citealt{Drout14}). Here we adopt the ``FBOT'' acronym.  

The non-radioactive sources of energy needed to explain FBOTs fall into two broad categories: (i) Interaction of the explosion's shock wave with a dense circumstellar environment or extended progenitor atmosphere (\citealt{ChevalierI2011,Balberg11,Ginzburg14}). This class of models has been applied to a variety of FBOTs with and without direct evidence for interaction in their spectra  (e.g., \citealt{Ofek10,Drout14,Pastorello15,Shivvers16,Rest18}. In this scenario the high luminosities of FBOTs are the result of  efficient conversion of ejecta kinetic
energy into radiation, as the explosion shock interacts with a dense external shell, while the rapid time-scale is attributed to the relatively compact radius of the shell.  (ii) Models involving prolonged energy injection from a central compact object, such as a magnetar with a millisecond rotation period (\citealt{Yu13,Metzger14,Hotokezaka17}), an accreting neutron star (NS; e.g. following a WD-NS merger; \citealt{Margalit16}), or an accreting stellar-mass (\citealt{Kashiyama15}) or supermassive black hole (BH e.g., \citealt{Strubbe09,Cenko2012MNRAS}).

Until recently, progress in understanding the intrinsic nature of FBOTs was hampered by their low discovery rate and typically large  distances  ($d\ge 500$ Mpc),  which limited opportunities for spectroscopic and multi-wavelength follow-up observations. Here we present extensive radio-to-$\gamma$-ray observations of the Astronomical Transient AT\,2018cow over its first $\sim$100 days of evolution. AT\,2018cow was discovered on June 16, 2018 by the ATLAS survey as a rapidly evolving transient located within a spiral arm of the dwarf star-forming galaxy CGCG 137-068 at 60 Mpc (\citealt{Smartt+2018a,Prentice+2018}). \cite{Prentice+2018,Perley18,Rivera18} and \cite{Kuin18} presented the UV/optical/NIR and soft X-ray properties of AT\,2018cow (as detected by \emph{Swift}) in the first $\sim 50$ days since discovery.  We present our UV/optical/NIR photometry and spectroscopy in  \S\ref{SubSec:UVOpticalNIRPhot} and \S\ref{SubSec:OpticalNIRSpec}. Broad-band soft-to-hard X-ray data from coordinated follow up with INTEGRAL, NuSTAR, \emph{Swift}-XRT and XMM are presented and analyzed in \S\ref{SubSec:XRTXMM}, \ref{SubSec:NuSTAR} and \S\ref{SubSec:Xrayjoint}, while our radio observations with VLA and VLBA are described in \S\ref{SubSec:radio}. We present the search for prompt $\gamma$-ray emission from AT\,2018cow with the Inter-Planetary Network in \S\ref{SubSec:IPN}. In \S\ref{Sec:inferences} we derive multi-band inferences on the physical properties of AT\,2018cow and we discuss the intrinsic nature of AT\,2018cow in \S\ref{Sec:interpretation}. We conclude in \S\ref{Sec:conclusiosn}.

Uncertainties are provided at the $1\,\sigma$ confidence level (c.l.) and we list $3\,\sigma$ c.l. upper limits unless explicitly stated otherwise. Throughout the paper we refer times to the time of optical discovery, which is  2018-06-16 10:35:02 UTC, corresponding to MJD 58285.44 \citep{Smartt+2018a,Prentice+2018}. AT\,2018cow is located in the host galaxy CGCG 137-068 ($z=0.0141$) and we adopt a distance of 60 Mpc as in \cite{Smartt+2018a,Prentice+2018,Perley18}. We assume $h=0.7$, $\Omega_M=0.3$, $\Omega_{\Lambda}=0.7$.

\section{Observations and data analysis}
\subsection{UV-Optical-NIR Photometry}
\label{SubSec:UVOpticalNIRPhot}
\begin{figure*}
\includegraphics[width=\textwidth]{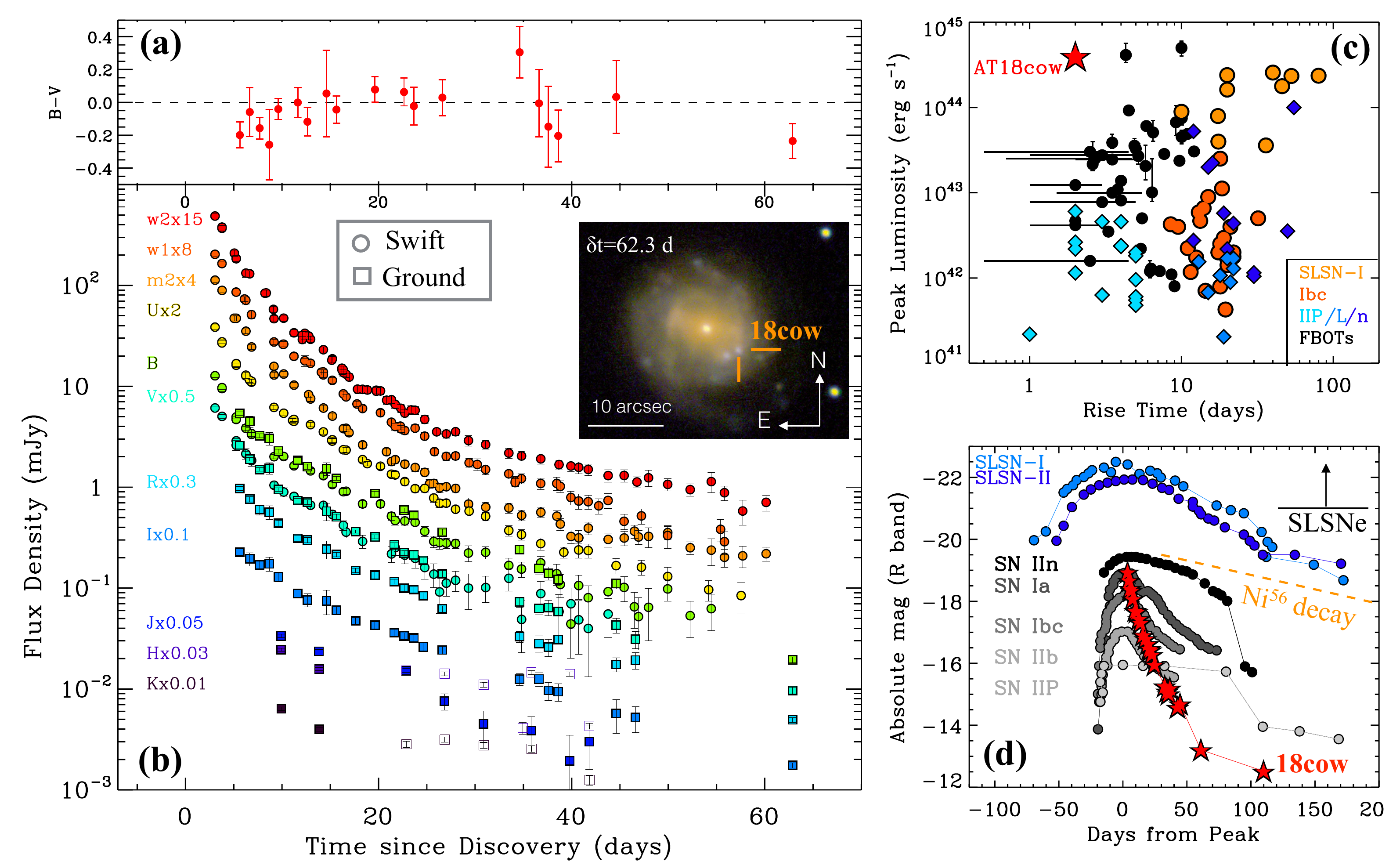}
\caption{\emph{Panel (a)}: AT\,2018cow maintains observed blue colors $(B-V)<0$ until late times, while the UV/optical/NIR flux rapidly decays. \emph{Panel (b)}: Filled circles: extinction-corrected, host-galaxy subtracted flux densities derived from  \emph{Swift}-UVOT observations. Filled squares: extinction-corrected flux densities derived from our CTIO photometry (\textit{BVRI} at $\delta t<50$ days), Keck photometry (\textit{BVRI} at $\delta t>50$ days), UKIRT and WIYN photometry (\textit{JHK}). For the NIR bands, empty symbols mark the times when significant contamination from the host galaxy emission is present. \emph{Inset:} RGB false-color image of AT\,2018cow and its host galaxy obtained on 2018 August 17 with DEIMOS mounted on Keck-II.  \emph{Panels (c-d)}: optical light-curve properties of AT\,2018cow in the context of other stellar explosions and FBOTs from the literature. AT\,2018cow  shows an extremely rapid rise time of a few days (as constrained by \citealt{Perley18,Prentice+2018}), and a decay significantly faster than $^{56}$Ni-powered decays (orange dashed line in panel(d)). AT\,2018cow rivals in luminosity the most luminous \emph{normal} SNe in the \textit{R}-band (d-panel), but it is more luminous than some SLSNe when its bolometric output is considered (panel (c)) due to its remarkably blue colors. Following \cite{GalYam12}, we show in panel (d) prototypical events for each class: PTF\,09cnd (SLSN-I, \citealt{Quimby11}), SN\, 2006gy (SLSN-II), 
``Nugent template'' for normal type-Ia SN, SN\,2005cl (SN IIn, \citealt{Kiewe12}), the average type Ibc light curve from \cite{Drout11}, SN\,2011dh (SN IIb, \citealt{Arcavi11,Soderberg12}), and the prototypical type II-P SN\,1999em (\citealt{Leonard02}). Other references: \cite{Hamuy03,Campana06,Taubenberger06,Valenti08,Botticella09,Cobb10,Kasliwal10,Ofek10,Poznanski10,Andrews11,Chomiuk11,Arcavi12,Bersten12,Valenti12,Drout+2013,Inserra13,Lunnan13,Drout14,Margutti1409ip,Vinko15,Nicholl16,Arcavi16,Pursiainen18}.}
\label{Fig:Optical}
\end{figure*}

\begin{figure*}
\includegraphics[scale=0.6]{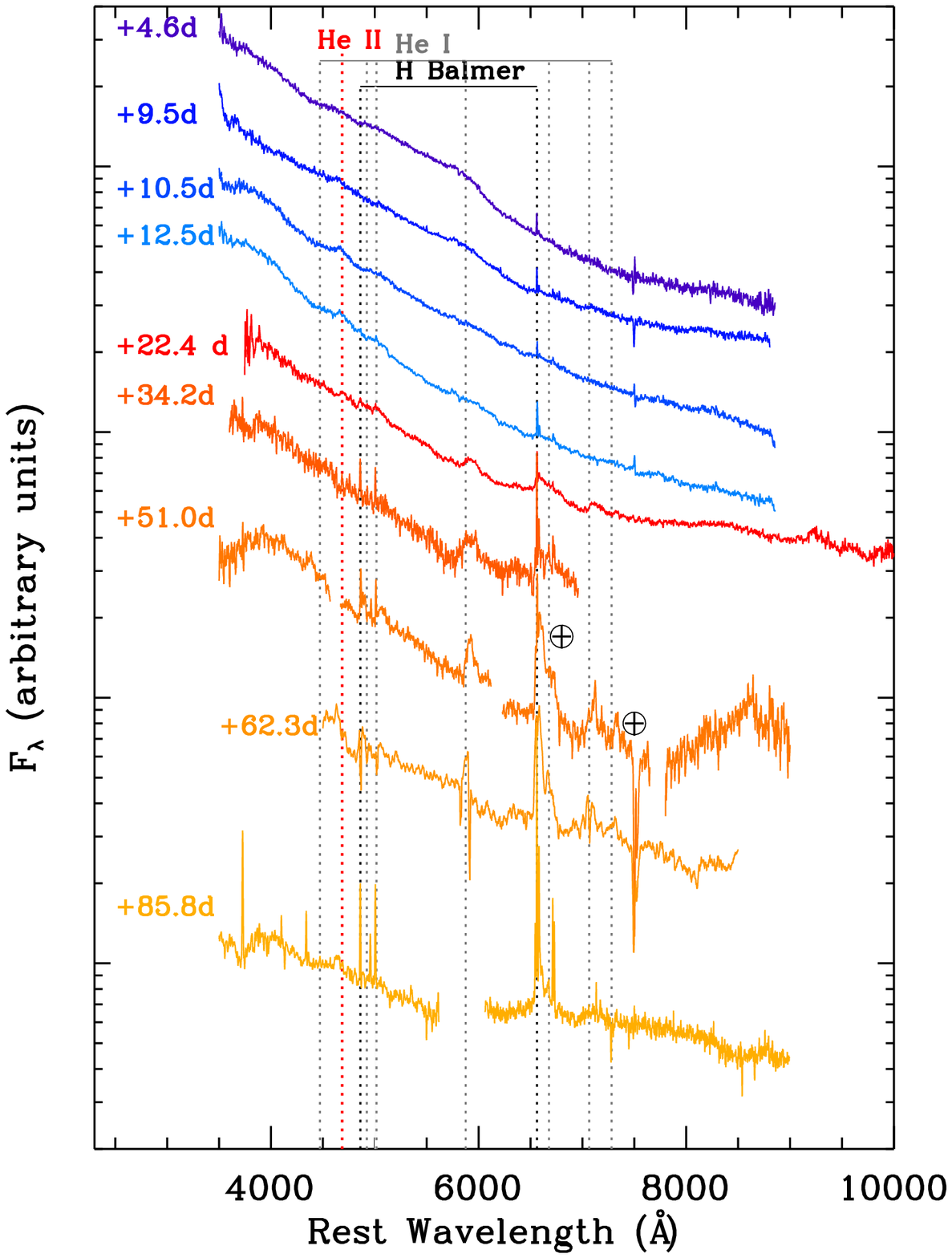}
\includegraphics[scale=0.6]{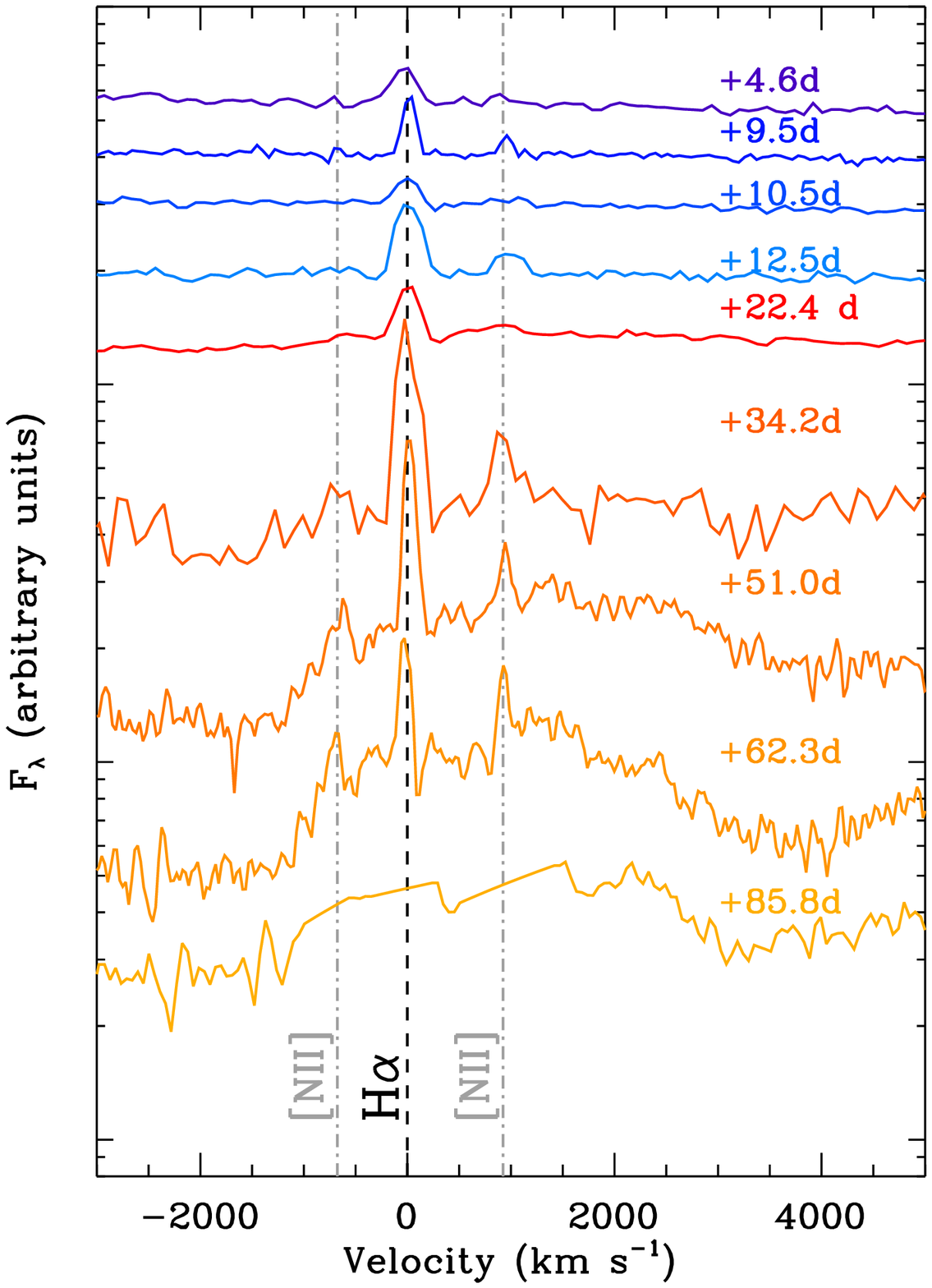}
\caption{Optical spectral evolution of AT\,2018cow (\emph{left panel}), with a zoom-in to the H$\alpha$ region of the spectrum in velocity space (\emph{right panel}). At $\delta t\lesssim$ 20 days the spectrum exhibits only extremely broad features with $v\sim0.1c$, in addition to narrow emission lines from the host galaxy. At $\delta t>$ 20 days He I and H features start to develop with velocities of a few 1000~\kms\ and a redshifted line profile. In the H$\alpha$ panel on the right, we clipped the strong narrow line emission from the host galaxy in our latest spectrum at $\delta t=85.8$ days for display purposes. }
\label{Fig:OpticalSpec}
\end{figure*}

\begin{figure}
\center{\includegraphics[scale=0.4]{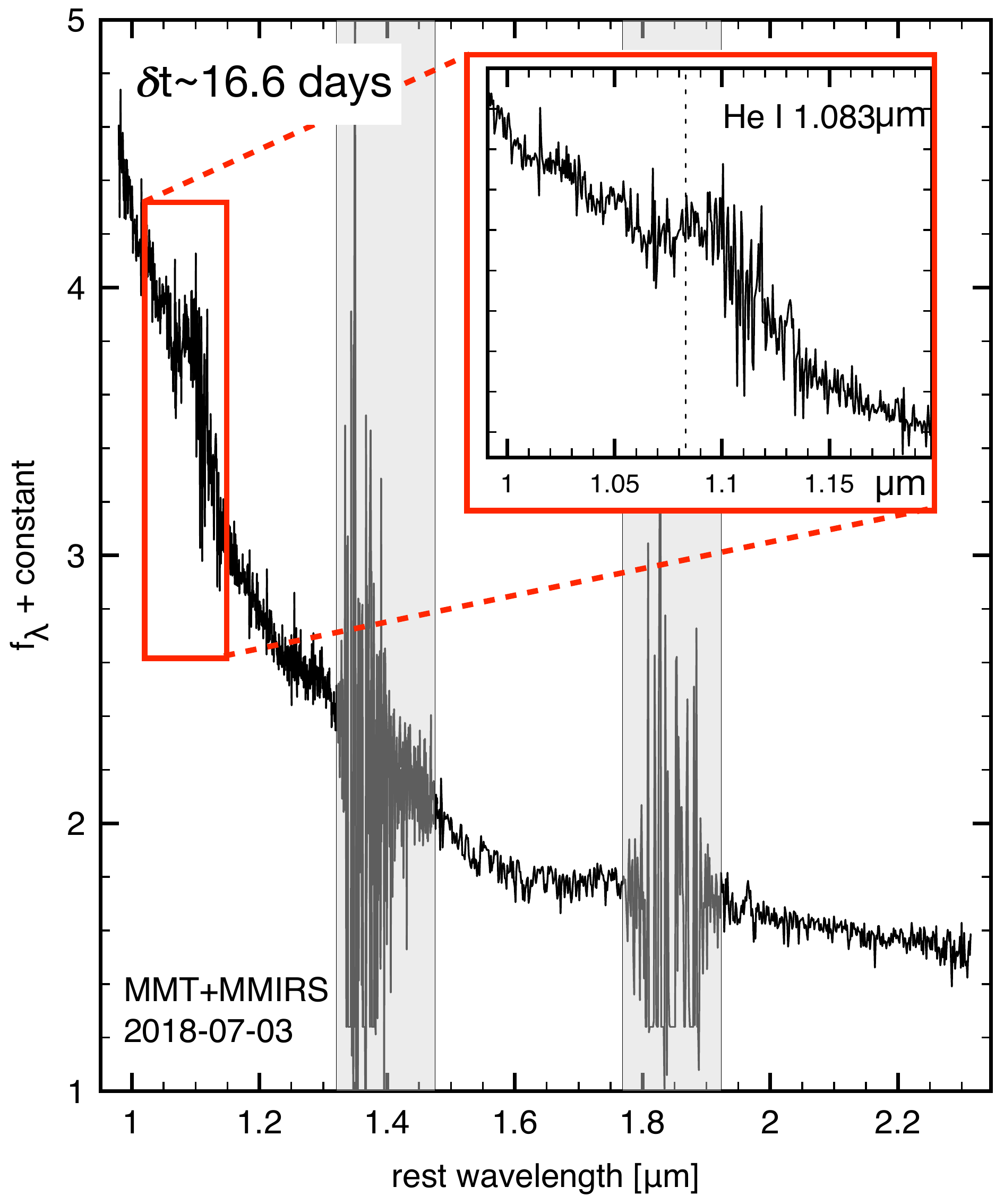}}
\vskip -0.2 cm
\caption{A NIR spectrum of AT\,2018cow acquired at $\sim17$ days shows the emergence of \ion{He}{1} emission with a characteristic redshifted line profile, as observed at optical wavelengths (Fig.~\ref{Fig:OpticalSpec}). The gray bands mark regions of strong telluric absorption.}
\label{Fig:NIRSpec}
\end{figure}

The UV Optical Telescope (UVOT, \citealt{Roming05}) on board the Neil Gehrels \emph{Swift} Observatory \citep{Gehrels04} started observing AT\,2018cow on 2018 June 19 ($\sim3$ days since discovery) with six filters $v$, $b$, $u$, $w1$, $w2$ and $m2$, in the wavelength range $\lambda_c=1928$ \AA\, (\textit{w2} filter) -- $\lambda_c=5468$  \AA\, (\textit{v} filter, central wavelength). We extracted aperture photometry following standard prescriptions by \citet{Brown09}, with the updated calibration files and revised zero points by \cite{Breeveld11}.  Each individual frame has been visually inspected and quality flagged. Observations with insufficient exposure time have been merged  to obtain higher signal-to-noise ratio (S/N) images from which we extracted the final photometry. We used a 3\arcsec\ source region of extraction to minimize the effects of the contamination from the underlying host-galaxy flux and we manually corrected for imperfections of the astrometric solution of the automatic UVOT pipeline re-aligning the frames. 
In the absence of template images, we estimated the host galaxy contribution by measuring the host galaxy emission at a similar distance from the nucleus. The results from our method are in excellent agreement with \cite{Perley18}. We note that at $\delta t>50$ days this method is likely to overestimate the UV flux of the transient, as the images show the presence of a bright knot of UV emission underlying AT\,2018cow that can only be properly accounted for with template images obtained in the future. 

Ground-based optical photometry has been obtained from ANDICAM, mounted on the 1.3-m telescope\footnote{Operated by the SMARTS Consortium.} at Cerro Tololo Interamerican Observatory (CTIO), the Low Resolution Imaging Spectrometer (LRIS; \citealt{Oke1995}), and the DEep Imaging Multi-Object Spectrograph (DEIMOS; \citealt{Faber2003}), mounted on the Keck telescopes. Images from the latter were reduced following standard bias and flat-field corrections. Data from ANDICAM, instead, came already reduced by their custom pipeline\footnote{https://github.com/SMARTSconsortium/ANDICAM}. Instrumental magnitudes were extracted using the point-spread-function (PSF) fitting technique, performed using the \textsc{SNOoPY}\footnote{Cappellaro, E. (2014). \textsc{SNOoPY}: a package for SN photometry, \url{http://sngroup.oapd.inaf.it/snoopy.html}} package.  Absolute calibrations were achieved measuring zero points and color terms for each night, estimated using as reference the magnitudes of field stars, retrieved from the Sloan Digital Sky Survey\footnote{http://www.sdss.org} \citep[SDSS;][]{York2000} catalog (DR9). SDSS magnitudes of the field stars were then converted to Johnson/Cousins, following \cite{Chonis2008}. Our \textit{BVRI} PSF photometry agrees well with the host-galaxy subtracted photometry presented by \cite{Perley18}.

We obtained near-IR imaging observations in the $JHK$-bands with the Wide-field Camera (WFCAM; \citealt{caa+07}) mounted on the 3.8-m United Kingdom Infrared Telescope (UKIRT) spanning $\delta t\sim10-42$~days. We obtained pre-processed images from the WFCAM Science Archive \citep{hcc+08} which are corrected for bias, flat-field, and dark current by the Cambridge Astronomical Survey Unit\footnote{http://casu.ast.cam.ac.uk/}. For each epoch and filter, we co-add the images and perform astrometry relative to 2MASS using a combination of tasks in Starlink\footnote{http://starlink.eao.hawaii.edu/starlink} and \texttt{IRAF}. 
For $J$-band, we obtain a template image from the UKIRT Hemispheres Survey DR1 \citep{dlr+18}, and use the HOTPANTS software package \citep{bec15} to perform image subtraction against this template to produce residual images. We perform aperture photometry using standard tasks in \texttt{IRAF}, photometrically calibrated to 2MASS.  
In the absence of a template image in $H$ and $K$-bands, we performed aperture photometry of the transient and host galaxy complex centered on the core of the host galaxy. We used standard procedures in \texttt{IRAF} and $2.5$ full-width half-maximum apertures.  At $\delta t<15$ days the host galaxy contribution is negligible, but dominates  the $HK$ photometry at $\delta t>30$ days. Single epochs of \textit{JHK}-band photometry were obtained 2018 June 26 ($\delta t\sim 9.86$ days) using the WIYN High-resolution Infrared Camera (WHIRC; \citealt{Meixner10}) mounted on the 3.5 m WIYN telescope, and 2018 August 31 ($\delta t\sim 75.7$ days) with the MMT and Magellan Infrared Spectrograph (MMIRS;  \citealt{McLeod2012}), mounted on the MMT telescope. These data were processed using similar methods. AT\,2018cow is not detected against the host-galaxy NIR background in our final observation.  After subtracting the bright sky contribution we estimated the instrumental NIR magnitudes  via 
PSF-fitting. We calibrate our NIR photometry relative to 2MASS\footnote{http://www.ipac.caltech.edu/2mass/} \citep{Skrutskie2006}. No color term correction was applied to the NIR data. 

UV, optical, and NIR photometry have been corrected for Galactic extinction with $E(B-V)=0.07$ mag, \citep{Schlafly11} and no extinction in the host galaxy. Our final photometry is presented in Tables \ref{Tab:GroundPhotOpt},\ref{Tab:GroundPhotNIR},\ref{Tab:UVOT1},\ref{Tab:UVOT2}. The UV/optical/NIR emission from AT\,2018cow is shown in Fig.~\ref{Fig:Optical}. 

\subsection{Optical  and NIR Spectroscopy}
\label{SubSec:OpticalNIRSpec}

We obtained 5 spectra of AT\,2018cow  using the Goodman spectrograph \citep{clemens04} mounted on the SOAR telescope in the time range $\delta t\sim 4.6-34.2$ days. We used the red camera and the 400 lines mm$^{-1}$ and 600 lines mm$^{-1}$ gratings, providing a resolution of $\sim 5$ \AA\ and $\sim 3$ \AA\ at 7000 \AA, respectively. We reduced Goodman data following usual steps including bias subtraction, flat fielding, cosmic ray rejection \citep[see][]{vandokkum01}, wavelength calibration, flux calibration, and telluric correction using our own custom \texttt{IRAF}\footnote{\texttt{IRAF} is distributed by the National Optical Astronomy Observatories, which are operated by the Association of Universities for Research in Astronomy, Inc., under cooperative agreement with the National Science Foundation.} routines.

On 2018 July 9 ($\delta t\sim 21.4$ days), we acquired a spectrum with the Low Dispersion Survey Spectrograph (LDSS3) mounted on the 6.5 m Magellan Clay telescope using the VPH-all grism and a 1$\arcsec$ slit. We obtained a spectrum with the Inamori-Magellan Areal Camera and Spectrograph (IMACS) mounted on the 6.5 m Magellan Baade telescope on 2018 August 6 ($\delta t\sim 51$ days), using the f/4 camera and 300 l/mm grating with a 0.9\arcsec\mbox{} slit. 
The data were reduced using standard procedures in \texttt{IRAF} and \texttt{PyRAF} to bias-correct, flat-field, and extract the spectrum. Wavelength calibration was achieved using HeNeAr comparison lamps, and relative flux calibration was applied using a standard star observed with the same setup.  

We observed AT\,2018cow on 2018 August 29 ($\delta t\sim 74$ days) with DEIMOS. We used a 0.7\arcsec\ slit and the 600~lines~mm$^{-1}$ grating with the GG400 filter, resulting in a $\sim3$~\AA\ resolution over the range $4500-8500$~\AA. We acquired a spectrum with LRIS on 2018 September 9 ($\delta t\sim$85.8 days). We used the 1.0\arcsec\ slit with the 400~lines~mm$^{-1}$ grating, achieving a resolution of $\sim6$ \AA\ and spectral coverage of 3200--9000 \AA. Due to readout issues, we lost a portion of the spectrum between 5800 and 6150~\AA. Reduction of these spectra were done using standard \texttt{IRAF} routines for bias subtraction and flat-fielding. Wavelength and flux calibration were performed comparing the data to arc lamps and standard stars respectively, acquired during the night and using the same setups. A final epoch of \textit{BVRI} photometry was acquired with LRIS on  2018 October 5 ($\delta t$ $\sim112$ days). 

We acquired one epoch of low-resolution NIR spectroscopy spanning $0.98-2.31\,\mu$m with the MMT using MMIRS on 2018 July 3 ($\delta t\sim16.6$ days). Observations were performed using a 1$^{\prime\prime}$ slit width in two configurations: zJ filter (0.95 - 1.50 $\mu$m) + J grism ($R \sim 2000$), and HK3 filter (1.35-2.3 $\mu$m) + HK grism ($R \sim 1400$). For each of the configurations the total exposure time was 1800 s, and the slit was dithered between individual 300 s exposures. We used the standard MMIRS pipeline \citep{Chilingarian15} to process the data and to develop wavelength calibrated 2D frames from which 1D extractions were made.

Figures \ref{Fig:OpticalSpec}--\ref{Fig:NIRSpec} show our spectral series. These figures show the drastic evolution of AT\,2018cow  from an almost featureless spectrum around optical peak with very broad features, to the clear emergence of H and He emission with asymmetric line profiles skewed to the red and significantly smaller velocities of a few 1000~\kms.  
In Table \ref{spec_tab} we summarize our NIR/optical spectroscopic observations.

\subsection{Soft X-rays: Swift-XRT and XMM}
\label{SubSec:XRTXMM}

\begin{figure}
\includegraphics[scale=0.48]{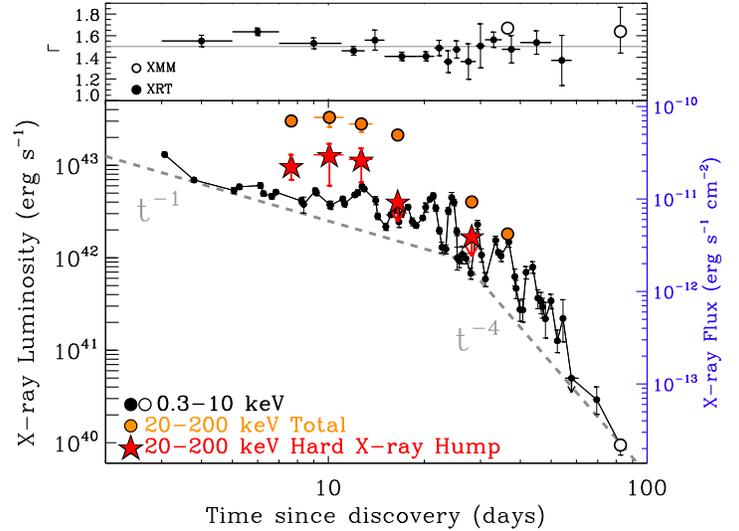}
\caption{Temporal evolution of AT\,2018cow at soft (black, 0.3-10 keV) and hard (orange and red, 20-200 keV) X-ray energies, as captured by \emph{Swift}-XRT, XMM, NuSTAR and INTEGRAL.  Soft X-rays are well modeled with a power-law spectrum with photon index $\Gamma\sim1.5$ and limited temporal evolution (upper panel). Above $\sim$20 keV an additional transient spectral component appears at $t<15$ days. Orange dots: total luminosity in the 20-200 keV band. Red stars: contribution of the additional hard X-ray energy component above the extrapolation of the power-law component from lower energies. Dashed gray lines: reference $t^{-1}$ and $t^{-4}$ power-law decays to guide the eye.  }
\label{Fig:Xray}
\end{figure}

\begin{figure}
\hskip -0.5 cm
\includegraphics[scale=0.42]{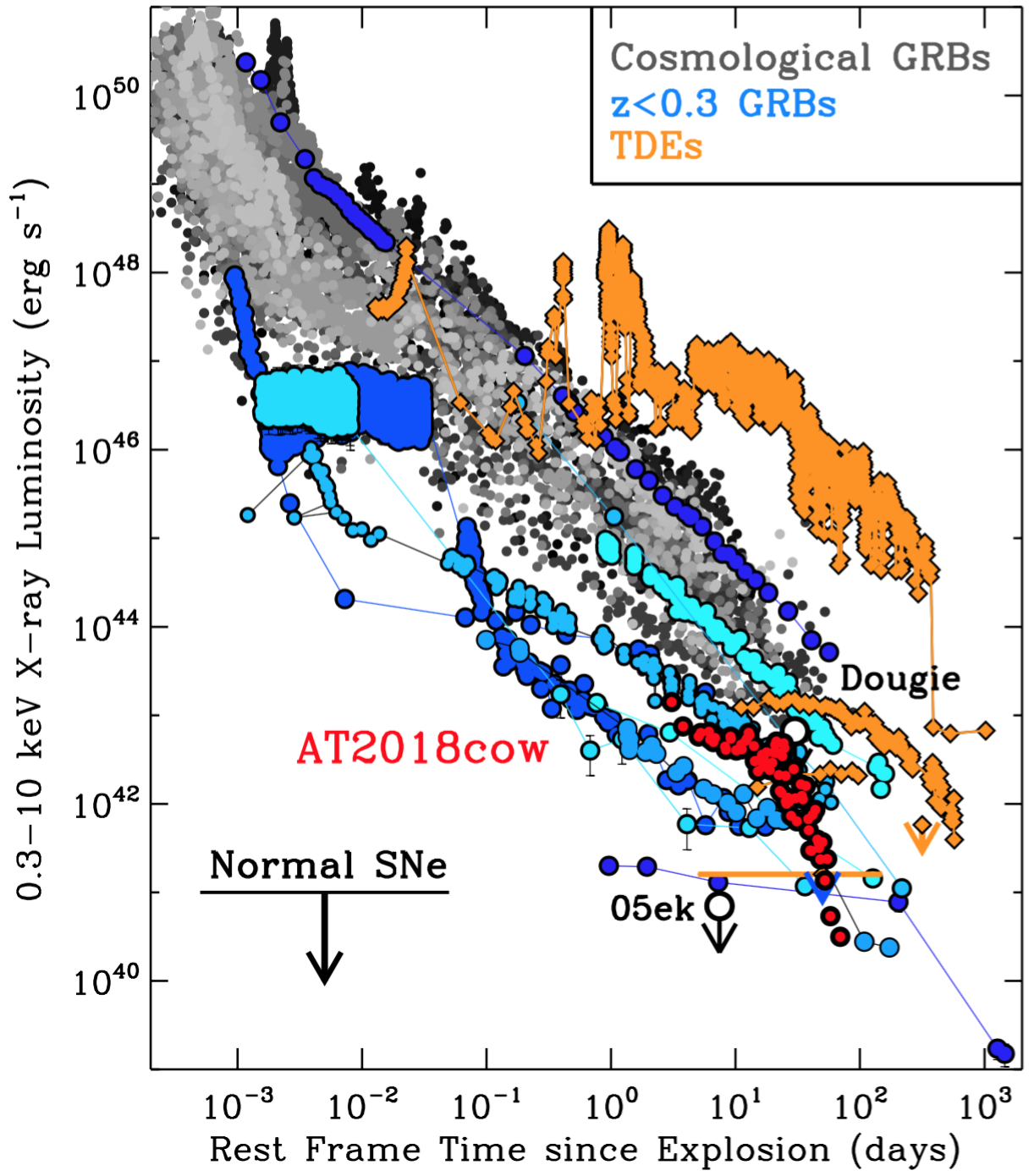}
\caption{X-ray emission from AT\,2018cow (red circles) in the context of long GRBs at cosmological distances (shades of gray), long GRBs in the local Universe (shades of blue), tidal disruption events (TDEs, orange diamonds), and normal core-collapse
SNe (black arrow and circle), which later show $L_{\rm X}<10^{41}\,\rm{erg\,s^{-1}}$. The upper limits on the X-ray emission from the very rapidly declining type-Ic SN\,2005ek and the fast-rising and luminous transient ``Dougie'' are marked with empty circles.  AT\,2018cow is significantly more luminous than normal SNe and competes in luminosity with local GRBs. References: \cite{Margutti13,Margutti100316D,Drout+2013,Vinko15,Margutti1715lh,Ross17,Eftekhari18}. }
\label{Fig:GRB}
\end{figure}

The X-Ray Telescope (XRT) on board the Neil Gehrels Swift Observatory \citep{Gehrels04,Burrows05} started observing AT\,2018cow on 2018 June 19 ($\sim3$ days following discovery). We reduced the \emph{Swift}-XRT data with HEAsoft v.~6.24 and corresponding calibration files, applying standard data filtering as in \cite{Margutti13}. A bright X-ray source is detected at the location of the optical transient, with clear evidence for persistent X-ray flaring activity on timescales of a few days (\S\,\ref{Subsec:Timing}), superimposed on an overall fading of the emission (Fig.~\ref{Fig:Xray}). 

A time-resolved spectral analysis reveals limited spectral evolution. Fitting the 0.3--10 keV data with an absorbed power-law model within XSPEC, we find that the XRT spectra are well described by a photon index $\Gamma\sim 1.5$ and no evidence for intrinsic neutral hydrogen absorption (Fig.~\ref{Fig:Xray}, upper panel). We employ Cash statistics and derive the parameter uncertainties from a series of MCMC simulations.  We adopt a Galactic neutral hydrogen column density in the direction of AT\,2018cow $\NHsub{MW}=0.05\times 10^{22} \,\rm{cm^{-2}}$  \citep{Kalberla05}. With a different method based on X-ray afterglows of GRBs, \citet{Willingale13} estimate $\NHsub{MW} =0.07\times 10^{22}\,\rm{cm^{-2}}$. In particular, the earliest XRT spectrum extracted between $3-5$ days since discovery can be fitted with $\Gamma=1.55\pm 0.05$ and can be used to put stringent constraints on the amount of neutral material in front of the emitting region, which is $\NHsub{int} < 6\times 10^{20}\,\rm{cm^{-2}}$ (we adopt solar abundances from \citealt{Asplund09} within XSPEC). The material is thus either fully ionized or absent  (\S\,\ref{SubSubSec:XrayEngine}). The results from the time-resolved \emph{Swift}-XRT analysis are reported in Table \ref{Tab:XRTcow}.
The total XRT spectrum collecting data in the time interval 3--60 days can be fitted with an absorbed power-law with $\Gamma=1.55\pm0.04$ and $\NHsub{int}<0.03\times 10^{22}\,\rm{cm^{-2}}$. From this spectrum we infer a 0.3--10 keV count-to-flux conversion factor of $4.3\times 10^{-11}\,\rm{erg\,cm^{-2}\,ct^{-1}}$ (absorbed), $4.6\times 10^{-11}\,\rm{erg\,cm^{-2}\,ct^{-1}}$ (unabsorbed), which we use to flux-calibrate the XRT light-curve (Fig.~\ref{Fig:Xray}). At the distance of $\sim 60$ Mpc, the inferred 0.3--10 keV isotropic X-ray luminosity at 3 days is $L_{\rm X}\sim10^{43}\,\rm{erg\,s^{-1}}$. AT\,2018cow is significantly more luminous than normal SNe and shows a luminosity similar to that of low-luminosity GRBs (Fig.~\ref{Fig:GRB}). The spectrum also shows evidence for positive residuals above $\sim8$ keV, which are connected to the hard X-ray component of emission revealed by NuSTAR and INTEGRAL (\S\ref{SubSec:NuSTAR}).

We triggered deep XMM observations of AT\,2018cow on 2018 July 23 ($\delta t\sim 36.5$ days, exposure time $\sim$32 ks, imaging mode, PI Margutti), in coordination with our NuSTAR monitoring. 
We reduced and analyzed the data of the European Photon Imaging Camera (EPIC)-pn data using standard routines in the Scientific Analysis System (SAS version 17.0.0) and the relative calibration files, and used MOS1 data as a validation check. After filtering data for high background contamination the net exposure times are 24.0 and 31.5 ks for the pn and MOS1, respectively. An X-ray source is clearly detected  
at the position of the optical transient. We extracted a spectrum from a circular region of 30\arcsec\ radius centered at the source position. Pile-up effects are negligible as we verified with the task \texttt{epatplot}. The background was extracted from a source-free region on the same chip. We estimate a 0.3--10 keV net count rate of $0.519 \pm 0.005$ c/s. The X-ray data were grouped to a minimum of 15 counts per bin.  
The 0.3--10 keV spectrum is well fitted by an absorbed power-law model with best-fitting $\Gamma=1.70\pm 0.02$ and marginal evidence for $\NHsub{int}\sim 0.02\times 10^{22}\,\rm{cm^{-2}}$ at the $3\,\sigma$ c.l.\ for $\NHsub{MW}=0.05\times 10^{22}\,\rm{cm^{-2}}$. Given that the uncertainty on \NHsub{MW} is also  $\sim 0.02\times 10^{22}\,\rm{cm^{-2}}$, we consider this value as an upper limit on \NHsub{int} at 36.5 days.

We acquired a second epoch of deep X-ray observations with XMM on 2018 September 6 ($\delta t\sim 82$ days, PI Margutti). The net exposure times are 30.5 ks and 36.8 ks, for the pn and MOS1, respectively. AT\,2018cow is clearly detected with net 0.3--10 keV count-rate $(6.0\pm0.7)\times 10^{-3}\,\rm{c\,s^{-1}}$. We used a source region of 20\arcsec\ to avoid contamination by a faint unrelated source located 36.8\arcsec south-west from our target (at earlier times AT\,2018cow is significantly brighter and the contamination is negligible).
The spectrum of AT\,2018cow is well fitted by a power-law model with $\Gamma=1.62 \pm0.20$ with unabsorbed 0.3--10 keV flux $\sim 2\times 10^{-14}\,\rm{erg\,cm^{-2}\,s^{-1}}$. 
We find no evidence for intrinsic neutral hydrogen absorption.  Finally we note that comparing the two XMM observations, we find no evidence for a shift of the X-ray centroid, from which we conclude that X-ray emission from the host galaxy nucleus, if present, is subdominant and does not represent a significant source of contamination.
The complete 0.3--10 keV X-ray light-curve of AT\,2018cow is shown in Fig.~\ref{Fig:Xray}.
\subsection{Hard X-rays: NuSTAR and INTEGRAL}
\label{SubSec:NuSTAR}

\begin{figure}
\includegraphics[scale=0.44]{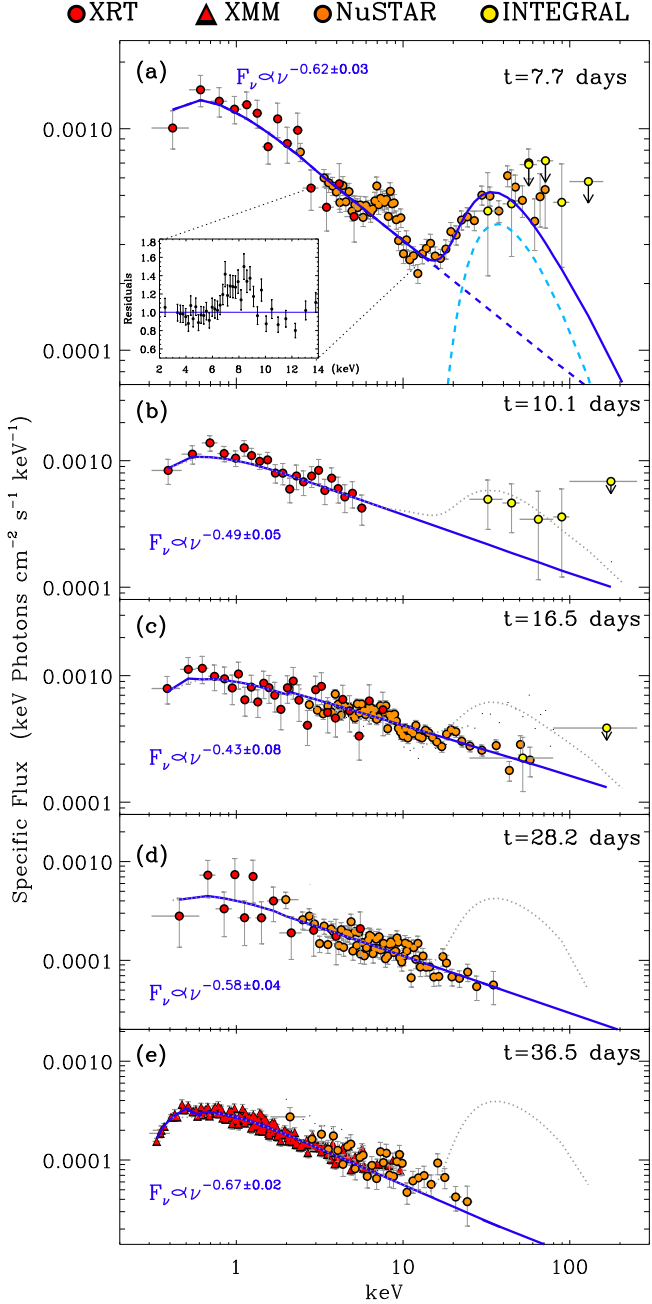}
\caption{Broad-band X-ray spectral evolution of AT\,2018cow. Coordinated observations of \emph{Swift}-XRT, XMM, INTEGRAL and NuSTAR revealed the presence of a hard X-ray emission component that dominates the spectrum above $\sim15$ keV at early times $\lesssim 15$ days (dashed light-blue line in panel (a)). The hard X-ray component later subsides. At $\delta t >15$ days the  broad-band X-ray spectrum is well described by an absorbed power-law with negligible intrinsic absorption (thick blue line in panels (b)-(d)). Black dotted line: early time hard X-ray ``bump''. Inset of panel (a): zoom-in into the region of positive residuals around 6-10 keV. }
\label{Fig:XraySpec}
\end{figure}

INTEGRAL started observing  AT\,2018cow on 2018 June 22 18:38:00 UT until July 8 04:50:00 UT ($\delta t\sim 6-22$ days)
as part of a public target of opportunity observation. The total on-source time is $\sim$900 ks (details are provided in Table~\ref{integral_tab}).  A source of hard X-rays is clearly detected at the location of AT\,2018cow at energies $\sim 30-100$ keV with significance 7.2 $\sigma$ at $\delta t\sim 6$ days. The source is no longer detected at $\delta t \gtrsim 24$ days (Fig.~\ref{Fig:XraySpec}). After reconstructing the incident photon energies with the latest available calibration files, 
we extracted the hard X-ray spectrum from the ISGRI detector \citep{isgri} on the IBIS instrument \citep{ibis} of INTEGRAL \citep{integral} for each of the $\sim$2\,ks long individual pointing of the telescope dithering around the source. We used the Off-line Scientific Analysis Software (OSA) with a sky model comprising only AT\,2018cow, which is the only significant source in the field of view. The energy binning was chosen to have 10 equally spaced logarithmic bins between 25 and 250 keV, the former being the current lower boundary of ISGRI energy window. We combined the spectra acquired in the same INTEGRAL orbit.  We use these spectra in \S\ref{SubSec:Xrayjoint} to perform a time-resolved broad-band X-ray spectral analysis of AT\,2018cow.

We acquired a detailed view of the hard X-ray properties of AT\,2018cow between 3--80 keV with a sequence of four NuSTAR observations obtained between 7.7 and 36.5 days (PI Margutti, Table \ref{Nustar_tab}). The NuSTAR observations were processed using \texttt{NuSTARDAS} v1.8.0 along with the NuSTAR CALDB released on 2018 March 12. We extracted source spectra and light curves for each epoch using the \texttt{nuproducts} FTOOL using a 30\arcsec\ extraction region centroided on the peak source emission. For the background spectra and light curves we extracted the data from a larger region ($\sim$85\arcsec) located on the same part of the focal plane. We produced response files (RMFs and ARFs) for each FPM and for each epoch using the standard \texttt{nuproducts} flags for a point source. 

AT\,2018cow is well detected at all epochs. The first NuSTAR spectrum at 7.7 days shows a clear deviation from a pure absorbed power-law model with $\Gamma\sim1.5$, and reveals instead the presence of a prominent  excess of emission above $\sim15$ keV, which matches the level of the emission captured by INTEGRAL, together with spectral features around 7--9 keV. By day 16.5 the hard X-ray bump of emission disappeared and the spectrum is well modeled by an absorbed power-law (Fig.~\ref{Fig:XraySpec}).
We model the evolution of the broad-band X-ray spectrum as detected by \emph{Swift}-XRT, XMM, NuSTAR, and INTEGRAL in \S\ref{SubSec:Xrayjoint}.

\subsection{Joint soft X-ray and hard X-ray spectral analysis}
\label{SubSec:Xrayjoint}
Our coordinated \emph{Swift}-XRT, XMM, NuSTAR and INTEGRAL monitoring of AT\,2018cow allows us to extract five epochs of broad-band X-ray spectroscopy ($\sim$0.3--100 keV) from 7.7 days to 36.5 days. We performed joint fits of data acquired around the same time, as detailed in Table \ref{Tab:specxray}. Our results are shown in Fig.~\ref{Fig:XraySpec}. We find that the soft X-rays at energies $\lesssim 7$ keV are always well described by an absorbed simple power-law model with photon index $\Gamma \approx$ 1.5--1.7 with no evidence for absorption from neutral material in addition to the Galactic value. Our most constraining limits from the time-resolved analysis are $\NHsub{int}<(0.03-0.04)\times 10^{22}\,\rm{cm^{-2}}$ (Table \ref{Tab:specxray}).

Remarkably, at $\sim$7.7 days, NuSTAR and INTEGRAL data at $E>15$ keV reveal the presence of a prominent component of emission of hard X-rays that dominates over the power-law component.\footnote{It is interesting to note in this respect the faint hard X-ray emission detected by \emph{Swift}-BAT in the first 15 days, with flux consistent with the NuSTAR observations, see Fig. 1 in \cite{Kuin18}.} We model the hard X-ray emission component with a strongly-absorbed cutoff power-law  model (light-blue dashed line in Fig.~\ref{Fig:XraySpec}, top panel). This is a purely phenomenological model that we use to quantify the observed properties of the hard emission component. A cutoff power-law is preferred to a simple power-law model, as a simple power-law would overpredict the highest energy data points at 7.7 days.
From this analysis the luminosity of the hard X-ray component at $\delta t\sim 7.7$ days is $L_{\rm x,hard}\sim 10^{43}\,\rm{erg\,s^{-1}}$ (20--200 keV).  A joint analysis of \emph{Swift}-XRT+INTEGRAL data at  $\delta t\sim 10.1$ days indicates that the component of hard X-ray emission became less prominent, and then disappeared below the level of the soft X-ray power-law  by $\delta t\sim 16.5$ days, as revealed by the coordinated \emph{Swift}-XRT, XMM and NuSTAR monitoring (Fig.~\ref{Fig:XraySpec}). We derive upper limits on the luminosity of the undetected hard X-ray emission component at $\delta t\ge 16.5$ days assuming a similar spectral shape to the one observed at $\delta t\sim 7-10$ days. 
As shown in Fig.~\ref{Fig:Xray}, the hard X-ray component fades quickly below the level of the power-law component that dominates the soft X-rays, which at this time evolves as $L_{\rm x}\propto t^{-1}$. The hard and soft X-ray emission components clearly show a distinct temporal evolution, suggesting that they originate from different emitting regions.  Table \ref{Tab:ene} lists the energy radiated by each component of emission.

We note that the first spectrum at 7.7 days shows positive residuals around $\sim$6--9 keV (Fig.~\ref{Fig:XraySpec}, inset). Typical spectral features observed in accretion disks (both around X-ray binaries and active galactic nuclei, AGNs) and interacting SNe are Fe K-alpha emission (between 6.4 keV and 6.97 keV depending on the ionization state) and the Fe K-band absorption edge. Typical interpretations of blueshifted iron line profiles include edge-on (or highly inclined) accretion disks or highly ionized absorption (e.g., \citealt{Reeves04}).  As a reference, interpreting the spectral feature detected in AT\,2018cow at $E\sim8$ keV with width $\sim1$ keV as Fe emission would require a blueshift corresponding to $v\sim0.1$\,c and Doppler broadening with similar velocity. We discuss possible physical implications  in \S\ref{SubSubSec:ComptonHump}.

\subsection{Radio: VLA and VLBA}
\label{SubSec:radio}
\begin{figure}
\includegraphics[scale=0.42]{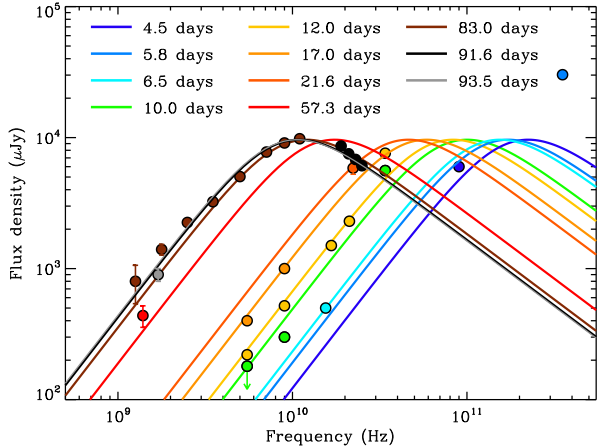}
\caption{The temporal evolution of the  radio spectrum of AT\,2018cow at $\nu<100$ GHz is well described by a smoothed broken power-law model with $F_{\nu}\propto \nu^{\beta}$ with $\beta\sim 1.7$ at $\nu<\nu_{br}$ and $\nu_{br}\propto t^{\alpha}$ with $\alpha\sim -1$. Above $\nu_{br}$ we find $F_{\nu}\propto \nu^{-1.2\pm0.1}$, consistent with the case of synchrotron emission from a population of fast-cooling electrons above the synchrotron self-absorption frequency $\nu_{sa}$ (\S\ref{SubSec:Radio}). In addition to our VLA and VLBA measurements at 21.6, 83.0 and 91.6 days, data have been collected from \citealt{deUgartePostigo18,Bright18,Smith18,Dobie+2018a, Dobie+2018b, Dobie+2018c, Nayana+2018, NayanaC2018, Horesh+2018,An18}. }
\label{Fig:RadioSED}
\end{figure}

\begin{figure}
\includegraphics[scale=0.9]{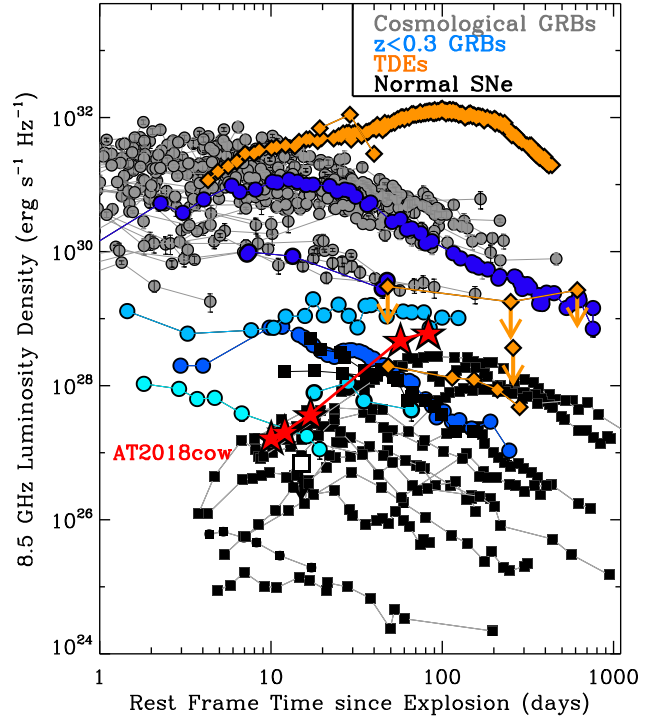}
\caption{Radio emission from AT\,2018cow (red stars), in the context of long GRBs (grey circles for GRBs at cosmological distances and shades of blue for GRBs in the local Universe),  TDEs (orange diamonds) and normal H-stripped core-collapse SNe including Ic-BL (black squares). The empty square marks the position of the extremely rapidly declining Ic-SN\,2005ek, which was not detected in the radio. 
The temporal evolution and  luminosity of AT\,2018cow is comparable to those of the most luminous normal SNe. References: \citet{Berger12,Cenko12,Chomiuk12,Chandra12,Zauderer13,Drout+2013,Chornock14,Margutti100316D,Nicholl16,Alexander16,Margutti1715lh,Eftekhari18}.}
\label{Fig:RadioALL}
\end{figure}


We observed AT\,2018cow with the Karl G. Jansky Very Large Array (VLA) on 2018 September 6 UT at $4-8$ GHz, on 2018 September 7 UT at $1-2$ GHz, $2-4$ GHz, and $8-12$ GHz, and on 2018 September 16 at $18-24$ GHz. The data were taken in the VLA's D configuration under program VLA/18A-123 (PI Coppejans). We reduced the data using the {\tt pwkit} package \citep{pwkit}, using 3C~286 as the bandpass calibrator and \objectname[]{VCS1 J1609+2641} as the phase calibrator. We imaged the data using standard routines in CASA \citep{casa} and determined the flux density of the source at each frequency by fitting a point source model using the {\tt imtool} package within {\tt pwkit}. This package uses a Levenberg-Marquardt least-squares optimizer to fit a small region in the image plane centered on the source coordinates with an elliptical Gaussian corresponding to the CLEAN beam. These data are shown together with the rest of our radio observations in Figure \ref{Fig:RadioSED} and Table \ref{Tab:Radio}.

We also obtained 22.3-GHz VLBI observations of AT\,2018cow with the High
Sensitivity Array of the National Radio Astronomy Observatory (NRAO)
on 2018 July 7 \citep{Bietenholz+2018}. The array consisted of the NRAO Very Long Baseline
Array with the exception of the North Liberty station ($9 \times 25$~m
diameter), and the Effelsberg antenna (100~m diameter).  We recorded both senses
of circular polarization at a total bit rate of 2048 Mbit~s$^{-1}$.
The observations were phase-referenced to the nearby compact source
\objectname[]{QSO J1619+2247}, with a cycle time of $\sim$100~s.  The
amplitude gains were calibrated using the system temperature
measurements made by the VLBA online system, and refined by
self-calibration on the calibrator sources, with the gains normalized
to a mean amplitude of unity to preserve the flux-density scale as
well as possible.  We will report on the VLBI results in more detail
in a future paper, but we include the total flux density observed with
VLBI here.  On 2018 July 7.96 (UT) we found $5850 \pm 610 \; \mu$Jy at
22.3~GHz.  The value was obtained by fitting a circular Gaussian
directly to the visibilities by least-squares.  As the source is not resolved the
nature of the model does not affect the flux density, and a value well
within our stated uncertainties is obtained if for example a circular disk model is used.  The array has good $u$-$v$ coverage down to baselines
of length $< 30 \; {\rm M}\lambda$,
therefore only flux density on angular scales $>8$~mas would be resolved out. At this epoch, the projected angular size of AT\,2018cow is $<8$~mas even in the case
of relativistic expansion, which implies that our measure of the radio flux density from AT\,2018cow is reliable. 
The uncertainty is the statistical
one with a 10\% systematic one added in quadrature.
Although at our observing frequency of
22.3~GHz some correlation losses might be expected, the visibility
phases were consistent from scan to scan, and we do not expect
correlation losses larger than our stated uncertainty.

We place our VLBA and VLA measurements into the context of radio observations from the literature \citep{deUgartePostigo18,Bright18,Smith18,Dobie+2018a, Dobie+2018b, Dobie+2018c, NayanaC2018, Horesh+2018,An18}. We find that at any given epoch, at $\nu<$ 100 GHz, the data are well described by an optically thick spectrum $F_{\nu}\propto \nu^{\beta}$ with $\beta\sim 1.7$ at $\nu<\nu_{br}$.
Similar to that of radio SNe (e.g., \citealt{Chevalier98, Soderberg05,Soderberg12}), the temporal evolution of the radio spectrum at $\nu<100$ GHz is well described by a broken power-law model with spectral break frequency $\nu_{br}(t)\propto t^{-1}$ above which the spectrum becomes optically thin (Fig.~\ref{Fig:RadioSED}). In Fig.~\ref{Fig:RadioSED} we show the best-fitting model in the case of constant spectral peak flux (which corresponds to a freely expanding blastwave). This model underpredicts the mm-wavelength data point at $\delta t=5.8$ days, which might be evidence for a slightly decelerating blastwave or the presence of an additional emission component, as discussed in \S\ref{SubSec:Radio}.  

The radio luminosity and temporal behavior of AT\,2018cow at $\sim9$ GHz are similar to those of the most luminous normal SNe, while AT\,2018cow is significantly less luminous than cosmological GRBs. The still rising light-curve at $\delta t\sim 80$ days makes it also distinct from low-energy GRBs in the local Universe (Fig.~\ref{Fig:RadioALL}). The radio flux-density measurements of AT\,2018cow are presented in Table \ref{Tab:Radio}.

\subsection{Search for prompt $\gamma$-rays with the IPN}
\label{SubSec:IPN}
The large X-ray luminosity of AT\,2018cow initially suggested a connection with long GRBs. Thus motivated, we searched for bursts of prompt $\gamma$-ray emission between the time of the last optical non-detection and the first optical detection of AT\,2018cow (i.e., between 2018 June 15 03:08 UT and June 16 10:35 UT, \citealt{Prentice+2018}). During this time interval one burst was detected on 2018-06-15 11:05:56 UT by the spacecraft of the InterPlanetary Network (IPN:  Mars Odyssey,
Konus-Wind, INTEGRAL SPI-ACS, Swift-BAT, and Fermi GBM). The burst localization by the IPN, INTEGRAL and the GBM however excludes at high confidence the location of AT\,2018cow, from which we conclude that there is no evidence for a burst of $\gamma$-rays associated with AT\,2018cow down to the IPN threshold (i.e., 10 keV - 10 MeV 3-s peak flux   $<3 \times 10^{-7}\,\rm{erg\,cm^{-2}\, s^{-1}}$ for a typical long GRB spectrum with Band parameters $\alpha=-1, \; \beta$=$-2.5$, and $E_p$=300 keV, e.g., \citealt{Band93}). For the time interval of interest the IPN duty cycle was $\sim$97\%. For AT\,2018cow the IPN thus rules out at $97\%$ c.l.\ bursts of $\gamma$-rays with peak luminosity $>10^{47}\,\rm{erg\,s^{-1}}$, which is the level of the lowest-luminosity GRBs detected \citep[e.g.,][]{Nava12}. 

We now examine which limits we can place on the probability of detection of weaker bursts, which would only trigger Fermi-GBM and/or Swift-BAT in the same time interval considerd above. Taking Earth-blocking and duty cycle into account, the joint non-detection probability by Fermi-GBM (8 - 1000 keV fluence limit of $\sim4 \times 10^{-8}\,\rm{erg\,cm^{-2}}$) and Swift-BAT (15-150 keV fluence limit of $\sim6 \times 10^{-9}\rm{erg\,cm^{-2}}$ based on the weakest burst detected within the coded field of view, FOV) is $\sim 29\%$. The non-detection probability of weaker bursts by Swift-BAT is $\sim71\%$ (within the coded FOV) and $\sim46\%$ (outside the coded FOV).

\subsection{Bolometric Emission and Radiated Energy}
\label{Subsec:Bol}
\begin{figure}
\hskip -0.8 cm
\includegraphics[scale=0.53]{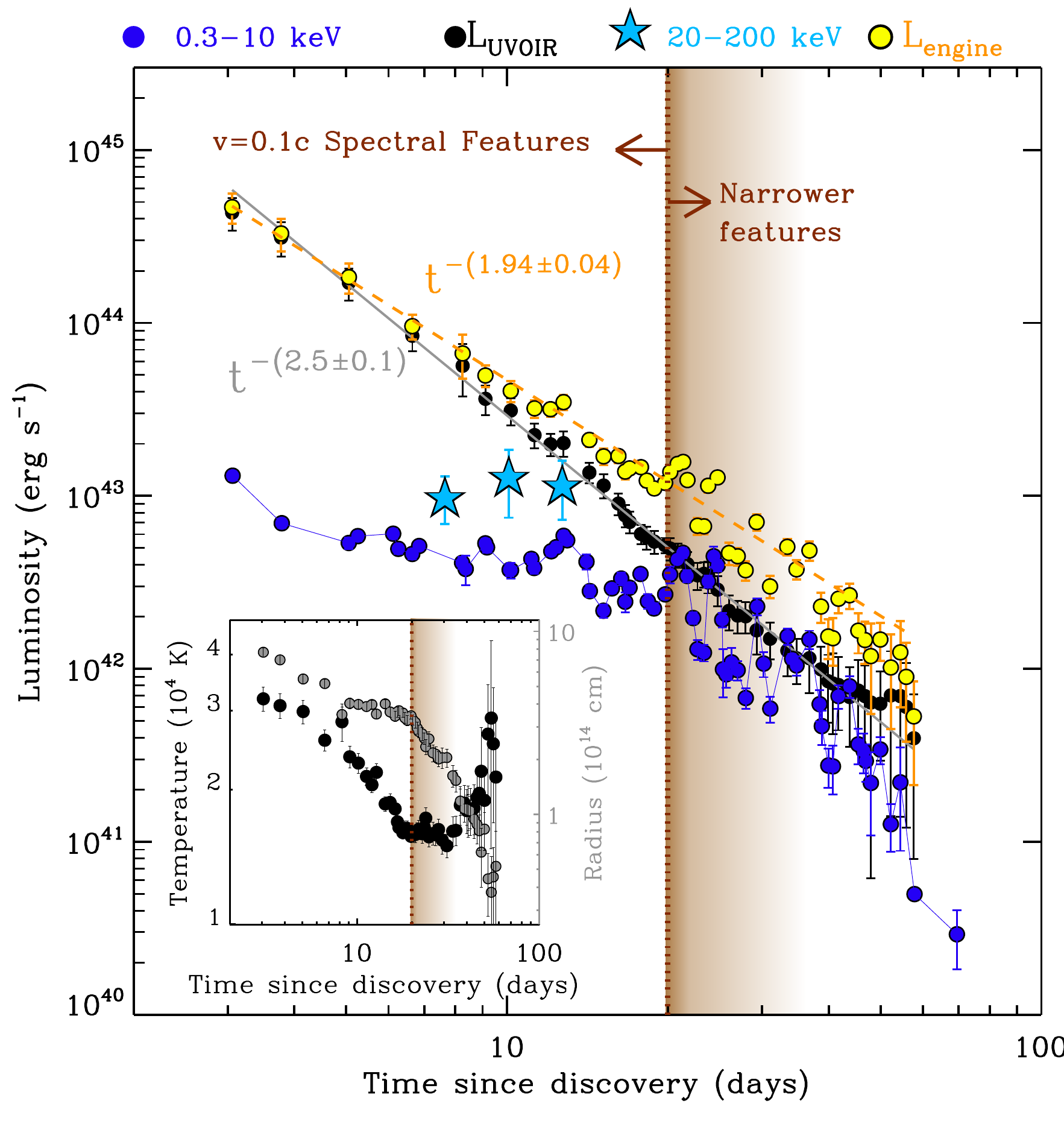}
\caption{Luminosity evolution of different components of emission: soft X-ray $0.3-10$ keV ($L_{\rm X}$, blue filled circles), hard X-ray bump of emission in the 20-200 keV band (light-blue stars), optical bolometric luminosity estimated from a black body fit to the UV/optical photometry ($L_{\rm{UVOIR}}\propto t^{-2.5\pm0.1}$, black circles). Yellow circles: ``engine'' luminosity ($L_{\rm engine}=L_{\rm X}+L_{\rm{UVOIR}}\propto t^{-1.94\pm0.04}$, where $L_{\rm X}$ has been integrated between 0.3-50 keV). Inset: evolution of the best-fitting blackbody temperature and radius. Brown shaded areas mark the approximate time of onset of narrower spectral features in the optical spectra. Interestingly, $\delta t\sim 20$ days marks an important time in the evolution of AT\,2018cow:  $L_{\rm X}\approx L_{\rm UVOIR}$; the rate of decay of $L_{\rm{X}}$ increases and the X-ray variability becomes more prominent; the blackbody temperature plateaus with  no evidence for cooling at $\delta t>20$ days, while the radius decreases at a faster rate; notably, around this time, broader H and He spectral features emerge in the optical spectra (Fig.~\ref{Fig:OpticalSpec}), while the hard X-ray spectral ``hump'' completely disappears.  }
\label{Fig:Lengine}
\end{figure}

\begin{deluxetable}{rrr}
\tablecolumns{3}
\tablewidth{0pc}
\tablecaption{Energy radiated by AT\,2018cow at $3$$<$$\delta t$$<$$60$ days. \label{Tab:ene}}
\tablehead{
\colhead{Component} & \colhead{Band} & \colhead{Radiated Energy (erg)}  }
\startdata
Power-law & 0.3-10 keV & $9.8^{+0.2}_{-0.1}\times 10^{48}$  \\
Power-law & 0.3-50 keV & $2.5^{+0.4}_{-0.3}\times 10^{49}$ \\
Hard X-ray bump & 20-200 keV & $\sim10^{49}$\\
Blackbody & UVOIR & $1.0^{+0.2}_{-0.2}\times 10^{50}$ \\
Non-thermal$^{a}$ & UVOIR  & $\sim 5\times 10^{48}$\\
\hline
Total & &$\sim 1.4\times 10^{50}\rm{erg}$\\
\enddata
\tablecomments{$^{a}$ Based on the analysis from \citet{Perley18}.}
\end{deluxetable}

Performing a self-consistent flux calibration of the UVOT photometry, and applying a dynamical count-to flux conversion that accounts for the extremely blue colors of the transient, we find that the UV+UBV  emission from AT\,2018cow is well modeled by a blackbody function at all times. We infer an initial temperature $T_{bb}\sim30000$ K and radius $R_{\rm bb}\sim8\times 10^{14}$~cm, consistent with \citet{Perley18}. $R_{\rm bb}$ and $T_{\rm bb}$ show a peculiar temporal evolution, with $R_{\rm bb}$ monotonically decreasing with time (with a clear steepening around 20 days), while the temperature plateaus at $\sim$15000 K, with no evidence for cooling at $\delta t>20$ days (Fig.~\ref{Fig:Lengine}). Indeed, $\delta t\sim 20$ days marks an important transition in the evolution  of AT\,2018cow: H and He features emerge in the spectra; the hard X-ray hump disappears; $L_{\rm X}$ approaches the level of the optical emission and later starts a steeper decline, while the soft X-ray variability becomes more pronounced with respect to the continuum.

After optical peak, we find that the resulting UV/optical bolometric emission is well modeled by a power-law decay $t^{-\alpha}$ with best-fitting $\alpha=2.50 \pm 0.06$ (Fig.~\ref{Fig:Lengine}), in agreement with \citet{Perley18}.  Also consistent with \citet{Perley18}, we find that the $R$, $I$, and NIR data from AT\,2018cow are in clear excess to the thermal blackbody emission and represent a different component.  In Fig.~\ref{Fig:Lengine} we show that the combined energy release of the thermal UV/optical emission and the soft X-rays (0.3-50 keV) follows a decay $\propto t^{-\alpha}$ with $\alpha=1.94 \pm 0.04$. This result is relevant if the thermal optical/UV and the soft X-rays are manifestations of the same physical component, like energy release from a central engine  (\S\ref{SubSubSec:engine}). 
Table \ref{Tab:ene} lists the energy radiated by each component of emission.
\subsection{Temporal Variability Analysis}
\label{Subsec:Timing}
We examined the 0.3--10~keV emission at $\delta t=3.5-55$ days for evidence of periodicity 
using the Lomb-Scargle periodogram (LSP; \citealt{Lomb76,Scargle82}), and the Fast $\chi^2$ algorithm\footnote{\url{http://public.lanl.gov/palmer/fastchi.html}} by \citet{Palmer09}, both of which are suitable for unevenly spaced series. As a first step we removed the overall trend of the time series, which was found to be best modeled by a simple exponential $\exp{(k-t/t^*)}$, with $k=-0.82\pm0.08$ and $t^*=14.7\pm0.9$~d. 
We applied the LSP and the Fast $\chi^2$  techniques to the resulting residuals.

\begin{figure}
\includegraphics[scale=0.6]{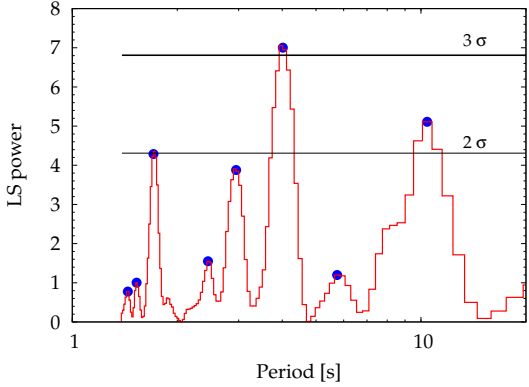}
\caption{Lomb-Scargle periodogram for the 0.3--10 keV  X-ray time series. Horizontal lines correspond to the significance levels (Gaussian). The most significant peak (3$\sigma$) is found on a period of $T_1=4.0$~d.  Two other peaks above 2$\sigma$ are at $T_2=10.4\pm2.1$~d ($2.6\sigma$), and one at $T_3=1.7\pm0.1$~d ($2.0\sigma$), respectively.}
\label{Fig:LSP}
\end{figure}

We calculated the LSP using the Numerical Recipes implementation \citep{NRC}, exploring the frequency range $0.005-0.65$~d$^{-1}$, corresponding to $1/(4T)$ and to an average Nyquist frequency, respectively, where $T$ is the total duration. To assess the significance of the peaks we detected, one must consider two issues: (i) the presence of red noise; (ii) the number of independent frequencies \citep[e.g.,][]{HorneBaliunas86}. We addressed (i) through a number of Monte Carlo simulations. We generated $5\times10^3$ time series with the same sequence of observing times, $t_i$, every time shuffling the observed count rates and associated uncertainties, so as to keep the same rate distribution and the same variance. For each of the simulated series we calculated the LSP under the same prescriptions as for the real one. We addressed (ii) through the identification of the peaks in all of the LSPs (both the true one and the ones from the shuffled data) 
by means of the peak-search algorithm {\sc mepsa} \citep{Guidorzi15}: given that only separate peaks are identified as independent structures, this properly accounts for the power associated with correlated frequencies. The significance of the peaks found in the real LSP were then compared against the distribution of peaks in the LSPs from the shuffled data. The result is shown in Figure~\ref{Fig:LSP}.

We further apply the Fast $\chi^2$ algorithm to detect periodic harmonic signals within unevenly spaced data affected by variable uncertainties. For a given number of harmonics and any given trial frequency, the Fast $\chi^2$ algorithm determines the solution that minimizes the $\chi^2$. This way, for a given number of harmonics, the best fundamental frequency along with its harmonics is given by the global minimum $\chi^2$. We applied the method to the observed X-ray de-trended light curve, each time allowing for one to six harmonics, as a trade-off between the need of providing a relatively simple modeling and the possibility of a rather complex periodic signal involving several harmonics. The explored frequencies are in the range $0.05-1$~d$^{-1}$. To assess the significance of our results we applied the Fast $\chi^2$ to the sample of synthetic light-curves and compared the results from the real time series. We find significant power ($\sim 3\sigma$ Gaussian) on a modulation timescale of $\sim4$~d in the first 40 days, consistent with the results from the LSP. We  note that a similar variability time-scale was also independently reported by \cite{Kuin18}.

Finally, we investigate whether there is correlated temporal variability between the X-ray and 
the UV/optical in AT\,2018cow. 
We fit a third-order polynomial to the soft X-ray, $w1$, $w2$ and $m2$ light-curves in the log-log space to remove the overall temporal decay trend. Our time series  consists of the ratios of the observed fluxes over the best-fitting ``continuum'', where uncertainties have been propagated following standard practice. We find that all the UV light-curves show a high degree of correlation with $P-$ values $<0.01$\% for either the Spearman Rank test or the Kendall Tau test. We also find a hint for correlated behavior between the UV-bands and the X-rays with limited significance corresponding to $P\gtrsim 5$\% (Spearman Rank test). The correlation is stronger at $\delta t<30$ days.

\begin{figure*}
\includegraphics[scale=0.33]{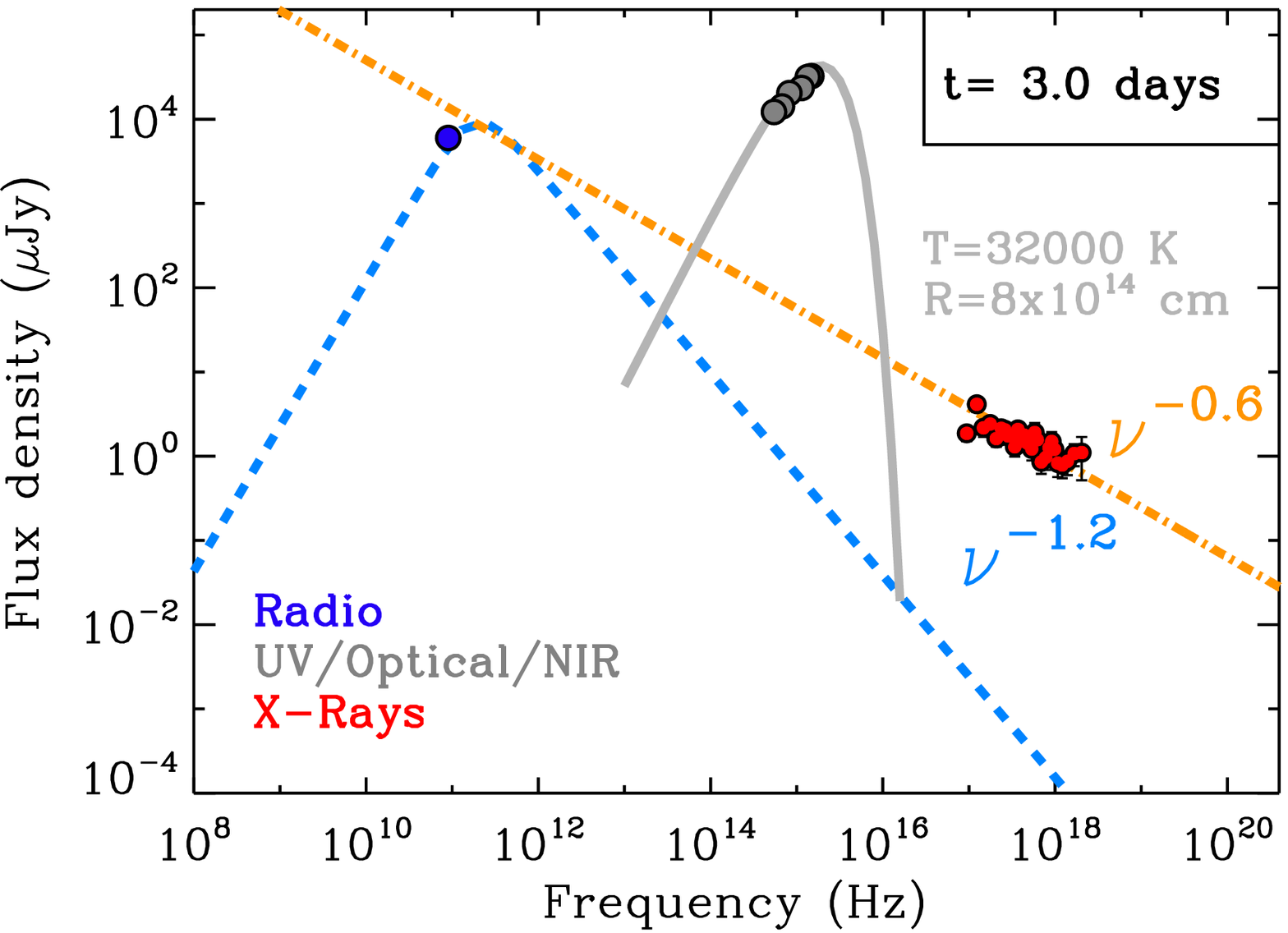}
\includegraphics[scale=0.33]{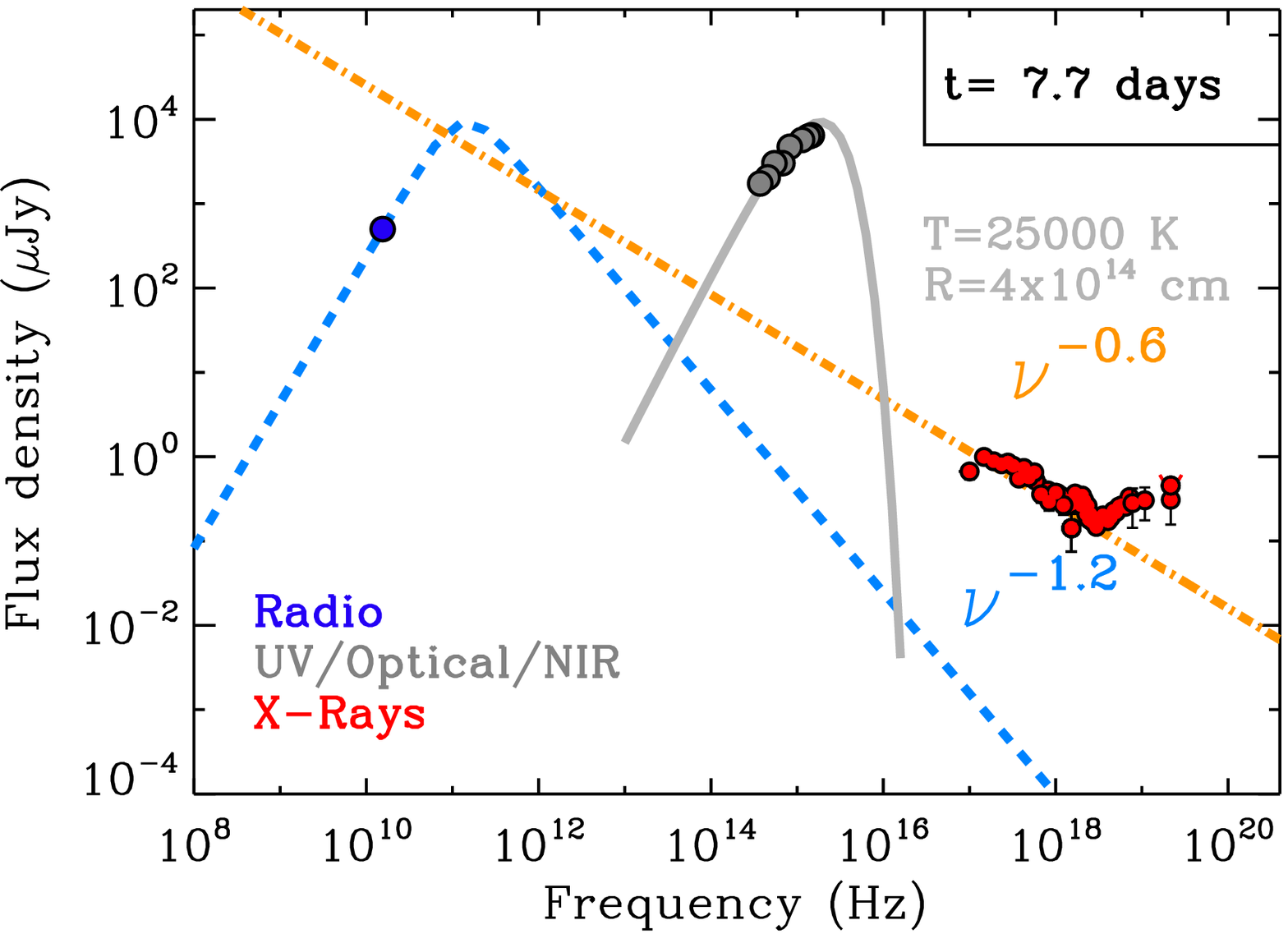}
\includegraphics[scale=0.33]{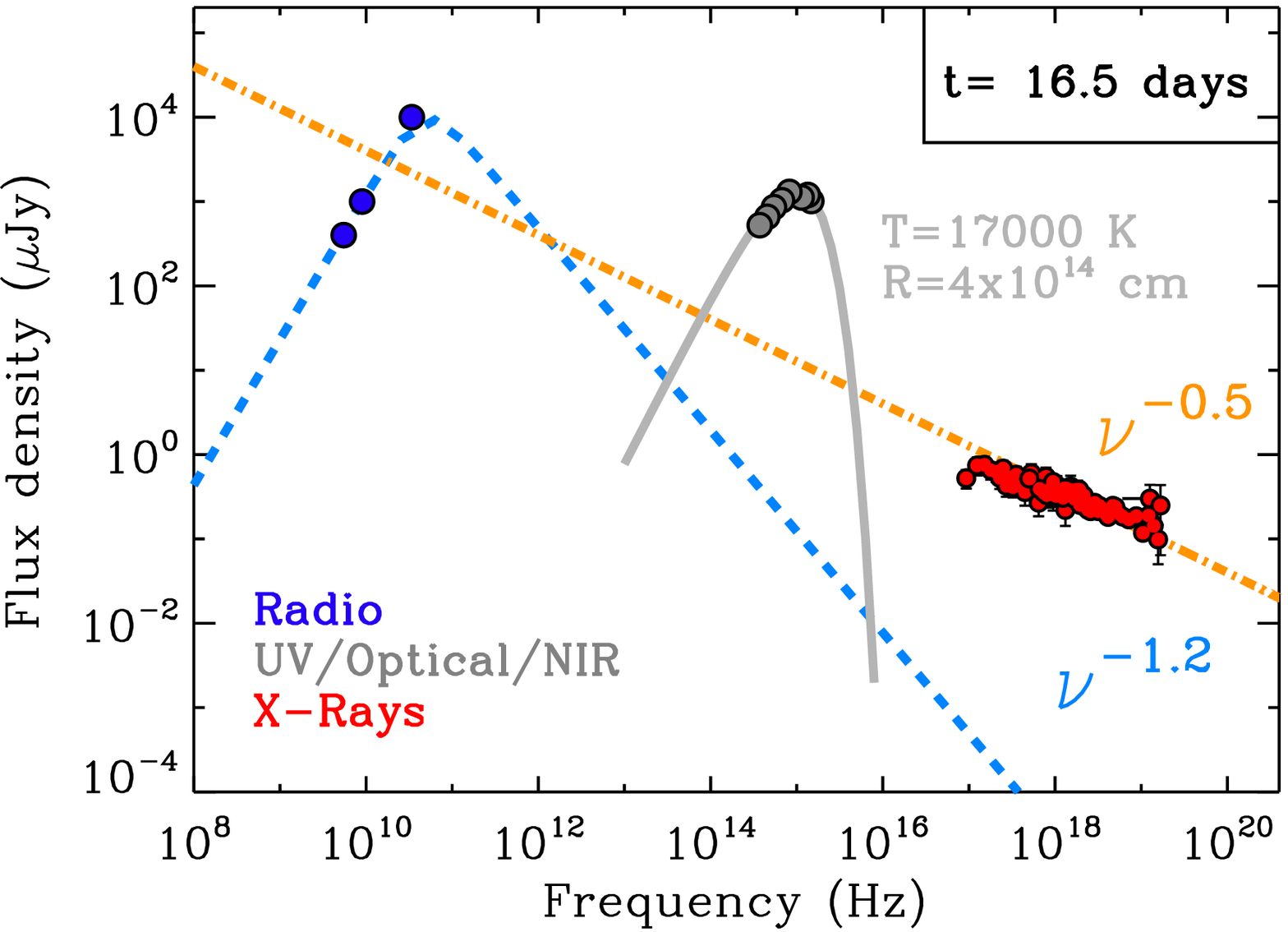}
\center{\includegraphics[scale=0.33]{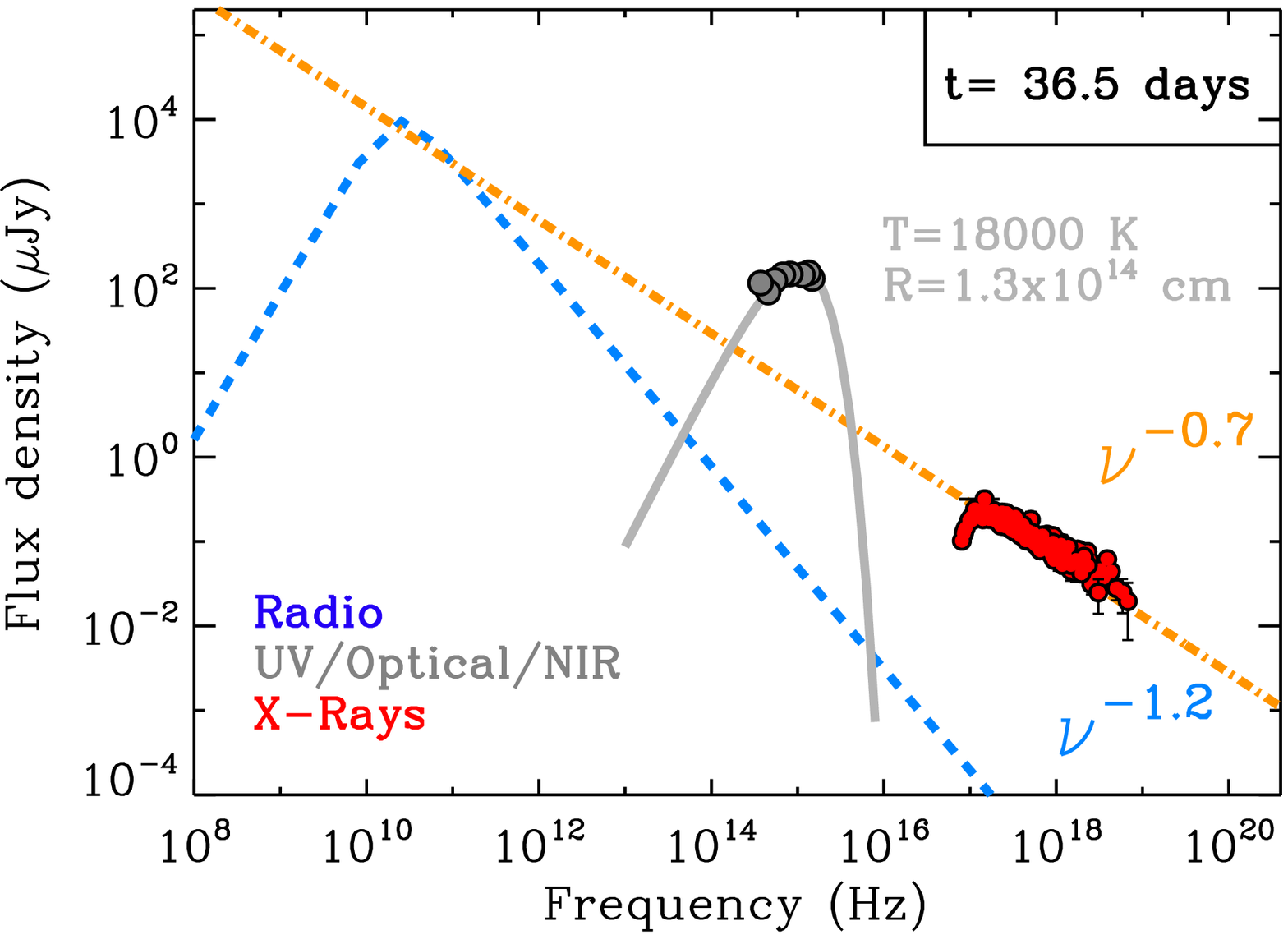}
\includegraphics[scale=0.33]{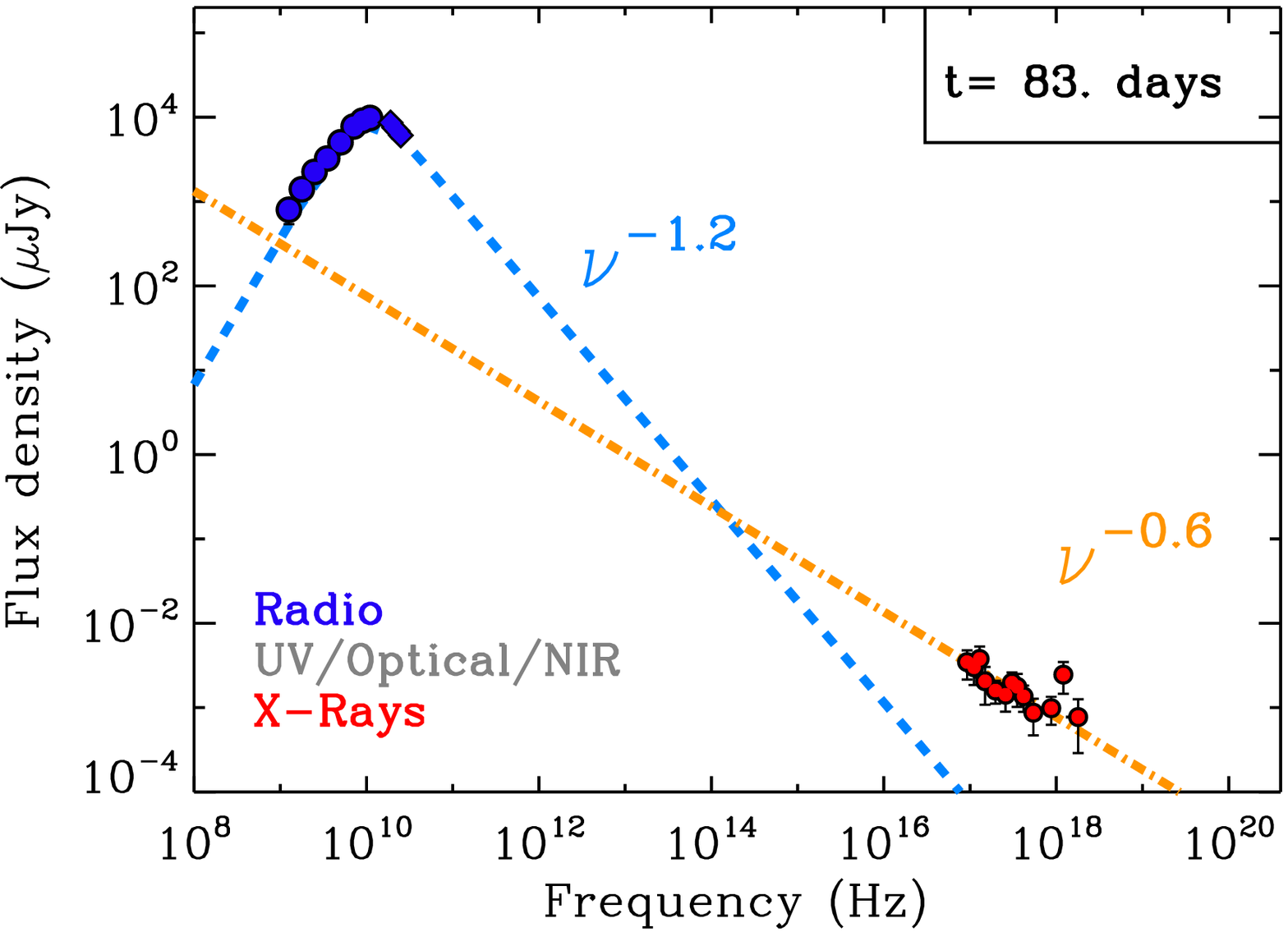}}
\caption{Evolution of the broad-band radio-to-hard-Xray spectrum of AT\,2018cow from around the time of optical peak at 3 days to the time of the last X-ray detection at 83 days. The presence of the excess of emission at hard X-rays is clearly apparent in the 7.7-day SED. Blue, gray and red filled circles mark radio, optical and X-ray observations, respectively. The radio data shown in the SEDs at $\delta t=3$, 7.7 and 16.5 days are from \cite{deUgartePostigo+2018,Bright18,Dobie+2018a}. Orange dot-dashed line: best fitting power-law model from \S\ref{SubSec:Xrayjoint}.  Blue dashed line: best fitting radio model from \S\ref{SubSec:radio}. Gray thick line: best fitting black body model from \S\ref{Subsec:Bol}. The X-ray, optical, and radio originate from three distinct emission components with different temporal and spectral evolution. }
\label{Fig:superSED}
\end{figure*}

\section{Multi-band inferences}
\label{Sec:inferences}
In this section we discuss basic inferences on the physical properties of AT\,2018cow based  on the information provided by each part of the electromagnetic spectrum individually, before synthesizing the information and speculate on the intrinsic nature of AT\,2018cow in \S\ref{Sec:interpretation}.
\subsection{Thermal UV-Optical Emission}
\label{SubSec:optical}
The key observational results are: (i) A very short rise time to peak, $t_{\rm rise}\sim$ few days \citep{Perley18,Prentice+2018}. (ii) Large bolometric peak luminosity, $L_{\rm pk,bol}\sim4\times 10^{44}\,\rm{erg\,s^{-1}}$, significantly more luminous than normal SNe and  more luminous than some SLSNe (Fig.~\ref{Fig:Optical}). (iii) Persistent blue colors, with lack of evidence for cooling at $\delta t\gtrsim30$ days (the effective temperature remains $>15000$ K). (iv) Large blackbody radius $R_{\rm bb}\sim 8\times 10^{14}$~cm inferred at $\delta t\sim 3$ days (Fig.~\ref{Fig:Lengine}). (v) Persistent optically thick UV/optical emission with no evidence for transition into a nebular phase at $\delta t<90$ days (Fig.~\ref{Fig:OpticalSpec}). (vi) The spectra evolve from a hot, blue, and featureless continuum around the optical peak, to very broad features with $v\sim0.1\,c$ at $\delta t\sim 4-15$ days
(Fig.~\ref{Fig:OpticalSpec}).  (vii) Redshifted H and He features emerge at $\delta t>20$ days with significantly lower velocities  $v\sim 4000$~\kms\ (Fig.~\ref{Fig:OpticalSpec}), implying an abrupt change of the velocity of the material which dominates the emission.
 The centroid of the line emission is offset to the red with $v\sim1000$~\kms\ (Fig.~\ref{Fig:OpticalSpec}).  (viii) There is evidence for a NIR excess of emission with respect to a blackbody model from early to late times, as pointed out by \citealt{Perley18} (Fig.~\ref{Fig:superSED}). 
\subsubsection{Engine-Powered Transient}
\label{SubSubSec:engine}
For optical/UV emission powered by the diffusion of thermal radiation from an initially compact opaque source the light curve rises and peaks on the  diffusion timescale \citep{Arnett82}:
\begin{equation}
t_{\rm pk} \approx \left(\frac{M_{\rm ej}\kappa}{4\pi v_{\rm ej}c}\right)^{1/2} \approx 2.7\,\,{\rm d}\left(\frac{M_{\rm ej}}{0.3M_{\odot}}\right)^{1/2}\left(\frac{v_{\rm ej}}{0.1c}\right)^{-1/2},
\label{eq:tdiff}
\end{equation}
where we use $\kappa = 0.1$ cm$^{2}$ g$^{-1}$ as an estimate of the effective opacity due to electron scattering or Doppler-broadened atomic lines.  For  $t_{\rm pk} \sim 2-4$ d (Fig.~\ref{Fig:Optical}) Eq.\ \ref{eq:tdiff} implies a low ejecta mass $M_{\rm ej} \sim 0.1-0.5 M_{\odot}$ and high ejecta velocity $v_{\rm ej} \sim 0.05-0.1 c$ to match the inferred blackbody radius at early times, corresponding to a kinetic energy of the optical/UV emitting material $E_{\rm k} \sim 10^{50.5}-10^{51.5}$ erg 
(consistent with the inferences by \citealt{Prentice+2018} and \citealt{Perley18}).  The low ejecta mass immediately excludes light curve models powered by $^{56}$Ni decay, which would require $M_{\rm{Ni}}>6\,M_{\sun}$ to reproduce the large peak luminosity of  AT\,2018cow, and instead demand another energy source. We note that $M_{\rm ej} \sim 0.1-0.5 M_{\odot}$ should be viewed as a constraint on the fast-moving ejecta mass that participates in the production of the high-luminosity peak. As we will see in the next sections, the phenomenology of AT\,2018cow  requires the presence of additional, slower material preferentially distributed in an equatorial belt. 

One potential central energy source is an ``engine'', such as a millisecond magnetar or accreting black hole, which releases a total energy $E_{\rm e}$ over its characteristic lifetime, $t_{\rm e}$.  The engine deposits energy into a nebula behind the ejecta at a rate:
\be
L_{\rm e}(t) = \frac{E_{\rm e}}{t_{\rm e}}\frac{(\alpha-1)}{(1 + t/t_{\rm e})^{\alpha}},
\label{eq:Le}
\ee
where $\alpha = 2$ for an isolated magnetar \citep{Spitkovsky06}, $\alpha = 2.38$ for an accreting magnetar \citep{Metzger18mag}, $\alpha = 5/3$ for fall-back in a TDE \citep[e.g.,][]{Rees88,Phinney89}, and $\alpha < 5/3$ in some supernova fall-back models \citep[e.g.,][]{Coughlin18} or viscously-spreading disk accretion scenarios \citep[e.g.,][]{Cannizzo89}.  In engine models, the late-time decay of the bolometric luminosity obeys $L_{\rm opt} \propto L_{\rm e} \propto t^{-\alpha}$, such that the measured value of $\alpha \approx 2-2.5$  (\S\ref{Subsec:Bol}, Fig.~\ref{Fig:Lengine}) would be consistent with a magnetar engine.\footnote{A precise measurement of the late-time optical decay will only be possible after AT\,2018cow has faded away (allowing us to accurately remove the host-galaxy contribution). Here we note that steeper decays of $L_{opt}$ would also be consistent with a magnetar engine shining through low ejecta mass, which become ``transparent'' and incapable of retaining and thermalizing the engine energy. A similar scenario was recently invoked by \cite{Nicholl18mag} to explain the rapid decay of the SLSN 2015bn at late times.} 
However, given the uncertainties in the bolometric correction, the TDE/supernova fall-back case ($\alpha = 5/3$) may also be allowed.  Finally, the ``engine'' may not be a compact object at all, but rather a deeply-embedded radiative shock, produced as the ejecta interacts with a dense medium.

As a concrete example, consider an isolated magnetar ($\alpha = 2$), which at times $t \gg t_{\rm e}$ obeys $L_{\rm e} \approx (E_{\rm e}t_{\rm e})/t^{2}$.  Assuming that most of the energy is released over $t_{\rm e} \ll t_{\rm pk}$ (as justified by the narrowly-peaked light curve shape) and that most of the engine energy is not radiated, but instead used to accelerate the ejecta to its final velocity $v_{\rm ej}$, then $E_{\rm e} \lesssim M_{\rm ej}v_{\rm ej}^2/2$, with $E_{\rm rad} \approx \int_{\rm t_{\rm pk}}^{\infty} L_{\rm e}dt \approx \frac{E_{\rm e}t_{\rm e}}{t_{\rm pk}}$, from which it follows that: 
\be
t_{\rm e} \gtrsim \frac{2E_{\rm rad}}{M_{\rm ej}v_{\rm ej}^{2}}t_{\rm pk} \sim 3\times 10^{3}{\rm s} \left(\frac{v_{\rm ej}}{0.1c}\right)^{-2}\left(\frac{M_{\rm ej}}{0.3M_{\odot}}\right)^{-1}.
\ee
Here we used the total radiated UVOIR energy, $E_{\rm rad} \sim  10^{50}$ erg from Table \ref{Tab:ene}. The engine is thus relatively constrained: it must release a total energy $E_{\rm e}$ that varies from $E_{\rm{rad}}\sim10^{50}\,\rm{erg}$ to $E_{\rm k} \sim 10^{50.5}-10^{51.5}$ erg, most of it over a characteristic lifetime $t_{\rm e} \sim 10^{3}-10^{5}$ s.

We conclude with constraints on the properties of a $\rm{Ni}$-powered transient that might be hiding within the central-engine dominated emission.  In the context of the standard \citet{Arnett82} modeling, modified following \citet{Valenti08}, and assuming a standard SN-like explosion with $M_{\rm ej}\sim$ a few $M_{\sun}$ and $E_{\rm k}\sim 10^{51-52}$~erg, we place a limit $M_{\rm{Ni}}<0.06\,\rm{M_{\sun}}$ not to overproduce the observed optical-UV thermal emission, consistent with \citet{Perley18}. The limit becomes significantly less constraining $M_{\rm{Ni}}<0.2-0.4\,\rm{M_{\sun}}$ if we allow for smaller $M_{\rm ej}\sim 0.5\,\rm{M_{\sun}}$, as in this case the diffusion time scale is shorter (i.e., the transient emission peaks earlier), the $\gamma$-rays from the $^{56}\rm{Ni}$ decay are less efficiently trapped and thermalized within the ejecta, and the transient enters the nebular phase earlier. However, the large $M_{\rm{Ni}}/M_{\rm ej}$ would result in red colors as the UV emission would be heavily suppressed via iron line blanketing, which is not observed. The observed blue colors (Fig.~\ref{Fig:Optical}) indicate instead that emission from a Ni-powered transient with small ejecta is never dominant, which implies $M_{\rm{Ni}}\lesssim 0.1\,\rm{M_{\sun}}$. 
\subsubsection{Shock Break-Out}
\label{SubSubSec:SBO}
Alternatively, we consider the possibility that the high luminosity and rapid-evolution of AT\,2018cow result from a shock break-out from a radially-extended progenitor star, an inflated progenitor star or thick medium (i.e., if the star experiences enhanced mass loss just before stellar death). Shock break-out scenarios have been invoked to explain some fast-rising optical transients \citep[e.g.,][]{Ofek10,Drout14,Shivvers16,Arcavi16,Tanaka16}. 

For a typical SN shock velocity  $v_{\rm sh} \approx 10^{4}$ \kms, and the observed peak time of AT\,2018cow the inferred stellar radius is $R_{\star} \approx v_{\rm sh}t_{\rm pk} \gtrsim 10^{14}$ cm $\sim 10$ AU, much larger than red supergiant stars.  Furthermore, the explosion of such a massive star is expected to be followed by a longer plateau phase not observed in the monotonically-declining light curve of AT\,2018cow. We conclude that shock break-out from a stellar progenitor is not a viable mechanism for AT\,2018cow.

The {\em effective} radius of a massive star could be increased just prior to its explosion by envelope inflation or enhanced mass loss timed with stellar death, as observed in a variety of SNe \citep[e.g.,][]{Smith14}.  Assuming an external medium with a wind-like density profile $\rho_{w} = \dot{M}_{w}/(4\pi v_{w}r^{2}) = A/r^{2}$ and radial optical depth $\tau_w = \int_{r}^{\infty}\rho_{\rm w}\kappa dr \simeq \rho_{w}r \kappa$, the photon diffusion timescale is:  
\be
t_{\rm pk,w} \sim \tau_w \frac{r}{c} \approx \frac{A\kappa}{c} \approx 1.9 \,{\rm d}\,\left(\frac{A}{10^{5}A_{\star}}\right),
\label{eq:trisew}
\ee
where $A_{\star} = 5\times 10^{11}$ g cm$^{-1}$ (i.e. $A=A_{\star}$ for the standard mass-loss rate $\dot{M}_{w} = 10^{-5}M_{\odot}$ yr$^{-1}$ and wind velocity $v_w = 10^{3}$ \kms, e.g., \citealt{Chevalier00}).  The luminosity of the radiative shock $L_{\rm sh} = (9\pi/8)\rho_{\rm w}v_{\rm sh}^{3}r^{2}$ at the break-out radius $r = ct_{\rm pk,w}/\tau_w$ is:
\begin{eqnarray}
L_{\rm sh}(t_{\rm pk,w}) \simeq  \frac{9\pi }{8}v_{\rm sh}^{3}\frac{c t_{\rm pk}}{\kappa_{\rm opt}} \approx 2\times 10^{44}{\rm erg\,s^{-1}}\left(\frac{t_{\rm pk}}{2{\rm d}}\right)\left(\frac{v_{\rm sh}}{10^{4}\,\kms}\right)^{3},
\label{eq:Lshw}
\end{eqnarray}
From Eq.~\ref{eq:Lshw} and \ref{eq:trisew} we conclude that a shock break-out from an extended medium with density structure corresponding to an {\em effective} mass-loss rate $A \sim 10^{5}A_{\star}$ can explain both the timescale and peak luminosity of the optical emission from AT\,2018cow. 
Following the initial break-out, radiation from deeper layers of the expanding shocked wind ejecta would continue to produce emission. 
However, such a cooling envelope is predicted to redden substantially in time \citep[e.g.,][]{Nakar14}, in tension with the observed persistently blue optical/UV colors and lack of cooling at $\delta t>20$ days (Fig.~\ref{Fig:Lengine}) 
We conclude that, even if a shock break-out is responsible for the earliest phases of the optical emission and for accelerating the fastest ejecta layers, a separate more deeply-embedded energy source is needed at late times to explain the properties of AT\,2018cow. 
\subsubsection{Reprocessing by Dense Ejecta and the Spectral Slope of the Optical Continuum Emission}
\label{SubSubSec:repro}
We argued in previous sections that the sustained blue emission from AT\,2018cow is likely powered by reprocessing of radiation from a centrally-located X-ray source embedded within the ejecta. Here we discuss details of the reprocessing picture and what can be learned about the ejecta structure of AT\,2018cow.

Late-time optical spectra at $\delta t>20$ days  (Fig.~\ref{Fig:OpticalSpec}) show line widths of $\sim4000$~\kms\ ($\sim0.01 c$), indicating substantially lower outflow velocities than at earlier times (when $v\sim0.1$c), and an abrupt transition from very high velocity to lower velocity emitting material (Fig.~\ref{Fig:Lengine}). While it might be possible to explain this  phenomenology in a spherically-symmetric model with a complex density profile, a more natural explanation is that the ejecta/CSM of AT\,2018cow is aspherical, e.g., with fast-expanding material along the polar direction, and slower expanding dense matter in the equatorial plane (Fig.~\ref{fig:ShockCartoon}). This picture is independently supported by the observed properties of X-ray emission discussed and by the emission line profiles, as discussed in \S\ref{SubSec:Xray} and \S\ref{SubSusbSec:Lines}.\footnote{Aspherical ejecta may be supported by the early detection of time-variable optical polarization ($p \sim 0.3-1\%$; \citealt{SmithATel18}).  However, the non-thermal NIR component identified by \citet{Perley18} could also explain this polarization, which is consistent with the claimed rise of the polarization into the red.}

Although we approximated the UV/optical spectral energy distribution (SED) with a blackbody function in \S~\ref{Subsec:Bol}, the true SED is likely to deviate from a single blackbody spectrum. The observed slope of the early optical SED, $L_{\nu} \propto \nu^{1.2}$ can be used to constrain the ejecta stratification in reprocessing models.  Neglecting Gaunt factors, and using the result from radiative transfer calculations that do not assume local thermodynamic equilibrium (LTE), which indicate that in the outer layers of the reprocessor the free electron temperature $T_e$ tends to level off to a constant value much greater than the effective temperature \citep[e.g.,][]{Hubeny00,Roth16}, the free-free emissivity in the optical-to-infrared is $j_{\nu}^{\rm ff} \propto n_e n_i \propto \rho^2$ (where $n_e$ and $n_i$ are the number-density of electrons and positive ions, respectively, and we used the fact that $h\nu\ll kT_e$). This result holds as long as the material is highly ionized. 

While the bound electrons are coupled to the radiation field and are likely to be out of the thermal equilibrium, the free electrons should be in LTE, so that $j_{\rm \nu}^{\rm ff} = \alpha^{\rm ff}_{\nu} B_{\nu}(T_e)$. For  $h\nu\ll kT_e$, we find $\alpha_{\nu}^{\rm ff} \propto \rho^2 \nu^{-2}$. We assume that near the surface of the emitting material the density can be locally modeled by a power-law in radius $\rho \propto r^{-n}$, for some $n > 1$. Due to the ionization from the engine, electron scattering dominates the opacity. The total optical depth (integrated from the outside in) is then wavelength-independent and 
$\tau_{\rm es} (r) \sim  \left(\frac{1}{n-1}\right)\,\, \rho_0 \kappa_{\rm es} \,\, r_0^n \, \,  r^{1 - n} $,  
where $r_0$ is some reference radius within the region where the power-law expression for the density holds, and $\rho_0\equiv \rho(r_0)$. Let $\alpha^{\rm es}$ and $\alpha_{\nu}^{\rm abs}$ denote the opacity coefficients from electron scattering and continuum absorption, respectively. We define an opacity ratio 
  $\epsilon_{\nu} = \frac{\alpha_\nu^{\rm abs}}{\alpha^{\rm es} + \alpha_{\nu}^{\rm abs}} \approx \frac{\alpha_\nu^{\rm abs}}{\alpha^{\rm es}} \propto \rho \nu^{-2}$. 

The effective optical depth to absorption is $\tau_{\rm eff}(\nu)\sim\sqrt{\epsilon_{\nu}} \tau_{\rm es}$, where  $\tau_{\rm es}$ is measured from the outside in, and we evaluate $\epsilon_\nu$ at the thermalization depth $r_{\nu,\rm therm}$, which is the radius where $\tau_{\rm eff}(\nu)=1$. We define $\tau_{\rm es}(r_{\nu,\rm therm})\equiv \tau_{\nu,\rm therm}$. It follows that 
$ \tau_{\nu,\rm therm}  \approx \frac{1}{\sqrt{\epsilon_\nu}} \propto \rho^{-1/2}\nu^{1}$, which implies that the thermalization radius scales with $\nu$ as $r_{\nu,\rm therm}\propto \nu^{\frac{2}{2 - 3 n}}$. Following \citet{Roth16}, we can approximate the \emph{observed} spectrum as $  L_{\rm \nu} \approx 4 \pi \int_{r_{\nu,\rm therm}}^\infty  j_{\rm \nu} \, 4 \pi r^2 dr$. Substituting the scalings above we find:\footnote{A related analysis by \citet{Shussman16} results in $L_\nu \propto \nu^{\frac{30n - 16}{21n - 8}}$ when converted to our notation, which has similar behavior as our result for large $n$. In that work, rather than assuming that $T_e$ levels off near the surface, the authors assume that $T_e \propto \tau^{1/4}$.} 
\begin{equation}                                   
  L_{\rm \nu} \approx \frac{\left( 4 \pi \right)^2}{2n - 3} \,\, j_\nu\left(r_{\nu,\rm therm}\right) r_{\nu,\rm therm}^3 \propto r_{\nu,\rm therm}^{3 - 2n} \propto  \nu^{\frac{4n - 6}{3n - 2}}    
  \label{Eq:Lnufreefree}
  \end{equation}
For $n = 2$ we have $L_{\nu}\propto \nu^{1/2}$, in reasonable agreement with the results by \citet{Roth16}. For large $n$ this tends toward an asymptotic scaling $L_{\nu} \propto \nu^{4/3}$, which is similar to the measured slope of the \emph{optical} continuum of AT\,2018cow. 

We conclude that in AT\,2018cow optical continuum radiation is reprocessed in a layer with a steep density gradient $n\gg 1$. Our derivation also indicates that the spectral slope should be roughly independent of the luminosity of the engine, as is observed, as long as the high ionization state is maintained.

\subsubsection{Spectral Line Formation}
\label{SubSusbSec:Lines}
In AT\,2018cow no clear spectral lines are apparent at early times, which can be understood as the result of a high degree of ionization and the low contrast of broad spectral features with very large velocities $\sim 0.1c$.
H and He lines with $v\sim 4000$~\kms\ emerge at $\delta t>20 $ days (Fig.~\ref{Fig:OpticalSpec}, 
see also \citealt{Perley18}) with asymmetric line profiles in which the red wing extends farther than the blue wing. This spectral line shape emerges naturally in radiative transfer calculations of line formation in an optically thick, expanding atmosphere \citep{Roth18}. The line photons must scatter several times before escaping, and in the process they do work on the gas and lose energy in proportion to the volume-integrated divergence of the radial velocity component. However, these calculations also predict a net \emph{blueshift} for the centroid of the line, which is not observed in AT\,2018cow. In AT\,2018cow emission lines possess redshifted centroids (Fig.~\ref{Fig:OpticalSpec}).  The redshift of the line centroids is hard to accommodate in spherical models and points to asphericity in the ejecta of AT\,2018cow. A potential geometry of the expanding ejecta that would be consistent with the observed redshifted line centroids is that of an equatorially-dense reprocessing layer and a low density polar region, where the projected area of the photosphere on the receding side is larger than on the approaching side, due to the angle the observer makes with the equatorial plane.  A schematic diagram of this geometry is shown in Fig.~\ref{fig:ShockCartoon}.

\subsection{Radio Emission at $\nu<100$ GHz}
\label{SubSec:Radio}
The key observational results are: (i) An optically-thick
spectrum with $F_{\nu}\propto \nu^{2}$ at $\nu<100$ GHz for weeks after discovery.  (ii) The spectral peak frequency cascades down with time  following $\nu_{\rm br}\propto t^{-1}$. (iii) With $L_{\nu}\sim 4\times 10^{27}\,\rm{erg\,s^{-1}\,Hz^{-1}}$ at $\nu\sim8$ GHz around $\delta t\sim 20$ days and a rising emission with time, the radio emission from AT\,2018cow is markedly different from GRBs in the local universe and cosmological GRBs and more closely resembles that of the most luminous normal radio SNe (Fig.~\ref{Fig:RadioALL}).  
\subsubsection{Radio Emission from External Shock Interaction}
\label{SubSubSec:radio}
\begin{figure}
 \includegraphics[width=0.9\columnwidth]{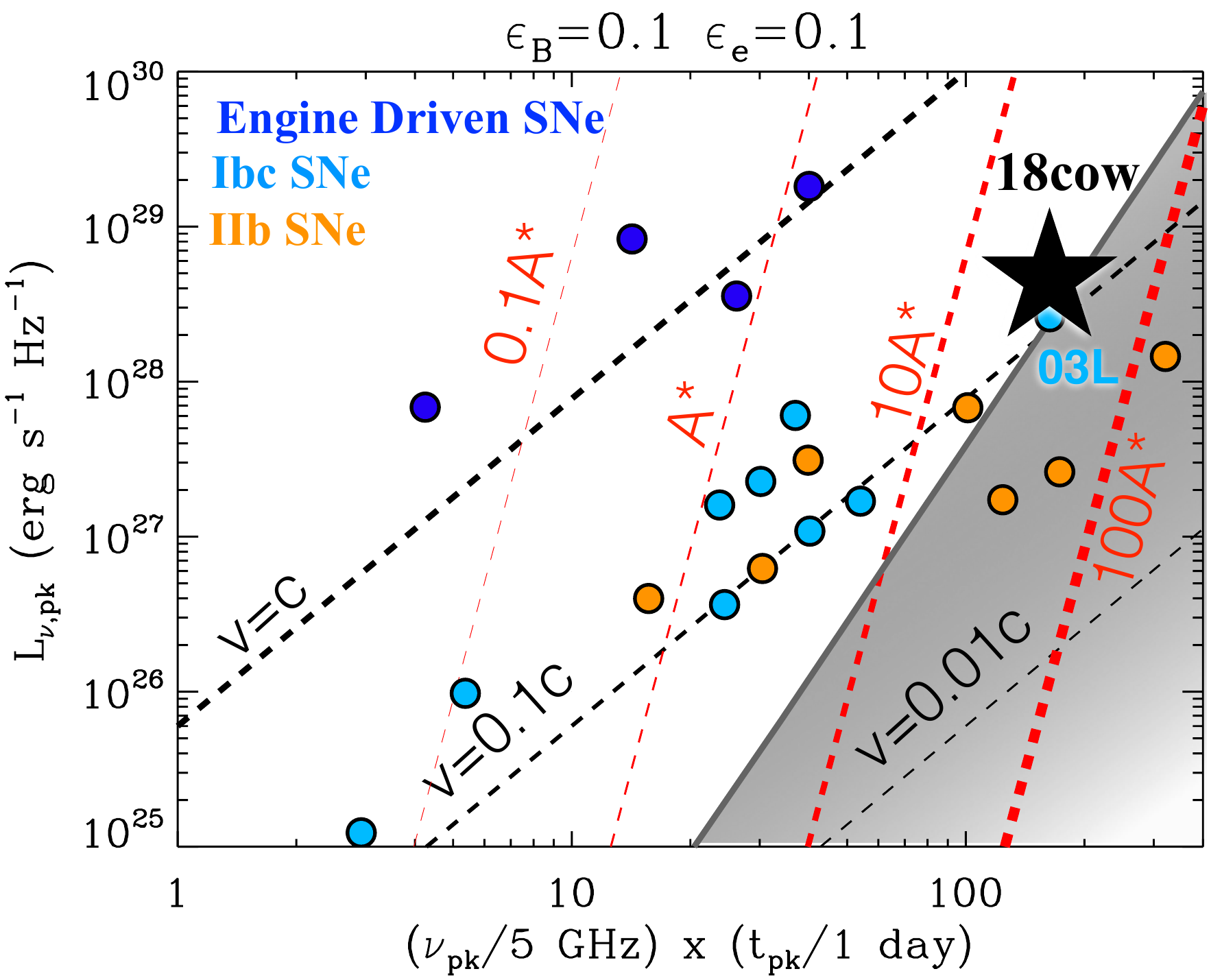}
 \includegraphics[width=0.9\columnwidth]{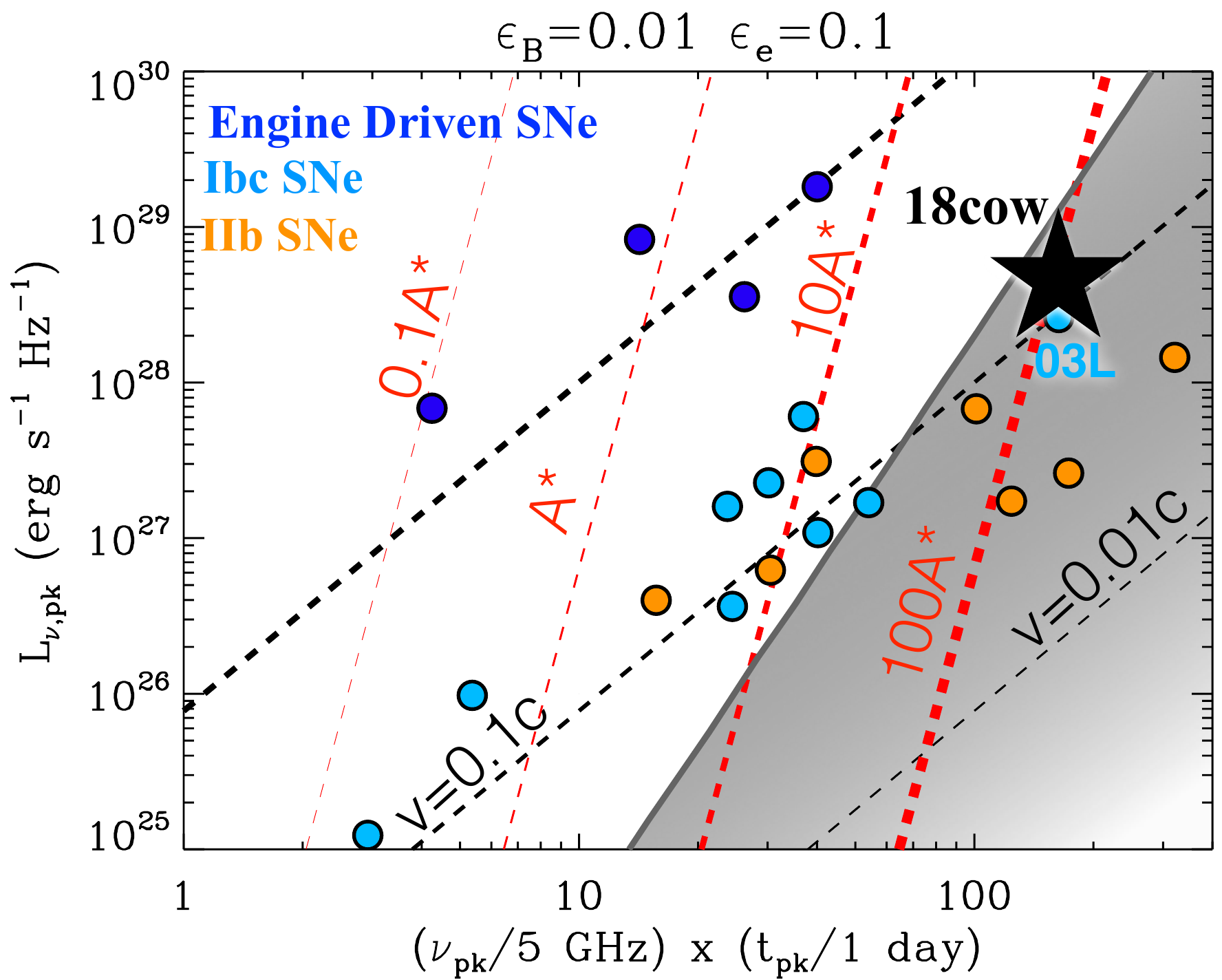}
  \caption{Phase space of radio observables: peak radio luminosity ($L_{\nu,\rm pk}$), peak spectral frequency ($\nu_{\rm pk}$) and peak time ($t_{\rm pk}$). Black (red) dashed lines: lines of constant shock velocity (constant mass-loss rate) following the standard formulation of synchrotron self-absorbed radio emission from a freely expanding blast wave in a wind-like environment \citep[e.g.,][]{Chevalier98,Soderberg05,Soderberg12}.  Black stars: AT\,2018cow at $t = 83$ days, when we estimate $\nu_{\rm pk}\sim 10$ GHz and  $L_{\nu,{\rm pk}}\sim4\times 10^{28}\,\rm{erg\,s^{-1}cm^{-2}}$ (Fig.~\ref{Fig:RadioSED}).
Grey area: region of the parameter space ruled out based on the lack of evidence for free-free absorption (\S\ref{SubSubSec:radio}, Eq.~\ref{eq:ff}). In both panels we assume $\epsilon_e=0.1$. The upper (lower) panel is for $\epsilon_B=0.1$ ($\epsilon_B=0.01$). Blue, light-blue, and orange filled circles mark the position of engine-driven SN with relativistic ejecta (i.e., GRBs and relativistic SNe), normal H-stripped core-collapse SNe, and type-IIb SNe, respectively, from \citet{Soderberg12}.
  }
  \label{Fig:RadioChevalier}
\end{figure}

The observed radio emission from AT\,2018cow in the first weeks ($F_{\nu}\propto \nu^{2}$ at $\nu<100$ GHz, Fig.~\ref{Fig:RadioSED}) is consistent with being self-absorbed synchrotron radiation likely produced from an external shock generated as the ejecta interacts with a dense external medium, as observed in radio SNe \citep[e.g.,][]{Soderberg05,Soderberg12}. The evolution of the spectral break frequency $\nu_{\rm br}\propto t^{\alpha}$ with $\alpha\sim-1$, interpreted as the synchrotron self-absorption frequency $\nu_{\rm sa}$, is consistent with that expected of ejecta undergoing free expansion \citep[$\alpha = -1$, e.g.,][]{Chevalier98}.

In the context of self-absorbed synchrotron emission from a freely expanding blast wave propagating into a wind medium, the observed peak luminosity $L_{\nu,{\rm pk}}$, spectral peak frequency $\nu_{\rm pk}$, and peak time $t_{\rm pk}$ directly constrain the environment density, blast wave velocity, and kinetic energy. Following \citet{Chevalier98} and \citet{Soderberg05}, in Fig.~\ref{Fig:RadioChevalier} we show a peak flux $F_{\nu, \rm pk}\sim 10$ mJy (corresponding to $L_{\nu,{\rm pk}}\sim 4\times 10^{28}\,\rm{erg\,s^{-1}\,Hz^{-1}}$) and  $\nu_{\rm sa} \approx 10$ GHz around $\delta t\sim83$ days imply a shock/outer ejecta velocity $v_{\rm sh} \sim 0.1c$ interacting with a dense wind with $A \sim 10-100A_{\star}$ (depending on $\epsilon_B=0.1-0.01$). The corresponding wind density is:
\be
n_w = \frac{\rho_w}{m_p} = \frac{\dot{M}}{4\pi v_w r^{2} m_p} \approx 9\times 10^{6}{\rm cm^{-3}}\,\,\left(\frac{A}{100A_{\star}}\right)v_{0.1}^{-2}t_{\rm wk}^{-2},
\ee
where we have taken $r = vt$ as the shock radius, $v_{0.1} \equiv v_{\rm sh} /0.1c, t_{\rm wk} \equiv t/{\rm week}$. For these parameters, the equipartition energy (which is a lower limit to the kinetic energy of radio emitting material) is $E_{\rm eq}\sim  5\times 10^{47}$ erg.

Shock velocities $\sim0.1\,c$ are common among normal stripped-envelope  radio SNe (Fig.~\ref{Fig:RadioChevalier}). The values $A \sim 10-100A_{\star}$ needed to explain the luminosity of the radio emission of AT\,2018cow are similar to those inferred in previous radio-bright supernovae (e.g., SN\,2003L, \citealt{Chevalier06}), but substantially smaller than the values $A \sim 10^{5}A_{\star}$ needed on smaller radial scales to explain the early optical peak if the latter is powered by shock break-out from a wind (Eqs.~\ref{eq:trisew}, \ref{eq:Lshw}).  The velocity of the fastest ejecta inferred from the radio $\sim 0.1 c$ is consistent with that needed to explain the rapid optical rise time (Eq.~\ref{eq:tdiff}).

We can further constrain the environment density using the lack of evidence for a low frequency cut-off in the radio spectrum (Fig.~\ref{Fig:RadioSED}) due to free-free absorption \citep[e.g.,][]{Weiler+2002}, as follows.  The optical depth of the forward shock to Thomson scattering and free-free absorption are given, respectively, by
\be
\tau_{\rm T} \simeq \frac{\dot{M}\kappa_{\rm es}}{4\pi v_w r} \approx 0.01 \left(\frac{A}{100A_{\star}}\right) v_{0.1}^{-1} t_{\rm wk}^{-1}
\ee
\be
\tau_{\rm ff} \simeq \frac{\alpha_{\rm ff}r}{3} \approx 10\left(\frac{\nu}{10{\rm GHz}}\right)^{-2}\left(\frac{T_{\rm g}}{10^{4}{\rm K}}\right)^{-3/2}\left(\frac{A}{100A_{\star}}\right)^{2}v_{0.1}^{-3}t_{\rm wk}^{-3},
\ee
where we have taken $\kappa_{\rm es} = 0.38$ cm$^{2}$ g$^{-1}$ for fully ionized solar-composition ejecta and $\alpha_{\rm ff} \approx 0.03 n_w^{2}\nu^{-2}T_{\rm g}^{-3/2}$ cm$^{-1}$ as the free-free absorption coefficient, where $T_{\rm g}$ is the temperature of the gas, normalized to a value $T_{\rm g} \gtrsim 10^{4}$ K typical of photoionized gas.

From Fig.~\ref{Fig:RadioSED}, the optically thick part of the radio spectrum, which scales as $F_{\nu}\propto \nu^2$, without any evidence for free-free absorption, demands  $\tau_{\rm ff}($15\,GHz$) \ll 1$ at $t = 6.5$ d and $\tau_{\rm ff}($5\,GHz)$\ll1$ at $t = 12$ d. These limits translate into similar upper limits on the environment density:
\be
\frac{A}{A_{\star}} \lesssim 30 v_{0.1}^{3/2}\left(\frac{T_{\rm g}}{10^{4}{\rm K}}\right)^{3/4}
\label{eq:ff},
\ee
consistent with values $A \sim 10-100A_{\star}$ for $v_{\rm sh} \sim 0.1c$ (Fig.~\ref{Fig:RadioChevalier}).

We conclude with considerations about the shock microphysics parameters and the properties of the distribution of electrons responsible for the radio emission. The radio emission is produced by relativistic electrons accelerated into a power-law distribution at the forward shock, e.g., $dN/d\gamma_e \propto \gamma_e^{-p}$ with $p \ge 2$.  The Lorentz factor $\gamma_e = \gamma_\nu$ of the electrons which contribute at radio frequency $\nu = 0.3(eB\gamma_e^{2}/2\pi m_e c)$ is
\be
\gamma_{\nu} \simeq 4.6\left(\frac{m_e c\nu}{eB}\right)^{1/2} \approx 153\epsilon_{\rm B,-2}^{-1/4}\,t_{\rm wk}^{1/2}\left(\frac{\nu}{100{\rm GHz}}\right)^{1/2}\left(\frac{A}{100A_{\star}}\right)^{-1/4},
\label{eq:gammanu}
\ee
where we have estimated the magnetic field behind the shock as
\be
B_{\rm sh} = \left(6\pi \epsilon_{B}\rho_{\rm w}v_{\rm sh}^{2}\right)^{1/2} \approx 5.1{\rm G}\epsilon_{\rm B,-2}^{1/2}\,t_{\rm wk}^{-1}\left(\frac{A}{100A_{\star}}\right)^{1/2}.
\label{eq:Bsh}
\ee
From Fig.~\ref{Fig:RadioChevalier}, for $\epsilon_{B}\ll 0.01$ we would require larger values $A \gtrsim 100A_{\star}$, incompatible with the observed lack of free-free absorption (eq.~\ref{eq:ff}).

As shown in \S\ref{SubSubSec:XrayShock}, electrons with $\gamma_e \sim \gamma_{\rm \nu}$ responsible for the optically-thin radio emission $\nu \gtrsim \nu_{\rm sa}$ cool rapidly due to inverse Compton (IC) emission in the optical/UV radiation of AT\,2018cow.  Therefore, above $\nu \gtrsim \nu_{\rm sa}$ the fast-cooling scaling holds \citep[e.g.,][]{Granot02}:
\be
F_{\nu} = \nu^{-p/2} \underset{p = 2.3}\approx \nu^{-1.15} .
\ee
Our radio SED at $92$ days, with $F_{\nu}\propto \nu^{-1.2\pm0.1}$ above $\nu_{sa}$  confirms this inference (Fig.~\ref{Fig:RadioSED}).
The predicted luminosity in the optical/NIR band ($\nu \gtrsim 10^{14}$ Hz $\gg \nu_{\rm sa}$) is thus similar to, or smaller than, the radio luminosity $\nu L_{\nu} \sim$ a few $\times 10^{40}$ erg s$^{-1}$ and thus insufficient to explain the possible non-thermal NIR excess of luminosity $\sim 10^{42}$ erg s$^{-1}$ identified by \citet{Perley18}.  Any non-thermal component in the IR/optical range cannot be from the same synchrotron source as the forward shock that produces the radio emission at $\nu<100$ GHz.

The radio emission at $\nu>100$ GHz reported at early times is also more luminous than predicted by our model (Fig.~\ref{Fig:RadioSED}). This observation might suggest  a slightly decelerated blastwave where $F_{\rm{pk}}\propto t^{-0.2}$ (e.g., \citealt{Chevalier98}), or that the radio data at $\nu>100$ GHz might be dominated by a separate emission component (e.g., reverse shock) if physically associated with AT\,2018cow (Fig.~\ref{Fig:RadioSED}). 

\subsubsection{Constraints on off-axis relativistic jets}
\label{SubSubSec:jets}
\begin{figure}
 \includegraphics[width=1.03\columnwidth]{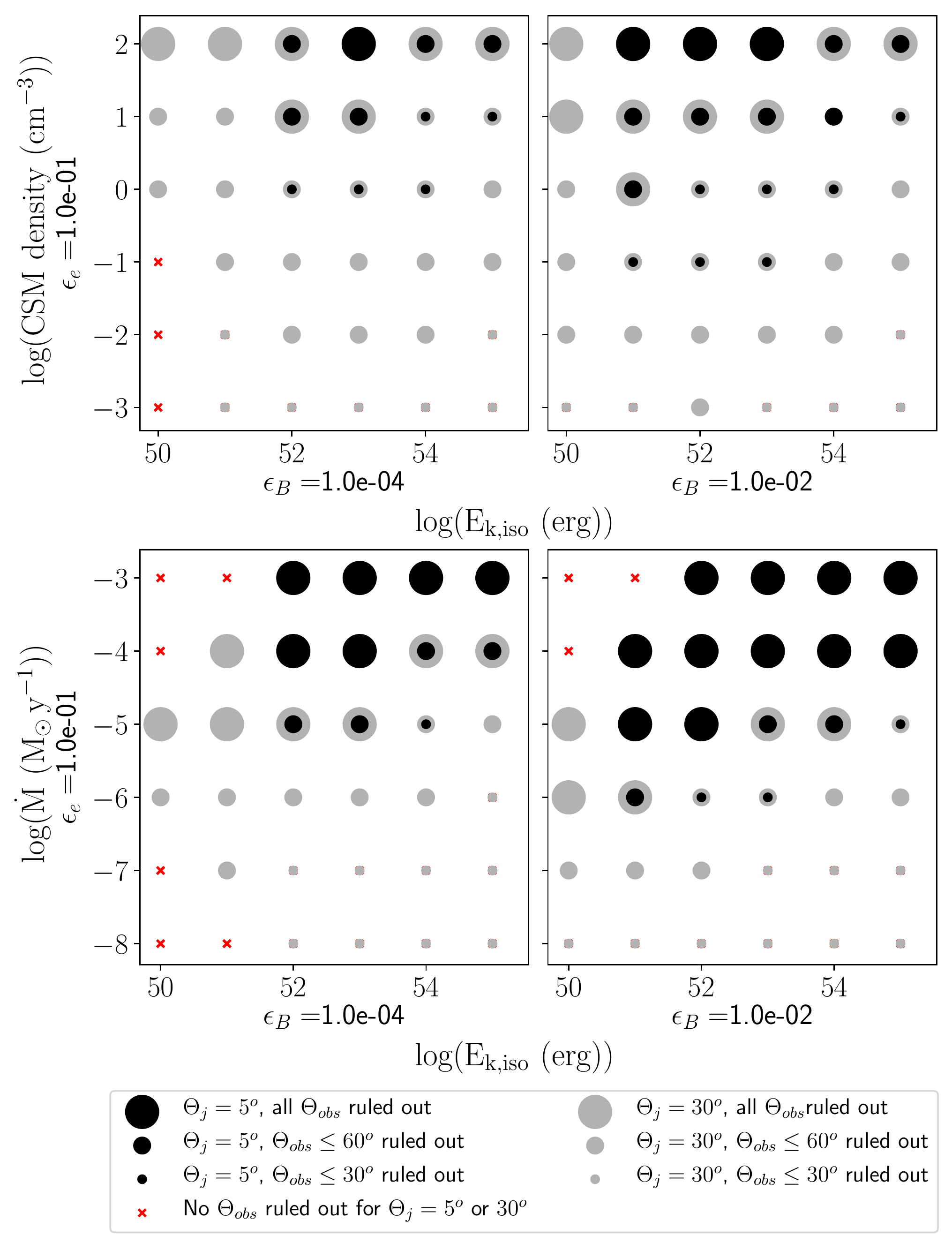}
  \caption{Constraints on off-axis relativistic jets in AT\,2018cow, for a range of microphysical parameters ($\epsilon_e$ and $\epsilon_B$), jet opening angles $\theta_j$, off-axis angle $\theta_{\rm obs}$, jet isotropic equivalent kinetic energies $E_{\rm k,iso}$ and environment densities. Upper (lower) panel: ISM (wind) environment. We assume a wind velocity $v_w=1000$~\kms\ (i.e., $\dot M=10^{-5}\,\rm{M_{\sun}\,yr^{-1}}$ corresponds to $A=A_{*}$).}
  \label{Fig:jets}
\end{figure}

The observed radio emission is consistent with arising from non-relativistic ejecta with  velocity similar to that of normal SNe ($v_{\rm sh}\sim0.1$c) interacting with dense circumstellar medium (CSM) with $A \sim 10-100 A_{\star}$.  No high energy prompt emission was detected in association with AT\,2018cow (\S\ref{SubSec:IPN}). However, AT\,2018cow showed evidence for broad spectral features in the optical emission, with velocities comparable and even larger than those seen in broad-lined Ic SNe associated with GRBs (\citealt{Modjaz16}, Fig.~\ref{Fig:OpticalSpec}). In this section we constrain the properties of an off-axis jet in AT\,2018cow. Emission from a collimated outflow originally pointed away from our line of sight becomes detectable as the blast wave decelerates into the environment and relativistic beaming of the radiation becomes less severe with time \citep[e.g.,][]{Granot02bis}. 

The observed radio emission from an off-axis jet primarily depends on the jet opening angle ($\theta_j$), off-axis angle ($\theta_{\rm obs}$), jet isotropic-equivalent kinetic energy ($E_{\rm k,iso}$), environment  density (parametrized as $n$ and $\dot M$ for an ISM and wind-like medium, respectively) and shock microphysical parameters ($\epsilon_B$, $\epsilon_e$).  We employ realistic simulations of relativistic jets propagating into an ISM and wind-like medium to capture the effects of lateral jet spreading with time, finite jet opening angle and transition into the non-relativistic regime. We ran the code BOXFIT (v2; \citealt{vanEerten10,vanEerten12}) at 1.4 GHz, 9 GHz, 15 GHz, and 34 GHz, for a range of representative parameters of long GRB jets ($E_{\rm k,iso}=10^{50}-10^{55}$~erg, $\epsilon_B=10^{-4}-10^{-2}, \epsilon_e=0.1$) and environment densities ($n=10^{-3}-10^{2}\,{\rm cm}^{-3}, \dot M=10^{-8}-10^{-3}\,\rm{M_{\sun}\,yr^{-1}}$). We use $p=2.5$ and explore the parameter space for two jets of $\theta_j=5\arcdeg$ and $\theta_j=30\arcdeg$, representative of strongly collimated and less collimated outflows, respectively (as found for normal  GRBs and low-energy GRBs, \citep[e.g.,][]{Racusin09,Ryan15,Margutti100316D}. 

With reference to Fig.~\ref{Fig:jets} we find that 
less-collimated outflows with $\theta_j=30\arcdeg$ are ruled out in the ISM case for large densities $n>1\,\rm{cm^{-3}}$. For a wind-type medium with $\dot{M}_w \sim 10^{-3}-10^{-4} M_{\odot}$ yr$^{-1}$, consistent with the values $A \sim 10-100 A_{\star}$ inferred for the forward shock radio emission, jets with $E_{\rm k,iso} \ge 10^{52}$ erg are presently ruled out for $\epsilon_e=0.1$, $\epsilon_B=0.01$ and jet opening angles $\theta_{\rm j} \approx 5-30\arcdeg$, corresponding to beaming-corrected jet energies $E_{\rm k} \ge 4\times 10^{49}$ erg (for $\theta_{\rm j} = 5\arcdeg)$ or $E_{\rm k} \ge 10^{51}$ erg (for $\theta_{\rm j} = 30\arcdeg)$.   Successful jets with 
$E_{\rm j}<4\times 10^{49}\,\rm{erg}$ ($\theta_j=5\arcdeg$) and $E_{\rm j}<10^{51}\,\rm{erg}$ ($\theta_j=30\arcdeg$) propagating into a wind medium with $\dot M< 10^{-4}\,\rm{M_{\sun}yr^{-1}}$ are allowed.

\subsection{Hard and Soft X-ray Emission}
\label{SubSec:Xray}
The key observational results are: (i) Luminous X-ray emission discovered at the level of $L_{\rm X}\sim 10^{43}\,\rm{erg\,s^{-1}}$. $L_{\rm x}$ is significantly larger than seen in normal SNe, and similar to the values seen in GRBs in the local universe (Fig.\ \ref{Fig:GRB}).  (ii) Persistent X-ray flaring with short variability timescales of a few days 
superimposed on a secular decay, which is initially gradual $\propto t^{-1}$ but then steepens around $\delta t\sim 25$ days to a faster decay $\propto t^{-4}$ around the same time as the appearance of narrow optical features.
(iii) Presence of two X-ray components of emission with distinct temporal evolution and spectral properties: a persistent source in the $>0.1$ keV range, as well as a transient component of hard X-ray emission at energies $>10 $ keV detected at $\delta t \sim 8$ days and which disappears by $\delta t \sim 17$ days (Fig.\ \ref{Fig:XraySpec}).  (iv) The persistent X-ray spectral component of emission is well modeled by $F_{\nu}\propto \nu^{-\beta}$ with $\beta\sim 0.5$ with  no evidence for intrinsic neutral hydrogen absorption (Fig.\ \ref{Fig:Xray}).

Below we discuss the physical origin of the X-ray emission associated with AT\,2018cow.
\subsubsection{X-ray Emission from External Shock Interaction}
\label{SubSubSec:XrayShock}

\begin{figure}
\hskip -0.7 cm
 \includegraphics[width=1.1\columnwidth]{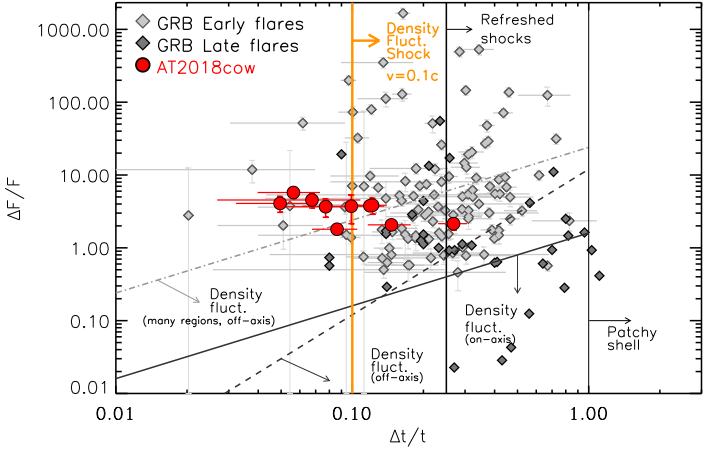}
  \caption{Flux contrast $\Delta F/F$ vs.\ relative variability $\Delta t/t$ for X-ray flares in AT\,2018cow (red filled circles) and long GRB flares at early $t<1000$ s and late times $t>1000$ s (gray diamonds). Kinematically allowed regions of the parameter space in the context of relativistic collimated outflows are marked with black/gray solid, dashed and dot dashed lines \citep[details in][]{Ioka05}.  A slower shock with $v_{\rm sh }\sim0.1-0.2\,c$ like the one powering the radio emission in AT\,2018cow is expected to produce $\Delta t/t\ge 0.1-0.2$ if the overdensity  region covers a large fraction of the solid angle (orange line). Flares in AT\,2018cow have been fitted using the same procedure as used for long GRBs adopting a \citet{Norris05} profile. The width of a flare $\Delta t$ is measured at $1/e$ the flare peak value. Variability in AT\,2018cow violates the expectations from density fluctuations in relativistic jets and slower shocks. Long GRB flare data from \cite{Chincarini10,Margutti10,Bernardini11}. }
  \label{Fig:flares}
\end{figure}

We first consider the possibility that the X-rays originate from the same forward shock responsible for the radio synchrotron emission ($\S\ref{SubSubSec:radio}$).  The kinetic luminosity of the radio-emitting forward shock,
\be
L_{\rm sh} = 4\pi r^{2}\frac{9}{32}v_{\rm sh}^{3}\rho_{\rm w} \approx  5\times 10^{42}{\rm erg\,s^{-1}}v_{0.1}^{3}\left(\frac{A}{100 A_{\star}}\right),  
\ee
is close to the X-ray luminosity of AT\,2018cow (Fig.~\ref{Fig:Xray}).  This suggests a picture in which the X-rays are IC emission from optical/UV photons upscattered by relativistic electrons accelerated at the forward shock.  Further supporting this scenario, the radio-to-X-ray luminosity ratio
$\frac{\nu_{\rm sa}L_{\nu_{\rm sa}}}{L_{\rm X}} \sim 0.02$
is comparable to the ratio of the magnetic energy density $U_{B} = B_{\rm sh}^{2}/8\pi$ (Eq.~\ref{eq:Bsh}) to the optical/UV photon energy density $U_{\gamma} = L_{\rm opt}/(4\pi c r_{\rm sh}^{2})$, where $r_{\rm sh} = v_{\rm sh}t$,
\be
\frac{U_{\rm B}}{U_{\gamma}} \simeq 0.02\epsilon_{\rm B,-2}^{1/2}v_{0.1}^{2}t_{\rm wk}^{2}\left(\frac{A}{A_{\star}}\right)^{1/2},
\ee
where $L_{\rm opt} = 8\times 10^{43}t_{\rm wk}^{-2}$ erg s$^{-1}$.

However, the IC forward shock model cannot naturally explain the $F_{\nu} \propto \nu^{-0.5}$ spectrum of the persistent X-ray component, offers no consistent explanation of the transient hard X-ray component, and has difficulties accounting for the observed short-time scale variability, as we detail below.  

Electrons heated or accelerated at the shock cool in the optical radiation field on the expansion timescale for electron Lorentz factors above the critical value,  
\be
\gamma_{\rm c} \simeq \frac{3m_e c}{4 \sigma_{\rm T}U_{\gamma}t} = \frac{3\pi  r^{2} m_e c^{2}}{\sigma_{\rm T}L_{\rm opt}t} \underset{r = vt}\approx 1.0 v_{0.1}^{2}t_{\rm wk}^{3},
\ee
where $\sigma_{\rm T} \simeq 6.6\times 10^{-25}$ cm$^{2}$ is the Thomson cross section.  The electrons responsible for upscattering optical/UV seed photons of energy $\epsilon_{\rm opt} \simeq 3kT_{\rm opt} = 5{\rm eV}(T_{\rm opt}/2\times 10^{4}{\rm K})$ eV to X-ray energy $E_{\rm X} = (4/3)\epsilon_{\rm opt}\gamma_{\rm X}^{2}$ must possess Lorentz factors
\be
\gamma_{\rm X} \approx \left(\frac{3E_{\rm X}}{4\epsilon_{\rm opt}}\right)^{1/2} \approx 12\left(\frac{E_{\rm X}}{\rm 1 keV}\right)^{1/2} 
\ee
The values $\gamma_{\rm X} \approx 6-40$ needed to populate the XRT bandpass 0.3--10 keV, though lower than those producing the millimeter radio emission (eq.~\ref{eq:gammanu}), are in the fast-cooling regime $\gtrsim \gamma_{\rm c}$ for the first few weeks of evolution.  Thus, while for slow-cooling electrons the observed $F_{\nu}\propto \nu^{-0.5}$ spectrum would match the expectation $F_{\nu} \propto \nu^{-(p-1)/2}\approx \nu^{-0.5}-\nu^{-1}$ for $p=2-3$, it is {\em incompatible with the fast-cooling expectation,} $F_{\nu} \propto \nu^{-p/2}$, which gives a much softer spectrum than observed $F_{\nu}\approx \nu^{-1}-\nu^{-1.5}$ for $p=2-3$. 

We now consider an IC origin of the transient hard X-ray component of emission, which shows a rising slope of $F_{\nu} \propto \nu^{0.5}$ (Fig.~\ref{Fig:XraySpec}). This emission is too hard to be free-free or synchrotron radiation (it violates the ``synchrotron death line"; e.g., \citealt{Rybicki79}), possibly hinting at an IC origin.  In addition to accelerating electrons into a non-thermal distribution, the forward shock is also predicted to {\em heat} electrons \citep{Sironi15}, generating a relativistic Maxwellian particle distribution with a mean {\em thermal} Lorentz factor
\be
\gamma_{\rm th} \approx f_e\frac{m_p}{2m_e}\left(\frac{v_{\rm sh}}{c}\right)^{2} \approx 4.6(f_e/0.5) v_{0.1}^{2},
\label{eq:gamma_th}
\ee
where $f_e = 0.5$ is the fraction of the shock energy imparted to the electrons.\footnote{Further Coulomb heating of the electrons by ions downstream of the shock (e.g.,~\citealt{Katz11}) is inefficient given the low densities of the forward shock.}  Thus, it may be tempting to associate the transient hard X-ray ``bump'' with IC emission by a relativistic Maxwellian distribution of electrons.  However, the expected spectral peak would occur at an energy,
\be
E_{\rm X,th} = \frac{4}{3}\gamma_{\rm th}^{2}\epsilon_{\rm opt} \approx 0.14(f_e/0.5)v_{0.1}^{2}\left(\frac{T_{\rm opt}}{2\times 10^{4}{\rm K}}\right){\rm keV}
\ee
which is a factor $\lesssim 100$ smaller than the observed peak $E_{\rm X} \approx 50$ keV.

A final problem of the forward shock model is  the rapid and persistent X-ray variability, which in this model has to be attributed to density inhomogeneities in the environment (e.g., a series of thin shells or ``clumps'').  The shortest allowed variability time $\Delta t$ if the ejecta covers a large fraction of the solid angle is the light crossing time, which for shock of radius $r_{\rm sh} = v_{\rm sh}t$ with $v_{\rm sh}\approx 0.1-0.2\,c$ (\S\ref{SubSubSec:radio}) is constrained to obey
\be
\frac{\Delta t}{t} \gtrsim \frac{r_{\rm sh}}{c t} \sim \frac{v_{\rm sh}}{c} \approx 0.1-0.2
\ee
We measure the properties of the X-ray flares in AT\,2018cow following the same procedure as is
used for long GRBs, adopting a \citet{Norris05} profile. We find that the X-ray flares in AT\,2018cow show much faster variability and violate this expectation, and furthermore show no evidence for a linear increase of their duration as the blast wave expands, contrary to expectations (Fig.~\ref{Fig:flares}). Instead our analysis in \S\ref{Subsec:Timing} suggests the presence of a dominant time scale of variability of a few days. In Fig.~\ref{Fig:flares} we show that the X-ray variability observed in AT\,2018cow is also not consistent with the expectations from density fluctuations encountered by a relativistic jet (with an observation either on-axis or off-axis).   Differently from \citet{Rivera18}, we thus conclude that density fluctuations in the CSM environment of AT\,2018cow are unlikely to be the physical cause of the observed X-ray variability. The very rapid turn-off of the X-ray emission  as $L_{\rm X}\propto t^{-4}$ at $\delta t>$ 20 days (Fig.~\ref{Fig:Xray}) is also difficult to accommodate in models where the X-ray emission is powered by an external shock (the typical $L_{\rm X}$ decline is $\propto t^{-1}$ for a spherical blastwave and $\propto t^{-2}$ for a collimated outflow after ``jet-break'', e.g., \citealt{Granot02}).\footnote{The hint for a correlation of the UV and X-ray variability of \S\ref{Subsec:Timing}  also supports an ``internal'' origin of the X-ray emission.}

\subsubsection{X-rays from a Central Hard X-ray Source}
\label{SubSubSec:XrayEngine}
We consider an alternative scenario in which the observed X-rays originate primarily from an {\em internal} hard radiation source (either in the form of shocks or a compact object, Fig.~\ref{fig:ShockCartoon}), embedded within aspherical, potentially bipolar ejecta shell. The asphericity of the ejecta is a key requirement to explain the observed X-ray properties.
The high-density material at lower latitudes (blue region in Fig.~\ref{fig:ShockCartoon}) is opaque to X-rays below $\sim 15$~keV due to bound-free absorption. The observed X-rays in this energy range either escape directly through the highly-ionized low-density polar ejecta (lighter gray shaded region in Fig.~\ref{fig:ShockCartoon}) and/or are scattered into the line of sight by this material. The X-rays absorbed by the dense equatorial shell are reprocessed to lower frequencies and are powering the optical light curve.
The polar cavity is initially narrow and grows with time as the ejecta dilutes, expands, and becomes progressively transparent.

This scenario provides a natural explanation of the transient hard X-ray spectral component of Fig.~\ref{Fig:XraySpec}
which is created by the combined effect of photoelectric absorption at soft X-ray energies $\lesssim 10$ keV, and Compton downscattering of very hard X-rays at $> 50-100$ keV. Here we assume transmission and reflection through neutral gas, and leave the discussion of Fe-line formation to the next section. The power-law spectrum that dominates at soft X-ray energies $E<10$ keV is instead produced by X-ray photons that reach the observer without being absorbed or Compton downscattered (i.e., these photons provide a direct view of the central engine). At high energies the spectral shape of the observed spectrum is controlled by the Thomson optical depth along our line of sight $\tau_T$. At early times, close to the optical peak at $t_{\rm{pk}}\sim t_{\rm diff}\approx 3$ days (Eq.~\ref{eq:tdiff}), $\tau_{\rm T} \sim (c/v_{\rm ej})(\kappa_{\rm es}/\kappa) \sim 20-40$, where $v_{\rm ej} \sim 0.1-0.2c$ and $\kappa_{\rm es} \sim 4 \kappa$. At this time nearly all of the UV/X-ray radiation of the central X-ray source is absorbed by the shell and reprocessed into optical/UV radiation. However, as the ejecta expands with time, its optical depth decreases $\tau_{\rm T} \propto t^{-2}$, reducing the fraction of the central X-rays being absorbed. At the time of our first NuSTAR/INTEGRAL observations at $\delta t\sim 8$ days, $\tau_{\rm T}\sim3$, resulting in moderate downscattering and attenuation of radiation at $E \ge 511 {\rm keV}/\tau_{\rm T}^2\sim 50$~keV (Fig.~\ref{Fig:XraySpec}). For the same parameters, we calculate  $\tau_{\rm T}\lesssim 1$ at $\delta t \sim 17$ days, by which time Compton downscattering plays a negligible role.
This prediction is  consistent with our observation of an uninterrupted power-law spectrum extending from 0.3 keV to $\sim 70$ keV at $\delta t\ge 17$ days (Fig.~\ref{Fig:XraySpec}).

\begin{figure}
 \includegraphics[scale=0.43]{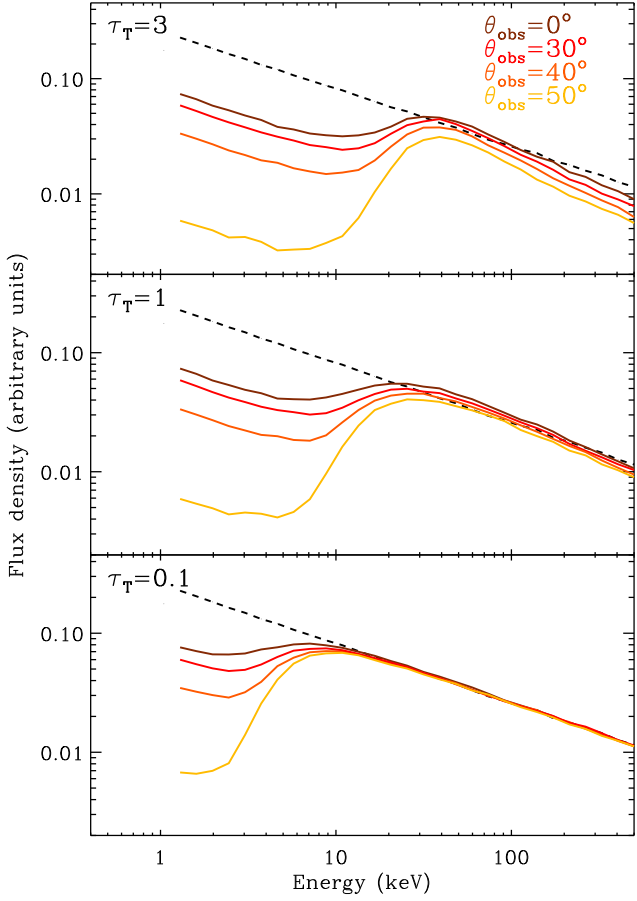}
 \caption{Transmitted X-ray spectrum from a central source with intrinsic spectrum $F_{\nu}\propto \nu^{-0.5}$ (black dashed line) for different viewing angles $\theta_{\rm obs}$. A shell of material of optical depth $\tau_{\rm T}$, radius $R_{\rm{sh}}$ and thickness $2\,R_{\rm{sh}}$ with a polar cavity of opening angle $\theta_{\rm{sh}}=30\arcdeg$ surrounds the source (motivated by the geometry shown in Fig.~\ref{fig:ShockCartoon}). At larger optical depths and intermediate viewing angles, the transmitted spectrum shows a ``hump'' of emission above $\sim$10 keV that becomes less prominent and disappears at lower $\tau_{\rm{T}}$, as observed in AT\,2018cow.}
  \label{Fig:diffusion}
\end{figure}

The observed soft X-ray spectral shape is controlled by the bound-free optical depth $\tau_{\rm{bf}}$. Most opacity in the keV range is the result of photoelectric absorption by CNO elements, particularly the usually abundant oxygen.  The bound-free opacity of oxygen is
\be
\kappa_{\rm bf} = \frac{f_{\rm n}X_{\rm O}\sigma_{\rm bf}}{16 m_p},
\ee
where $X_{\rm O}$ is the oxygen mass fraction, $f_{\rm n}$ is its neutral fraction and
\be
\sigma_{\rm bf} = \sigma_{\rm th}(\nu/\nu_{\rm th})^{-3}, \nu \gtrsim \nu_{\rm th}
\ee
where for K-shell electrons of oxygen we have $\sigma_{\rm th} = 2-1\times 10^{-19}$ cm$^{-2}$ and $h\nu_{\rm th} = 0.74-0.87$ keV.  Following \citet{Metzger17KN}, the neutral fraction in an ejecta shell of mass $M_{\rm ej}$ and radius $R_{\rm ej} = v_{\rm ej}t$ due to photoionization by X-ray luminosity $L_{\rm X} \approx \nu_{\rm th}L_{\nu_{\rm th}}$ is 
\begin{eqnarray}
f_{\rm n} &\approx& \frac{\alpha_{\rm rec}M_{\rm ej}h\nu_{\rm th}}{m_p L_{\rm X}v_{\rm ej}t\sigma_{\rm th}} \approx \frac{4\pi \alpha_{\rm rec}\tau_{\rm T} (v_{\rm ej}t)h\nu_{\rm th}}{m_p L_{\rm X} \sigma_{\rm th}\kappa_{\rm es}} 
\end{eqnarray}
where for K-shells of oxygen $\alpha_{\rm rec} \approx 2-4\times 10^{-11}$ cm$^{3}$ s$^{-1}$ \citep{Nahar97}.  The X-ray optical depth at $\nu_{\rm th} \sim 1$ keV is thus
\begin{eqnarray}
\tau_{\rm bf} &=& \tau_{\rm T}\frac{\kappa_{\rm bf}}{\kappa_{\rm es}} \sim \frac{\pi}{4}\frac{\alpha_{\rm rec} }{\kappa_{\rm es}^{2}}\frac{\tau_{\rm T}^{2}X_{\rm O}}{m_p^{2}}\frac{h\nu_{\rm th}}{L_{\rm X}}v_{\rm ej}t\nonumber \\
&\sim& 0.1\left(\frac{X_{\rm O}}{0.01}\right)\tau_{\rm T}^{2}\left(\frac{t}{{\rm 1 week}}\right)v_{0.1}\left(\frac{L_{\rm keV}}{10^{43}\,{\rm erg\,s^{-1}}}\right)^{-1}
\end{eqnarray}
For ejecta with an oxygen abundance close to solar abundance ($X_{\rm O} \sim 0.01$), $\sim$ keV X-rays of luminosity $\sim 10^{43}$ erg s$^{-1}$ ionize the ejecta sufficiently to escape unattenuated on timescales of a couple weeks when the Thomson column along the low density polar region is also low $\tau_{\rm T} \lesssim 1$.  However, at earlier times we expect $\tau_{\rm bf} \gtrsim 1$ when $\tau_{\rm T}$ is larger, i.e., around the time of the NuSTAR/INTEGRAL spectral ``hump'' (Fig.~\ref{Fig:XraySpec}).

We quantitatively explore the predictions of our model by performing a series of Monte Carlo calculations where we follow the escape of photons as they propagate through a uniform shell of radius $R_{\rm sh}$ and thickness $2\,R_{\rm sh}$ with a polar cavity of opening angle $\theta_{sh} = 30\arcdeg$ carved into it. We assume an isotropic source with intrinsic spectrum $F_{\nu}\propto \nu^{-0.5}$ as observed, and self-consistently account for photoelectric absorption and Compton scattering. Figure~\ref{Fig:diffusion} shows the results for the transmitted X-rays for different lines of sight $\theta_{\rm{obs}}$ and optical depth $\tau_{\rm{T}}$. Polar  observing angles (i.e., small $\theta_{\rm obs}$) receive a larger fraction of ``direct'' X-rays (including X-rays reflected off the cavity walls, Fig.~\ref{fig:ShockCartoon})  at any $\tau_{\rm{T}}$, while more equatorial views with larger $\theta_{\rm obs}$ are associated with more prominent ``humps'', as a larger fraction of X-rays intercepts absorbing/scattering material. However, as $\tau_{\rm T}$ drops with time as a result of the shell expansion, X-rays become detectable from a larger range of viewing angles while the ``hump'' moves to lower energies to eventually disappear.

In our model:
(i) The X-ray variability is \emph{intrinsic} to the central source, rather than being a consequence of inhomogeneities in the external medium.
(ii) The soft X-rays $\lesssim 10$ keV, which originate directly from the engine, may show more pronounced time variability than those associated with the transient hard X-ray spectral component, which instead are diffusing through an optically thick shell.
(iii) The true luminosity evolution of the central source is the {\em sum} of the optical and X-ray luminosities ($L_{\rm{engine}}$ in Fig.~\ref{Fig:Lengine}); the fact that the X-ray light curve decreases less rapidly at early times than the optical light curve, as shown in Fig.~\ref{Fig:Lengine}, is a consequence of the increasing fraction of escaping X-ray radiation with time.
\subsubsection{The connection of AT\,2018cow to other astrophysical sources with Compton-hump spectra}
\label{SubSubSec:ComptonHump}

In the previous section we provided a proof-of-concept that interaction (in the form of scattering and absorption) of X-ray photons from a source located within expanding ejecta with  temporally-declining optical depth $\tau$ provides a natural explanation of the broad-band X-ray spectrum of AT\,2018cow. The model is agnostic with regard to the physical nature of both the X-ray radiation and the reprocessing medium. The incident hard X-ray radiation in AT\,2018cow might originate from an embedded shock, or from a central ``nebula''  (similar to a pulsar-wind nebula around a young magnetar),  or an X-ray ``corona'' around an accreting BH.

Consistent with the picture above,  reflection emission from the reprocessing of inverse-Compton photons off a thick accretion flow produces a ``Compton hump'' feature 
similar in shape to what we observed in AT\,2018cow.  Such emission is typical of X-ray binaries (XRBs) and AGNs (e.g., \citealt{Risaliti13,Tomsick14} for recent examples), where a high-energy power-law component associated with the Compton-upscattering of seed thermal photons from the BH accretion disk by a hot cloud of electrons --- the ``corona'' --- interacts with cold matter in the disk. Reflection emission is typified by a $\sim 30$~keV Compton hump along with prominent Fe K$\alpha$-band emission and Fe K-shell absorption edges (e.g., Fig.\ 1 of \citealt{Risaliti13, Reeves04}), which can all become broadened by relativistic effects.  It is tempting to associate the transient excess of emission around $\sim8$ keV detected in the first spectrum of AT\,2018cow (Fig.~\ref{Fig:XraySpec}, inset) with a Fe K-shell spectral feature (emission/absorption). The $\sim$8 keV excess of emission disappears by 16.5 days together with the hump, which supports the idea of a physical link between the two components and motivates our attempt below to model AT\,2018cow with standard disk reflection models. While the actual geometry of AT\,2018cow is likely to be more complicated than in standard accretion disks (for AT\,2018cow the reprocessing material might be rapidly expanding and diluting), the same physics of reprocessing of hard X-ray radiation  (including reflection and partial transmission) applies.

\begin{figure}
\hskip -0.6 cm
 \includegraphics[scale=0.52]{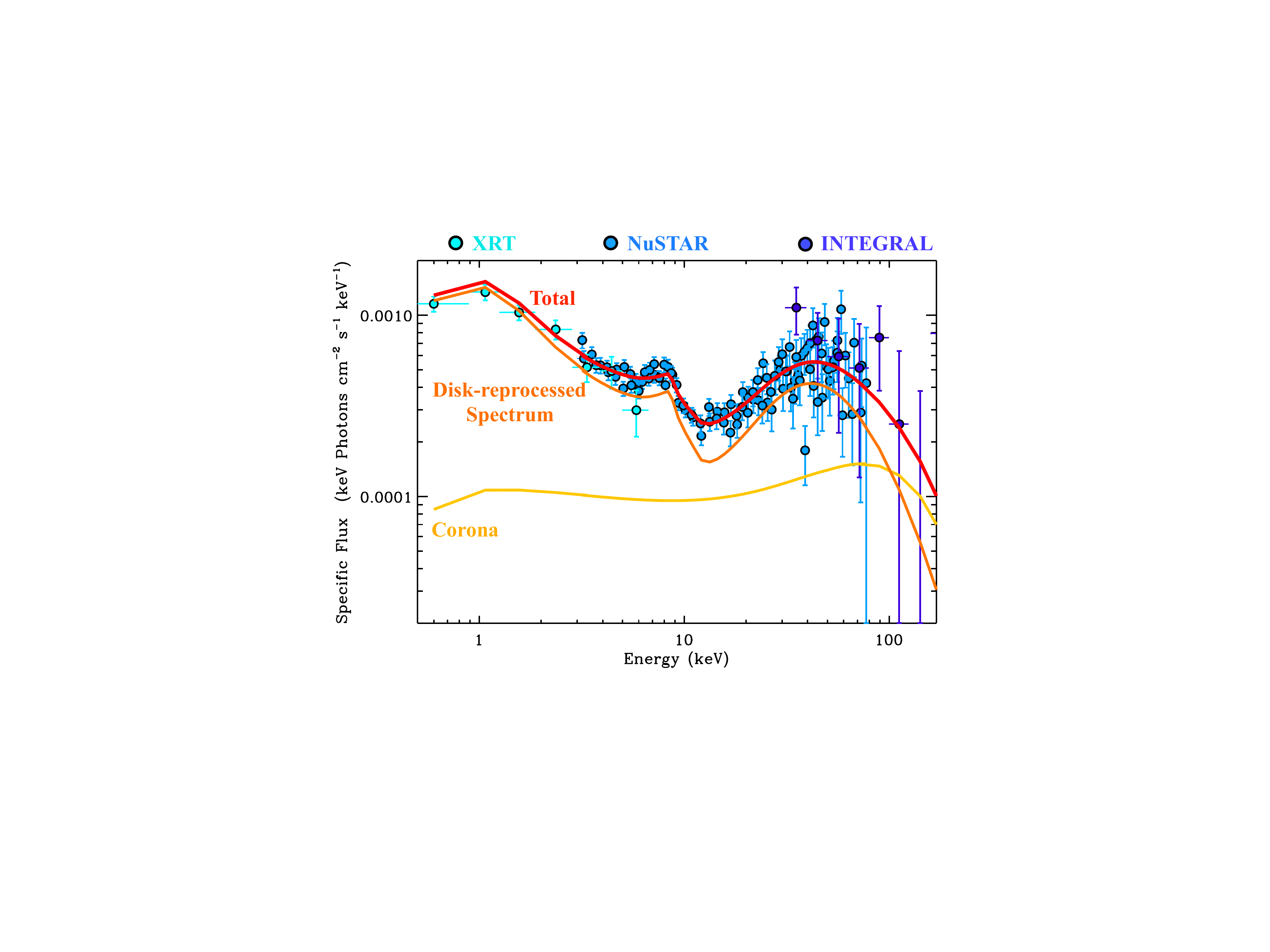}
 \caption{The broad-band X-ray spectrum of AT\,2018cow at $\delta t=7.7$ days is well fitted by a Comptonized disk-reflection model. This model naturally accounts for the 6-9 keV excess as Fe-K band fluorescence emission distorted  by the orbital and thermal motion of the gas.}
  \label{Fig:AccretionDisk}
\end{figure}

Fitting the broad-band X-ray data of AT\,2018cow at $\delta t=7.7$ days with a Comptonized disk-reflection model via \texttt{simpl} \citep{Steiner09} acting on a thermal component  and \texttt{relxill} \citep{Dauser14} produces a good fit to the data ($\chi^2/\rm{d.o.f.} \sim 1.0$), matching a $kT_e \geq 30$ keV corona, with reflection fraction $R_f \gtrsim 1$ (Fig.~\ref{Fig:AccretionDisk}).\footnote{$R_f$ is the ratio of Compton-scattered photons that illuminate the disk as compared to those reaching the observer.} The best fitting model predicts a moderate optical depth to the corona $\tau \sim 1-2$, which illuminates the walls of the reprocessing material in a funnel-like geometry. If sufficiently compact, the innermost corona may be Compton thick, which would produce a thermal feature at the coronal temperature, partially accounting for the hard X-ray excess. In this model the 7--9 keV excess is naturally explained as  Fe-K fluorescence emission originating away from the core along the funnel, where the ionization parameter drops below  $log~\xi \equiv L/nR^2 \lesssim 4$. In this model the Fe-K feature is primarily distorted  by the orbital and thermal motion of the gas, to produce the observed broad blueshifted Fe-K emission. In this scenario, the observed disappearance of the hard X-ray Compton hump and associated Fe emission can be explained by either (i) a decline in the accretion rate, which makes the funnel opening angle grow.  The corona is both less confined to a compact geometry and the walls of the funnel are extended, jointly resulting in less illumination and a diminished contribution from reflection. (ii) If the hard excess is in somewhat supplied by a Compton-thick coronal core with $kT_e \sim 30$~keV, then as $\dot{M}$ declines the region becomes optically thin, causing the high-energy thermal feature to drastically fade.

Finally, we address the question of the super-Eddington luminosity, which is of particular relevance if the $L_{\rm X} \gtrsim 10^{43}\,\rm{erg\,s^{-1}}$ in AT\,2018cow is powered by a stellar-mass BH (discussed in \S\ref{SubSec:IntFailedSN}). In the case of Intermediate Mass BHs (IMBHs) discussed in  \S\ref{SubSec:IntIMBH} the required accretion rate would only be mildly super-Eddington.
Theoretical studies of super-Eddington accretion flows have recently experienced a surge of interest thanks to the observational finding that numerous ultraluminous X-ray sources (ULXs), systems which are brighter than the Eddington limit of stellar-mass black holes, are in fact powered by pulsars (e.g., \citealt{Walton18a,Walton18b}).  Theoretical models predict that highly super-Eddington accretion produces a funnel geometry in the central flow \citep{Sadowski15}, not dissimilar from our model above. The emission is collimated and generally paired with powerful outflows, including radiatively powered jets \citep{Sadowski15}. 
In these systems, the seed X-ray luminosity can be ``boosted'' by a factor $\sim$ tens by scatterings by hot coronal electrons at $\tau \sim 1-2$. 
The underlying seed X-ray emission from the disk is then required to be $\sim 10^3\,L_{\rm{Edd}}$ for a BH of few $M_{\sun}$, in line with the observed super-Eddington emission in known neutron-star ULXs. 

\subsection{The excess of NIR Continuum Emission}
\label{SubSec:NIRcontinuum}
\citet{Perley18} identified an excess of NIR emission with respect to the UV/optical blackbody with $F_{\nu}\propto \nu^{-0.75}$, which they interpret as non-thermal synchrotron emission physically connected with the radio-mm emission at $\nu>100$ GHz. Our observations confirm the presence of the NIR excess (Fig.~\ref{Fig:superSED}). As shown in Fig.~\ref{Fig:superSED} the extrapolation of the model that best fits the radio observations at $\nu<100$ GHz severely underpredicts the NIR flux. From theoretical arguments we inferred in \S\ref{SubSubSec:radio} that electrons radiating at $>100$ GHz must be fast cooling, which would predict a steeper radio-to-NIR spectral slope than what is needed to connect the radio to the NIR band on the same synchrotron SED. Extrapolating the X-ray component to the NIR frequency range produces the same result of underpredicting the observed NIR emission.  We conclude that the NIR excess is unlikely to be directly related to the same populations of electrons that produce the non-thermal radio emission at $\nu<100$ GHz or the X-ray radiation. 

\cite{Kuin18} favor a different interpretation of the NIR excess as free-free emission from an expanding ``atmosphere'' with a shallow density gradient. The NIR emitting material would be located at larger distances than the optical/UV emitting material. This process is well known to produce a NIR excess of emission in hot stars surrounded by dense winds and Luminous Blue Variables \citep[see e.g.,][]{Wright75}, and has been invoked to explain the NIR excess in SN\,2009ip \citep{Margutti1409ip}. In this scenario, the spectral slope is directly connected with the density gradient of the NIR emitting material and it is not expected to evolve with time, as observed, as long as the high ionization state is maintained. From 
Eq.~\ref{Eq:Lnufreefree}, the measured spectral slope $F_{\nu}\propto \nu^{-0.75}$ suggests a medium with a shallow density gradient of ionized material $\rho\propto r^{-n}$ with $n<2$. In these conditions, matching the observed NIR luminosity requires large densities corresponding to an effective mass-loss rate $\gg100A^*$, which is inconsistent with our findings from the radio data  modeling (\S\ref{SubSubSec:radio}). More complicated geometries with a detached equatorial shell might provide a more consistent explanation. However, regardless of the geometry, this class of models does not naturally account for the NIR temporal variability reported by \cite{Perley18}.

We conclude that the observed NIR excess of emission is not directly related to the non-thermal X-ray and radio emission at $\nu<100$ GHz, and that an ``extended atmosphere'' model is unlikely to offer a quantitative explanation of the observed phenomenology.

\section{INTERPRETATION: the intrinsic nature of AT\,2018cow}
\label{Sec:interpretation}
 
\begin{figure*}
 \includegraphics[scale=0.45]{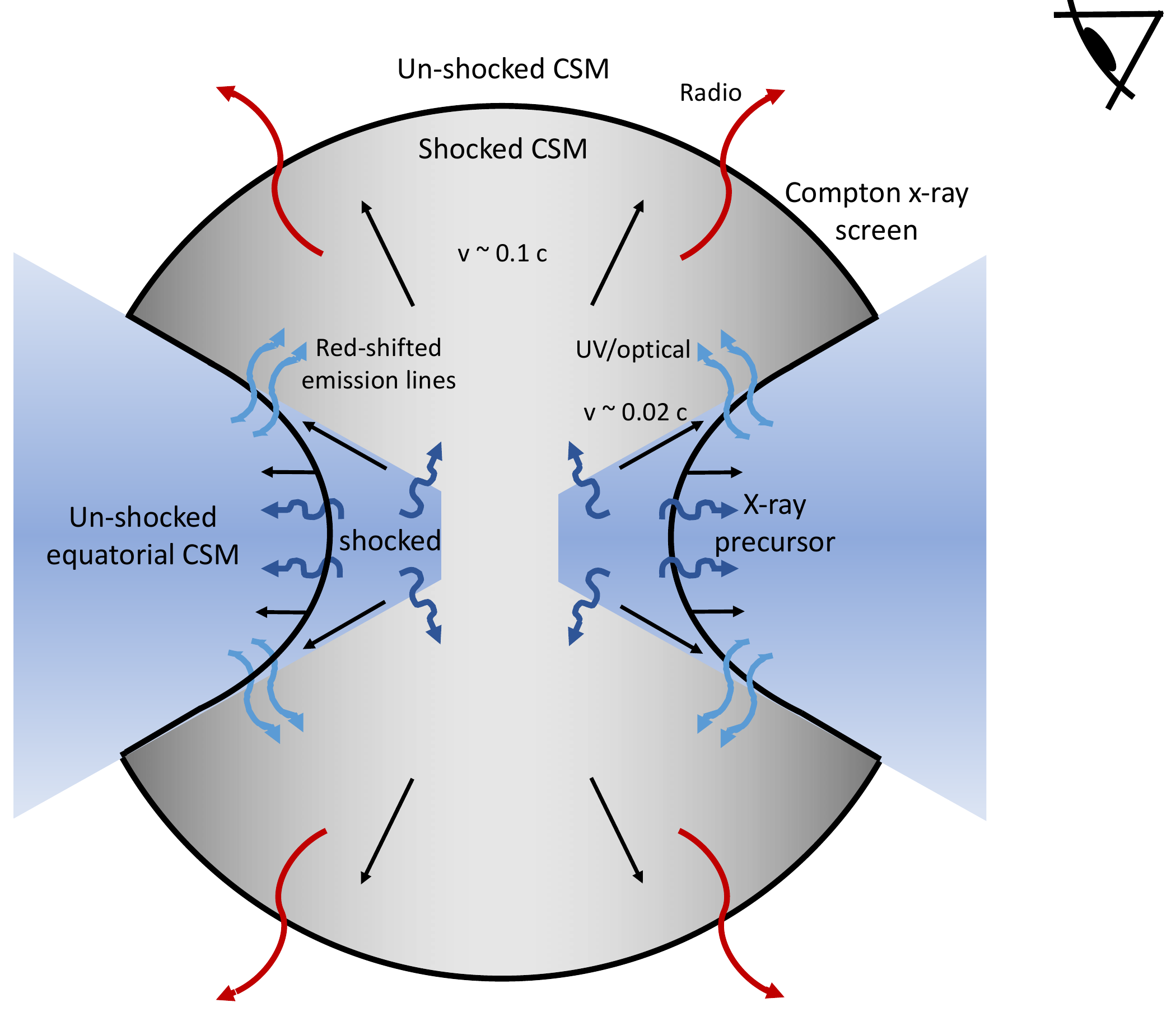}
 \includegraphics[scale=0.45]{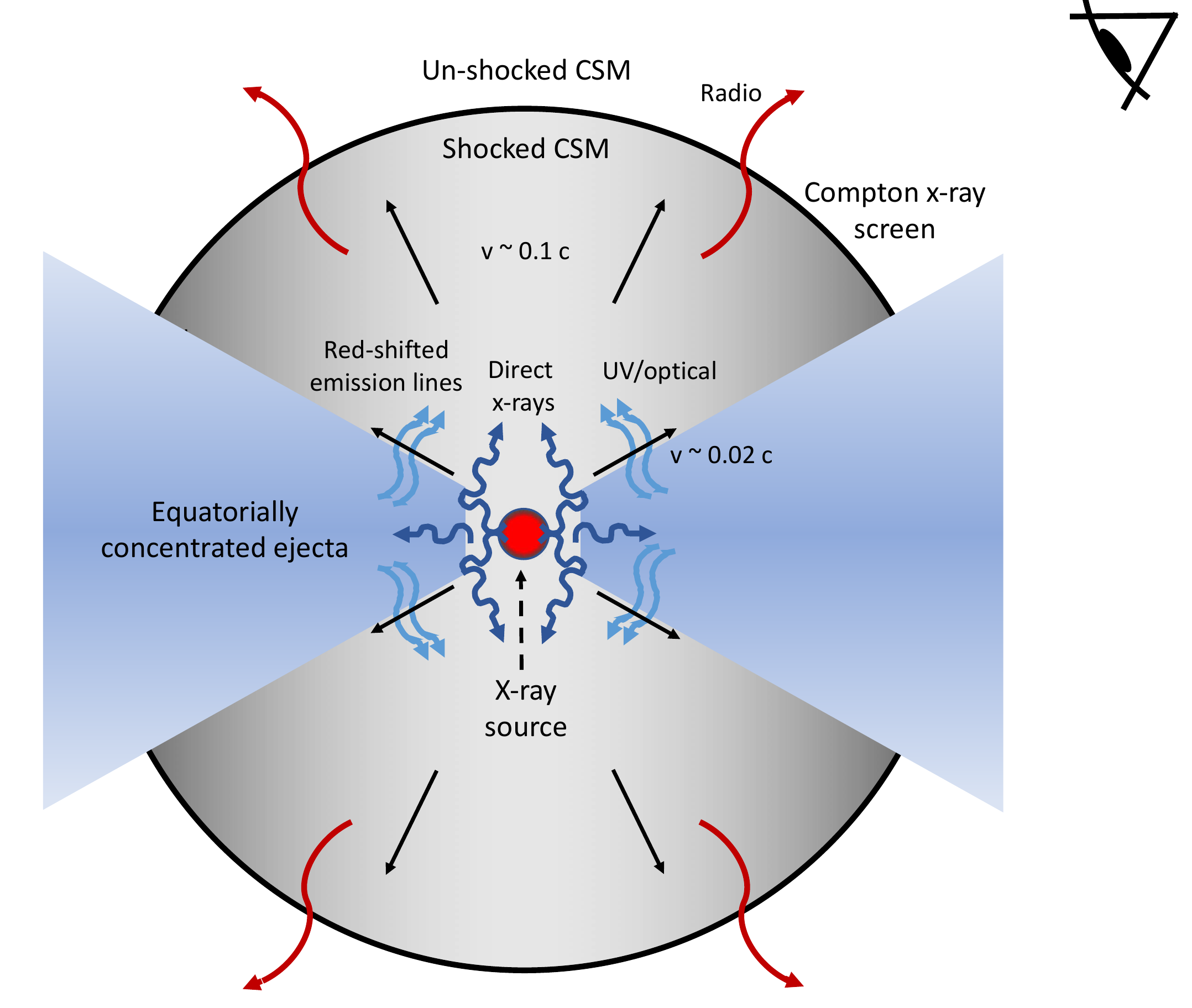}
 \caption{A shock-driven (left panel) or compact-object driven (right panel) origin of the emission from AT\,2018cow. \emph{Left Panel:} the source of radiative energy is interaction between the AT\,2018cow ejecta and the CSM, consisting of a dense equatorial disk and less dense polar regions. Optical/UV emission at early times is primarily from cooling of the fastest moving polar ejecta $v\sim0.1c$, which shocks the CSM leading to radio emission. The ejecta also drives a slower shock in the denser equatorial material, which leads to X-ray emission. For some lines of sight, the optical depth is large enough to modify the intrinsic spectrum and produce the hard X-ray energy ``hump''. The equatorial shock also launches a radial outflow with $v \sim 0.02 c$ that wraps around the disk. After $\sim 20$ days the optical/UV emission in AT\,2018cow is powered by the  reprocessing of precursor X-rays that penetrate ahead of the equatorial shock into the disk. 
The optical photosphere at this time is located within the $v \sim 0.02 c$ outflow. This transition from a situation where the photosphere resides in the fast polar matter at $\delta t<20$ days (and a large fraction of X-rays are reprocessed) to one in which we are directly seeing the inner shell/disk at $\delta t>20$ days is what causes the simultaneous change of the X-ray and optical properties of AT\,2018cow. The radio emission is unaffected as it originates from the external shock.
Our viewing angle presents us with a large emitting area from the receding side of the disk outflow and a limited view of the approaching side, leading to redshifted line peaks. \emph{Right Panel:} the central energy source of X-rays is a compact object (e.g., magnetar or accreting black hole). The X-rays are emitted by the engine and then scatter off the inner funnel walls toward the observer.
}
\label{fig:ShockCartoon}
 \end{figure*}

In this section we synthesize the previous discussion into a concordant picture to explain our multi-wavelength data and we speculate on the intrinsic nature of AT\,2018cow within a ``central X-ray source'' hypothesis.  Any model for the central X-ray source must at a minimum abide by the following constraints:

\begin{itemize}
\item{An ``engine'' that releases a total energy $E_{\rm e} \sim 10^{50}-10^{51.5}$ ergs, over a characteristic timescale $\sim 10^{3}-10^{5}$ s, with a late-time luminosity decay $L_{\rm e} \propto t^{-\alpha}$ with $\alpha \sim 2-2.5$.  The engine has a relatively hard intrinsic X-ray spectrum $F_\nu \propto \nu^{-0.5}$ and is responsible for the highly variable X-ray emission.}
\item{Presence of relatively dense CSM material extending to radii $\gtrsim 10^{16}$ cm.  Its radial profile is not well constrained, but the gaseous mass corresponds to that of an effective wind mass-loss parameter $A \sim 10-100\,A_{\star}$, similar to the CSM around luminous radio supernovae (Fig.~\ref{Fig:RadioALL} and \ref{Fig:RadioChevalier}).  The timescale for a stellar progenitor to lose a mass comparable to the ejecta mass $\sim 1 M_{\odot}$ for such parameters is only $\sim 10^{3}-10^{4}$ yr, necessitating a phase of stellar evolution that is short relative to the main sequence lifetime. }
\item{Asymmetric distribution of material in the vicinity of AT\,2018cow, with denser CSM/ejecta in the equatorial plane and less dense, fast expanding ejecta along the polar direction.}
\item{Presence of hydrogen and helium in the ejecta.}
\item{Limited amount of $^{56}$Ni synthesized, $M_{\rm Ni} \lesssim 0.04M_{\odot}$.}
\item{Low ejecta mass $\sim 0.1-1M_{\odot}$ with a wide range of velocities, from the fastest $v_{\rm ej} \approx 0.1-0.2 \, c$ (as needed to explain the early optical rise and radio emission) to the slowest $v_{\rm ej} \lesssim 0.02 c$ (needed to explain the persistent optically-thick photosphere and narrower late-time emission lines).  This range of velocities may be attributed to aspherical (e.g., bipolar) structure of the ejecta (Fig.~\ref{fig:ShockCartoon}).  A similar aspherical geometry is suggested by the shape of the optical emission lines (Fig.~\ref{Fig:OpticalSpec}) and by the emergent time-dependent spectrum of the central X-ray source (Fig.~\ref{Fig:diffusion}).}
\item{For a medium with $A \sim 10-100\,A_{\star}$ jets viewed off-axis with $E_{\rm k,iso} \ge 10^{52}$ erg are ruled out for jet opening angles $\theta_{\rm j} \approx 5-30\arcdeg$ corresponding to beaming-corrected jet energies $E_{\rm k} \ge 4\times 10^{49}$ erg (for $\theta_{\rm j} = 5\arcdeg)$ or $E_{\rm k} \ge 10^{51}$ erg (for $\theta_{\rm j} = 30\arcdeg)$ (Fig.~\ref{Fig:jets}). On-axis jets with $E_{\rm k,iso} \ge 10^{52}$ erg are ruled out for the entire range of environment densities considered, $\dot M \ge 10^{-8}\,\rm{M_{\sun}\,yr^{-1}}$, consistent with the non-detection of a prompt $\gamma$-ray signal by the IPN.}
\item{The engine model needs to be able to naturally accommodate the \emph{simultaneous} transition between two phenomenologically distinct phases of evolution in the X-ray \emph{and} optical bands, with the first phase  characterized by featureless optical spectra and the presence of the hard X-ray spectral hump, and the second phase at $\delta t>20$ days, (when $L_{\rm X}\sim L_{\rm UVOIR}$), characterized by the emergence of H and He emission in the spectra and a steeper X-ray decay with more pronounced variability. This transition could represent the photosphere radius receding from the fast polar ejecta to the slower equatorially-concentrated  reprocessing material, in the process at least partially exposing the central engine to our vantage point as high-latitude viewers.}
\item{Finally, any engine model needs to be naturally compatible with the location of AT\,2018cow within a faint spiral arm of the star-forming dwarf spiral galaxy CGCG 137-068 (Figure~\ref{Fig:Optical}; \citealt{Perley18}).}
\end{itemize}

 Figure~\ref{fig:ShockCartoon} illustrates the geometry of AT\,2018cow in the context of engine-driven models, where the engine is in the form of either a deeply embedded \emph{internal} shock (left panel) or a compact object  (right panel). In the case of a deeply embedded internal shock, the denser equatorial material pre-existed the AT\,2018cow event, while in the case of a compact-object central engine the thick equatorial torus might have been created by AT\,2018cow (i.e., is part of the AT\,2018cow ejecta).

Potential engine models are summarized in Table \ref{table:engine}.  Some models, like NS-NS, WD-NS mergers, and stripped-envelope supernovae, are immediately ruled out by the presence of H/He in the ejecta.  Below we discuss the more promising possibilities, some of which were already considered by \citet{Prentice+2018}, \citet{Perley18}, and \citet{Kuin18}, though in some cases we reach different conclusions.

\begin{table*}
\centering
\caption{Central X-ray ``Engine" Models for AT\,2018cow \label{table:engine}}
\begin{tabular}{cccccccc}
\hline
Model & Ejecta Mass/Velocity & Engine Timescale & CSM? & He? & H? & Reference \\
\hline
NS-NS Merger Magnetar & $\text{\sffamily X}$ & $\checkmark$ & $\text{\sffamily X}$ & $\text{\sffamily X}$ & $\text{\sffamily X}$ & 1&  \\
WD-NS Merger & \checkmark & \checkmark & $\text{\sffamily X}$ & $\text{\sffamily X}$ & $\text{\sffamily X}$ & 2 &  \\
IMBH TDE & \checkmark & Maybe$\dagger$ & $\text{\sffamily X}$ & \checkmark & \checkmark & 3 &  \\
Stripped-Envelope SN + Magnetar/BH & \checkmark & \checkmark & \checkmark & Maybe & $\text{\sffamily X}$ & 4 &  \\
Electron Capture SN + Magnetar & \checkmark & \checkmark & \checkmark & \checkmark & \checkmark & 5 &  \\
Blue Supergiant Failed SN + BH  & \checkmark & \checkmark & \checkmark & \checkmark & \checkmark & 6 &  \\
SN + Embedded CSM Interaction  & \checkmark & \checkmark & \checkmark & \checkmark & \checkmark & 7 &  \\
\hline \\
\end{tabular}
\\
$\dagger$ If circularization is efficient; (1) \cite{Yu13}, \cite{Metzger14}; (2) \cite{Margalit16}; (3) \cite{Chen18}; (4) \cite{Tauris15}; (5) E.g. \cite{Miyaji80,Nomoto82,Moriya14} and references there in; (6) \cite{Fernandez18}; (7) \cite{Andrews18}, \cite{Metzger17novae}. 
\end{table*}
\subsection{Millisecond Magnetar from Successful Supernova with Low Ejecta Mass}
\label{SubSec:IntMagnetar}

One possibility is a core collapse event with low ejecta mass giving birth to a rapidly spinning magnetar.  The presence of hydrogen in the ejecta favors low-mass stellar progenitors that have been theoretically linked to electron-capture SNe  (ecSNe, e.g., \citealt{Miyaji80,Nomoto82}), rather than an ultra-stripped massive star explosion \citep[e.g.,][]{Tauris15}.  ecSNe are expected to originate from progenitors with mass $\sim8-10\,\rm{M_{\sun}}$, and are predicted to have low explosion energy $E_{\rm k}\sim10^{50}\,\rm{erg}$, small $^{56}\rm{Ni}$ production ($\sim10^{-3}\,\rm{M_{\sun}}$) and ejecta mass of a few $M_{\sun}$.  The low $M_{\rm{Ni}}$ and explosion $E_{\rm k}$ are consistent with our inferences for AT\,2018cow, as most of the kinetic energy of the fast ejecta of AT\,2018cow might have been provided by the engine (as opposed to the initial explosion). The larger $M_{\rm{ej}}$ of ecSN models can also be consistent with AT\,2018cow, as our constraint $M_{\rm{ej}}<0.3\,\rm{M_{\sun}}$ of \S\ref{SubSubSec:engine} applies to the fastest polar ejecta only. Special circumstances are however required to create the aspherical ejecta distribution of AT\,2018cow and to produce a magnetar remnant. We speculate that the rapid rotation of the star, needed to endow the magnetar engine with its rapid rotation, or a jet/wind bubble (see below) could impart the ejecta with the needed equatorial-polar density asymmetry, and that these special requirements naturally explain why AT\,2018cow-like transients are much rarer than $8-10\,\rm{M_{\sun}}$ progenitor stars in the nearby universe.

To explain the required engine energy $E_{\rm e} \gtrsim 10^{50}-10^{51.5}$ erg as rotational energy of the magnetar, its initial spin period should obey $P_{0} \sim 3-20$ ms.  The duration of the engine for an isolated magnetar is given by the dipole spin-down time \citep[e.g.,][]{Spitkovsky06}, which for a $1.4\,M_{\odot}$ NS is given by,
\be
t_{\rm e} = t_{\rm sd} \approx 1.4\times 10^{4}{\rm s}\left(\frac{B_{\rm d}}{10^{15}{\rm G}}\right)^{-2}\left(\frac{P_{\rm 0}}{10\,{\rm ms}}\right)^{2},
\ee
Thus, to explain the engine timescale $t_{e} \sim 10^{3}-10^{5}$ s for $P_{0} \sim 3-20$ ms, we require a magnetar with $B_{\rm d} \approx 10^{15}$ G, in agreement with the findings by \citet{Prentice+2018}.  This strong  $B$ field is in the range of ``magnetars'' normally invoked as central engines of GRBs \citep{Metzger11}, but larger than those inferred for SLSNe, which typically require $B_{\rm d} \sim 10^{14}$ G \citep[e.g.,][]{Kasen10,Inserra13,Nicholl17}.  Alternatively, the intrinsic dipole magnetic field of the magnetar could be weaker, but the effective spin-down rate could be enhanced due to fall-back accretion from the progenitor star \citep{Metzger18mag}; such a scenario predicts a spin-down luminosity $L_{\rm e} \propto t^{-2.38}$, steeper than the usual $\propto t^{-2}$ for an isolated magnetar.

The source of the X-ray emission in this case is the ``nebula'' of hot plasma and magnetic fields inflated by the magnetar behind the expanding SN ejecta \citep[e.g.,][]{Metzger14}.  Though the details of the nebular spectrum are uncertain, the inferred $F_{\nu} \propto \nu^{-0.5}$ would be consistent with synchrotron radiation.  

It is natural to ask whether the same engine responsible for creating the nebula of luminous variable X-rays would also be expected to create a successful relativistic jet.  \citet{Margalit18} show that the central engine jet can escape homologously expanding ejecta if the energy of the jet exceeds a critical value
\be
E_{\rm j} \gtrsim 0.195(\gamma_{\rm j}/2)^{-4}E_{\rm k,0} \sim 10^{49}-10^{50}{\rm erg}
\ee
where $E_{\rm k,0} \gtrsim 10^{51}$ erg is the initial kinetic energy of the explosion and $\gamma_{\rm j}$ is the jet Lorentz factor while propagating through the star (values $\gamma_{\rm j} \sim 2-3$ are required to produce jet opening angles similar to GRBs, \citealt{Mizuta13}).  Thus, if a modest fraction of the engine energy goes into a collimated jet, a jet could break-through the star on a timescale $\lesssim t_{\rm e} \sim 10^{3}-10^{5}$ s \emph{prior to} the optical peak.

Our prompt $\gamma$-ray search with the IPN in \S\ref{SubSec:IPN} led to no evidence for prompt bursts of $\gamma$-ray emission in AT\,2018cow that might be associated with successful jets (in analogy with GRBs). Our analysis in \S\ref{SubSubSec:jets} limits the allowed parameter space to successful jets with $E_{\rm k,iso}<10^{52}\,\rm{erg}$ and $\epsilon_B<0.01$ propagating into a medium with $\dot M< 10^{-4}\,\rm{M_{\sun}yr^{-1}}$, corresponding to $E_{\rm j}<4\times 10^{49}\,\rm{erg}$ ($\theta_j=5\arcdeg$) and $E_{\rm j}<10^{51}\,\rm{erg}$ ($\theta_j=30\arcdeg$) . 

Regardless of whether a tightly collimated jet is created, a wider jet or wind bubble from the engine could impart the ejecta with the needed aspherical (e.g., bipolar) structure, even starting with spherical ejecta.  During this process, the secondary shock driven through the ejecta on timescale $\ll t_{\rm e}$ accelerates the outer layers of the ejecta to $v\sim0.1-0.2 c$ \citep[e.g.,][]{Kasen16,Suzuki16,Blondin17}, explaining the early optical rise and the non-thermal radio emission.
\subsection{Failed Explosion of a Blue Supergiant Star}
\label{SubSec:IntFailedSN}

Another possibility is a core collapse of a massive star that initially fails to explode as a successful SN, instead creating a black hole remnant (e.g.,  \citealt{Quataert12,Dexter13}). Blue supergiant stars have been recently invoked as progenitors of ultra-long GRBs (e.g.,\citealt{Quataert12,Gendre13,Nakauchi13,Wu13,Liu18, Perna18}). \cite{Fernandez18}  show that the failed explosion of a blue supergiant star ($M_{\star} \approx 25\,M_{\odot}$; $R_{\star} \approx 70-150\,R_{\odot}$), 
originating from the neutrino-induced mass loss that follows the formation of the neutron star \citep{Nadezhin80,lovegrove13, piro13,Coughlin18} results in the shock-driven ejection of $\sim 0.1M_{\odot}$ at a velocity $v \sim 0.02$ c.  The remaining star then accretes onto the newly-formed black hole.

If the envelope of the remaining bound star has sufficient angular momentum, it will form an accretion disk around the newly-formed BH, possibly producing the BH accreting scenario that well explains the broad-band X-ray spectrum of \S\ref{SubSubSec:ComptonHump}.
The disk will also produce wind ejecta that collides with the outflowing unbound shell, thermalizing its energy and accelerating it to a higher velocity $v_{\rm ej} \sim 0.1 c$ (e.g., \citealt{Dexter13}).  The timescale of the engine in this case is set by the gravitational free-fall time of the outer layers of the blue supergiant progenitor onto the central black hole, 
\be
t_{\rm e} \sim t_{\rm ff} \approx \left(\frac{R_{\star}^{3}}{GM_{\star}}\right)^{1/2} \approx 1.3{\rm d}\left(\frac{M_{\star}}{25M_{\odot}}\right)^{-1/2}\left(\frac{R_{\star}}{50R_{\odot}}\right)^{3/2},
\ee
consistent with the constraints on the engine lifetime $t_{\rm e}$.  The engine luminosity in this scenario would be expected to decay as the fall-back rate, $L_{e} \propto t^{-5/3}$ (however, see \citealt{Tchekhovskoy15}).

\subsection{Tidal Disruption by an Intermediate Mass Black Hole?}
\label{SubSec:IntIMBH}

Alternatively, \cite{Perley18} and \cite{Kuin18} suggested that AT\,2018cow could have been caused by the tidal disruption of a main sequence star by an intermediate mass black hole (IMBH).  For an IMBH of mass $M_{\bullet}$, the fall-back time of the mostly tightly bound debris following a TDE is given by (e.g., \citealt{Stone16,Chen18}):
\be
t_{\rm fb} \approx 4.1 {\rm d}\left(\frac{M_{\bullet}}{10^{4}M_{\odot}}\right)^{1/2}\left(\frac{M_{\star}}{M_{\odot}}\right)^{1/5},
\ee
where we have assumed a mass-radius relationship $M_{\star} \propto R_{\star}^{0.8}$ appropriate for the lower main-sequence. $t_{\rm fb}$ is similar to the observed rise time of AT\,2018cow.  Therefore, \emph{if} circularization of the debris is relatively efficient, such that the light curve rises on the initial fall-back time, then an IMBH mass of $10^{4}M_{\odot}$ could explain the short rise time of the emission.  We note however that the circularization time scale of the TDE debris is a highly debated point in the literature (see e.g., \citealt{Chen18}, who claim the circularization process could take years).  Efficient circularization could be particularly problematic for such a low-mass black hole, as the angle through which the stream precesses due to general relativistic effects is extremely small (unless the pericenter of the tidally-disrupted star was well within the tidal radius, which is unlikely).

One complication of this scenario is that the IMBH would be accreting at a rate exceeding its Eddington luminosity $L_{\rm Edd} \approx 10^{42}\,M_{\bullet,4}$ erg s$^{-1}$ by a large factor $\gtrsim 10-100$ at early times (Fig.~\ref{Fig:Lengine}).  However, recent radiation MHD simulations of super-Eddington flows by \cite{Jiang14} find that radiative efficiencies of $\sim 5\%$ (similar to thin disks) are possible for flows accreting up to 22$\dot{M}_{\rm Edd}$, in the range relevant here. \cite{Sadowski16} find a similar result for the kinetic luminosities from the disk.  If coming from the inner disk, the relevant radiation temperature would be close to that of the disk photosphere at the ISCO radius,
\be
kT_{\rm eff} \approx 0.11\,{\rm keV}\left(\frac{\dot{M}}{\dot{M}_{\rm Edd}}\frac{0.1}{\eta}\right)^{1/4}\left(\frac{M_{\bullet}}{10^{4}M_{\odot}}\right)^{-1/4},
\ee
While this is softer than the required emission of the central X-ray source powering AT\,2018cow, an additional process like Inverse Compton scattering of soft photons from the disk  by the corona electrons could create the necessary high-energy tail (as described in \S\ref{SubSubSec:ComptonHump}).

Perhaps a larger problem with the IMBH scenario is the origin of the dense external CSM responsible for the self-absorbed radio emission (Fig.~\ref{Fig:RadioChevalier}), which must then be present prior to the tidal disruption.  Interpreted as a wind of velocity $v_{\rm w} = 10^{4}$ \kms\ characteristic of AGN outflows, one requires a mass-loss rate of $\sim 10^{-2}-10^{-3}M_{\odot}$ yr$^{-1}$, in excess of the Eddington accretion rate of $\dot{M}_{\rm Edd} \approx L_{\rm Edd}/0.1 c^{2} \approx 10^{-4} M_{\odot}$ yr$^{-1}$ for a $10^{4}M_{\odot}$ BH.  Alternatively, if the CSM represents an {\it accretion flow} onto the IMBH, the required accretion rate on radial scales $r\sim 3\times 10^{15}$ cm achieved by the forward shock on timescales of weeks would be
\be
\dot{M} \sim \frac{M_{\rm enc}}{t_{\rm ff}} \sim 3\times 10^{-4}M_{\odot} {\rm yr^{-1}},
\ee
where $M_{\rm enc} \sim 3\times 10^{-6}M_{\odot}$ is the effective mass at radius $r$ for $A \sim 30A_{\star}$ (as needed to explain the radio observations) and
$t_{\rm ff} \sim (r^{3}/GM_{\bullet})^{1/2} \approx 2\times 10^{7}$ s is the free-fall time for $M_{\bullet} = 10^{4}M_{\odot}$.
Therefore, unless the IMBH was already embedded in a gas-rich AGN-like environment prior to the tidal disruption event, it is challenging to explain the inferred presence of the external CSM.  
\subsection{Embedded Equatorial CSM Shock}
\label{SubSec:IntShock}

Although we disfavor an external CSM shock as the origin of the observed X-ray emission from AT\,2018cow (\S~\ref{SubSubSec:XrayShock}), the central source could be powered by a deeply embedded, {\em internal shock} \citep[e.g.,][]{Andrews18}.  If the CSM were concentrated in an equatorial ring or sheet, this shock would be localized to the dense equatorial region (e.g., the central X-ray source would in fact be an X-ray {\em ring}, Fig.~\ref{fig:ShockCartoon}, left panel). One way to explain the steeply decaying luminosity of AT\,2018cow, $L_{\rm e} \propto t^{-\alpha}$ with $\alpha \approx 2-2.5$ would be if the shock were decelerating. In this class of models, the torus of dense material pre-existed the AT\,2018cow event. However, the very short rise-time of AT\,2018cow is directly related to its fast polar \emph{ejecta} and it is thus an intrinsic property of AT\,2018cow, independent from the denser equatorial ring.

Consider the CSM to have a radial density profile $\rho \propto r^{-\gamma}$.  If the shock is radiative, then its self-similar deceleration evolution is momentum conserving, and so, for a disk of vertical scale-height $H$ with a constant aspect ratio $H/r \sim constant$, we have $\rho v_{\rm sh} R_{\rm sh}^{3} \propto R_{\rm sh}^{4-\gamma}/t = constant$, i.e., $R_{\rm sh} \propto t^{1/(4-\gamma)}$ and $v_{\rm sh} \sim R_{\rm sh}/t \propto t^{(\gamma-3)/(4-\gamma)}$.  Therefore, the shock luminosity would evolve as
\be
L_{\rm e} = L_{\rm sh} \propto v_{\rm sh}^{3}R_{\rm sh}^{2}\rho  \propto t^{(2\gamma-7)/(4-\gamma)}
\ee
For a constant CSM density profile $\gamma = 0$, we find $L_{\rm sh} \propto t^{-7/4}$, similar to the luminosity decay of AT\,2018cow.  

Not addressed in this scenario is the relatively soft intrinsic spectrum of the X-rays escaping from the shock.  Given the high shock velocities, one would expect the shock-heated gas to emit through the free-free process, which for a single-temperature plasma predicts $F_{\nu} \propto \nu^{0}$ at frequencies $\nu \ll kT_{\rm sh}$, where $T_{\rm sh}$ is the post-shock temperature, flatter than observed for AT\,2018cow.  However, due to the aspherical shape of the shock front, and various hydrodynamical instabilities that radiative shocks are susceptible to, the post-shock gas will not be characterized by a single temperature (e.g.,~\citealt{Steinberg&Metzger18}).  As the velocity of the shock decreases with time, the temperature of the post-shock gas will also decrease.  This would result in a greater fraction of the shock power emerging at UV/soft X-ray frequencies, which are more easily absorbed by the ejecta, and thus could contribute to the observed rapid late-time decay in the X-ray luminosity $L_{\rm x} \propto t^{-4}$.  Future theoretical work is required to assess the X-ray emission emerging from shocks propagating into aspherical environments, as similar physics is at work in a variety of astrophysical transients (e.g., SNe IIn, luminous red novae, and classical novae; e.g.,~\citealt{Andrews18,Metzger17novae}). 

\section{Conclusions}
\label{Sec:conclusiosn}
In this first extensive radio-to-$\gamma$-ray study of an FBOT we uncovered a new class of astronomical transients that are powered by a central engine and are characterized by luminous and long-lived radio and X-ray emission.  Events similar to AT\,2018cow can be detected with current X-ray/radio facilities in the local universe at $z\le 0.2$. 

Our study highlights the importance of follow-up observations across the spectrum, including the hard X-ray range at $E>10$ keV, which are rarely performed.  This monitoring campaign led to the discovery of a new spectral component of hard X-ray emission at $E\ge10$ keV, with unprecedented properties among astronomical transients---but reminiscent of ``Compton humps'' and Fe K-shell emission observed in AGNs and XRBs--- which would have been entirely missed in the absence of NuSTAR/INTEGRAL observations. At the same time, observations of AT\,2018cow on the low frequency end of the spectrum at $\nu<100$ GHz revealed a non-relativistic blastwave propagating into a relatively dense environment, with properties not dissimilar from the brightest radio SNe. 

The X-ray and UV/optical emission of AT\,2018cow instead displays stark differences with respect to normal SNe, and points towards  a small amount of asymmetrically-distributed H/He-rich ejecta. Asymmetry is a key property of AT\,2018cow. The need for a departure from spherical symmetry independently arises from the peculiar velocity gradient of the optically emitting material, from the redshifted centroids of the optical/NIR emission line profiles, and from the X-ray temporal and spectral properties of the broad-band X-ray emission.
Our analysis furthermore identified two distinct phases of evolution of AT\,2018cow, marked by a simultaneous change of its optical and X-ray properties around $\sim 20$ days.  

The observed properties of AT\,2018cow rule out traditional models where the transient emission is powered by the radioactive decay of $^{56}$Ni and are not consistent with interaction-powered models, where the entire spectrum originates from the \emph{external} interaction of the blastwave with the environment (which was suggested by \citealt{Rivera18}). Our detailed modeling shows that the phenomenology of AT\,2018cow requires the presence of a central source of high-energy radiation shining through low-mass ejecta with pronounced equatorial-polar asymmetry. The ``central engine'' might be either in the form of a compact object (like a millisecond magnetar or black hole), or a deeply embedded \emph{internal} shock. We find that low-mass H-rich stars that have been predicted to end their lives as electron-capture SNe, or blue supergiant stars that fail to explode are viable progenitor systems of AT\,2018cow, and are consistent with the location of AT\,2018cow within a star-forming dwarf galaxy. The tidal disruption of a star by an off-center intermediate mass black hole suggested by \cite{Perley18} and \cite{Kuin18} is disfavored by the large environmental density that we infer. 

Panchromatic studies of future FBOTs will clarify if AT\,2018cow is a representative member of its class and  will  reveal the connection (or lack thereof) of FBOTs to other classes of explosive transients, like GRBs or TDEs (the only two types of transients that are known to show persistent and rapid X-ray variability so far). In this respect it is interesting to mention that hints for hard X-ray excesses have been reported in long GRBs (\citealt{Moretti+2008,Margutti+2008}), which are central-engine powered explosions, and in ultra-long GRBs (\citealt{Stratta+2013,Bellm+2014}), for which the connection to blue supergiant progenitors has already been suggested (e.g., \citealt{Quataert12,Gendre13,Nakauchi13,Wu13,Liu18}).

\acknowledgements
We are very grateful to the entire NuSTAR, INTEGRAL, \emph{Swift}, XMM,  VLA, and VLBA teams for making this observing campaign possible. 

Some of the observations reported here were obtained at the MMT Observatory, a joint facility of the University of Arizona and the Smithsonian Institution under programs (2018A-UAO-G15, 2018B-SAO-21, 2018B-UAO-G16 PIs Fong, Patnaude, Terreran).
Some of the data presented herein were obtained at the W. M. Keck Observatory, which is operated as a scientific partnership among the California Institute of Technology, the University of California and the National Aeronautics and Space Administration under program NW254 (PI Miller). The Observatory was made possible by the generous financial support of the W. M. Keck Foundation. The authors wish to recognize and acknowledge the very significant cultural role and reverence that the summit of Maunakea has always had within the indigenous Hawaiian community.  We are most fortunate to have the opportunity to conduct observations from this mountain. 
The Keck and MMT observations were supported by Northwestern University and the Center for Interdisciplinary Exploration and Research in Astrophysics (CIERA). This paper includes data acquired with UKIRT under program (U/18A/UA01, PI Fong). UKIRT is owned by the University of Hawaii (UH) and operated by the UH Institute for Astronomy; operations are enabled through the cooperation of the East Asian Observatory.
This paper includes data gathered with the 6.5 meter Magellan Telescopes located at Las Campanas Observatory, Chile.
Based in part on observations at Cerro Tololo Inter-American Observatory, National Optical Astronomy Observatory (NOAO Prop. 2018A-0343, 2018B-0210; PI: G. Terreran), which is operated by the Association of Universities for Research in Astronomy (AURA) under a cooperative agreement with the National Science Foundation. 
Based in part on observations obtained at the Southern Astrophysical Research (SOAR) telescope, which is a joint project of the Minist\'{e}rio da Ci\^{e}ncia, Tecnologia, Inova\c{c}\~{a}os e Comunica\c{c}\~{a}oes (MCTIC) do Brasil, the U.S. National Optical Astronomy Observatory (NOAO), the University of North Carolina at Chapel Hill (UNC), and Michigan State University (MSU). This paper uses data products produced by the OIR Telescope Data Center, supported by the Smithsonian Astrophysical Observatory. Support for this work was provided to MRD by NASA through Hubble Fellowship grant NSG-HF2-51373 awarded by the Space Telescope Science Institute, which is operated by the Association of Universities for Research in Astronomy, Inc., for NASA, under contract NAS5-26555.

Partly based on observations with INTEGRAL, an ESA project with
instruments and science data centre funded by ESA member states
(especially the PI countries: Denmark, France, Germany, Italy,
Switzerland, Spain) and with the participation of Russia and the USA. Partly based on observations obtained with XMM-Newton (IDs: 0822580401, 0822580501, AO-17, program \#82258, PI Margutti), an ESA science mission with instruments and contributions directly funded by ESA Member States and NASA. We acknowledge the use of public data from the Swift data archive. This work made use of data from the NuSTAR mission (IDs 90401327002, 90401327004, 90401327006, 90401327008), a project led by the California Institute of Technology, managed by the Jet Propulsion Laboratory, and funded by the National Aeronautics and Space Administration. This research has made use of the NuSTAR Data Analysis Software (NuSTARDAS) jointly developed by the ASI Science Data Center (ASDC, Italy) and the California Institute of Technology (USA).
Observations taken with the VLA (program 18A-123, PI Coppejans) were used in this research. The National Radio Astronomy Observatory is a facility of the National Science Foundation operated under cooperative agreement by Associated Universities, Inc.

B.~D.~M.~acknowledges support from NSF grant AST-1615084, NASA Fermi Guest Investigator Program grants NNX16AR73G and 80NSSC17K0501; and through the Hubble Space Telescope Guest Investigator Program grant HST-AR-15041.001-A.
P.~U. and S.~M. acknowledge financial support from ASI under contracts ASI/INAF 2013-025-R0.
D.G. acknowledges the financial support of the UnivEarthS Labex program at Sorbonne Paris Cit\'e (ANR-10-LABX-0023 and ANR-11-IDEX-0005-02). K.~D.~A.  acknowledges support provided by NASA through the NASA Hubble Fellowship grant HST-HF2-51403.001 awarded by the Space Telescope Science Institute, which is operated by the Association of Universities for Research in Astronomy, Inc., for NASA, under contract NAS5-26555.
I.C. acknowledges support by the Telescope Data Center, Smithsonian Astrophysical Observatory and the Russian Science Foundation grant 17-72-20119. D.~J.~P. acknowledges support through NASA Contract NAS8-03060.
LD acknowledges support by the Bundesministerium f\"ur Wirtschaft und 
Technologie and the Deutsches Zentrum f\"ur Luft und Raumfahrt through the grant FKZ 50 OG 1602.
N.R. acknowledges the support from the University of Maryland  through the Joint Space Science Institute Prize Postdoctoral Fellowship as well support from the Center for Research and Exploration in Space Science and Technology II.
M.~R.~D. acknowledges support from the Dunlap Institute at the University of Toronto.

\facilities{Swift, XMM, NuSTAR, INTEGRAL, Keck-I, Keck-II, WIYN, SMARTS, UKIRT, MMT, VLA, VLBA} 
\software{HEASoft, CIAO, OSA, SAS, CASA}

\newpage
\appendix
\renewcommand\thetable{A\arabic{table}}  
\begin{deluxetable*}{rrrrrr}
\tablecolumns{6}
\tablewidth{0pc}
\tablecaption{Log of NIR/optical spectroscopic observations\label{spec_tab}}
\tablehead{
\colhead{UT date} & \colhead{MJD} & \colhead{Phase} & \colhead{Instrument/} &
\colhead{Wavelength} & \colhead{Resolution} \\
\colhead{} & \colhead{(d)} & \colhead{(d)} & \colhead{telescope} &
\colhead{range (\AA)} & \colhead{(\AA)} }
\startdata
2018-06-21 & 58290.1 & +4.6 & Goodman red cam/SOAR & 3500--8980 & 5.2 \\
2018-06-26 & 58295.1 & +9.5 & Goodman red cam/SOAR & 3530--8930 & 3.0 \\
2018-06-27 & 58296.1 & +10.5 & Goodman red cam/SOAR & 3500--8980 & 5.2 \\
2018-06-29 & 58298.2 & +12.5 & Goodman red cam/SOAR & 3500--8980 & 5.0 \\
2018-07-03 & 58302.3 & +16.6 & MMIRS/MMT & 9800--23100 & 5--17 \\
2018-07-09 & 58308.2 & +22.4 & LDSS3/Magellan (Clay)  & 3800--10000 & 8.0 \\
2018-07-21 & 58320.1 & +34.2 & Goodman red cam./SOAR & 3500--7050 & 4.4\\
2018-08-06 & 58336.0 & +51.0 & IMACS/Magellan (Baade)  & 3500--9000  & 5  \\
2018-08-17 & 58347.7 & +62.3 &DEIMOS/Keck-II & 4500--8500 &3.0\\
2018-09-10 & 58371.2 & +85.8 & LRIS/Keck-I & 3200--9000 &6.0\\
\enddata
\end{deluxetable*}


\begin{deluxetable*}{rrrrrr}
\tablecolumns{6}
	\tablewidth{0pc}
	\tablecaption{\emph{Swift}-XRT time resolved spectral analysis. We model the 0.3-10 keV spectrum with an absorbed simple power-law spectrum (\texttt{ztbabs*tbabs*pow} within Xspec). The Galactic column density of neutral hydrogen in the direction of AT\,2018cow is $\NHsub{MW}=0.05\times 10^{22}\,\rm{cm^{-2}}$ \citep{Kalberla05}. \label{Tab:XRTcow}}
	\tablehead{ \colhead{Start} &
		\colhead{End} & \colhead{\NHsub{int}} & \colhead{$\Gamma$} & \colhead{Absorbed Flux} &\colhead{Unabsorbed Flux}  \\
    (days) & (days) & ($10^{22}\,\rm{ cm^{-2}}$) & & 0.3-10 keV $(\rm{erg\,s^{-1}\,cm^{-2}})$ & 0.3-10 keV $(\rm{erg\,s^{-1}\,cm^{-2}})$}
\startdata
3 & 5 & < 0.0624 & $1.55 \pm 0.053$ & $2.01^{+0.10}_{-0.095} \times 10^{-11}$ & $2.13^{+0.11}_{-0.10} \times 10^{-11}$ \\
5 & 7 & < 0.0586 & $1.63 \pm 0.035$ & $1.65^{+0.052}_{-0.052} \times 10^{-11}$ &  $1.77^{+0.056}_{-0.056} \times 10^{-11}$\\
7 & 11 & < 0.0919 & $1.52 \pm 0.050$ & $8.19^{+0.34}_{-0.38} \times 10^{-12}$ & $8.69^{+0.36}_{-0.40} \times 10^{-12}$ \\
11 & 13 & < 0.0538 & $1.46 \pm 0.040$ & $9.87^{+0.40}_{-0.36} \times 10^{-12}$ & $1.04^{+0.042}_{-0.038} \times 10^{-11}$ \\
13 & 15 & < 0.157 & $1.56 \pm 0.093$ & $5.64^{+0.54}_{-0.43} \times 10^{-12}$ & $6.00^{+0.57}_{-0.46} \times 10^{-12}$ \\
15 & 19 & < 0.0641 & $1.41 \pm 0.039$ & $5.26^{+0.19}_{-0.18} \times 10^{-12}$ & $5.52^{+0.20}_{-0.19} \times 10^{-12}$ \\
19 & 21.5 & < 0.0758 & $1.41 \pm 0.042$ & $8.47^{+0.33}_{-0.36} \times 10^{-12}$ & $8.89^{+0.35}_{-0.38} \times 10^{-12}$ \\
21.5 & 23 & < 0.127 & $1.49 \pm 0.071$ & $4.07^{+0.29}_{-0.26} \times 10^{-12}$ & $4.30^{+0.31}_{-0.27} \times 10^{-12}$ \\
23 & 24.5 & < 0.240 & $1.36 \pm 0.10$ & $4.84^{+0.54}_{-0.51} \times 10^{-12}$ & $5.06^{+0.56}_{-0.53} \times 10^{-12}$ \\
24.5 & 26 & < 0.167 & $1.47 \pm 0.080$ & $4.17^{+0.38}_{-0.30} \times 10^{-12}$ & $4.40^{+0.40}_{-0.32} \times 10^{-12}$ \\
26 & 29 & < 0.193 & $1.36 \pm 0.16$ & $2.73^{+0.48}_{-0.36} \times 10^{-12}$ & $2.85^{+0.50}_{-0.38} \times 10^{-12}$ \\
29 & 31 & < 0.224 & $1.50 \pm 0.21$ & $1.96^{+0.42}_{-0.36} \times 10^{-12}$ & $2.07^{+0.44}_{-0.38} \times 10^{-12}$ \\
31 & 35 & < 0.157 & $1.56 \pm 0.071$ & $2.32^{+0.17}_{-0.16} \times 10^{-12}$ & $2.47^{+0.18}_{-0.17} \times 10^{-12}$ \\
35 & 40 & < 0.298 & $1.47 \pm 0.13$ & $1.53^{+0.22}_{-0.16} \times 10^{-12}$ & $1.61^{+0.23}_{-0.17} \times 10^{-12}$  \\
40 & 50 & < 0.145 & $1.54 \pm 0.11$ & $9.46^{+1.0}_{-0.89} \times 10^{-13}$ & $1.00^{+0.11}_{-0.094} \times 10^{-12}$ \\
50 & 58 & < 1.02 & $1.37 \pm 0.23$ & $3.81^{+0.86}_{-0.70} \times 10^{-13}$ & $3.99^{+0.90}_{-0.73} \times 10^{-13}$ \\
\enddata
\end{deluxetable*}

\newpage
\begin{deluxetable*}{rrrrrr}
	\tablecolumns{6}
	\tablewidth{0pc}
	\tablecaption{Log of INTEGRAL observations \label{integral_tab}}
	\tablehead{ \colhead{Orbit} &
		\colhead{Start time (UT)} & \colhead{Stop time (UT)} & \colhead{Midtime (MJD)} & Phase &\colhead{ontime}  \\
     &  &  & (d) & (d) & (ks)  }
	\startdata
1968 &2018-06-22 18:39:55 & 2018-06-24 22:21:07 &58292.9&7.4& 164\\
1969 &2018-06-25 10:33:59 & 2018-06-27 14:10:12 &58295.5&10.1& 169\\
1970 &2018-06-28 01:32:03 & 2018-06-30 05:00:38 &58298.1&12.7& 176\\
1971 &2018-06-30 18:05:43 & 2018-07-02 17:38:49 & 58300.7&15.3& 156\\
1972 &2018-07-03 10:08:38 & 2018-07-04 04:50:23 &58302.8&17.4& 59\\
1973 &2018-07-06 00:56:16 & 2018-07-08 04:58:20 &58306.1&20.7& 175\\
	\enddata
\end{deluxetable*}


\begin{deluxetable*}{rrrrrr}
	\tablecolumns{6}
	\tablewidth{0pc}
	\tablecaption{Log of NuSTAR observations \label{Nustar_tab}}
	\tablehead{ \colhead{ID}
		& \colhead{Start time (UT)} & \colhead{Stop time (UT)} & \colhead{Midtime (MJD)} & \colhead{Phase} &\colhead{Exposure Time}  \\
        & & & (d) & (d) & (ks)  }
	\startdata
90401327002 &2018-06-23 17:31:09	  & 2018-06-24 11:01:09 & 58293.1 & 7.7 & 27.9\\
90401327004 & 2018-07-02 14:00:12 & 2018-07-03 07:30:00 & 58301.9 & 16.5 & 30.0 \\
90401327006 & 2018-07-14 06:20:09 & 2018-07-14 23:35:00 & 58313.6 & 28.2 & 31.2 \\
90401327008 & 2018-07-22 13:56:09  & 2018-07-23 09:06:09 & 58321.8 & 36.5 & 13.9 \\
	\enddata
\end{deluxetable*}


\begin{deluxetable*}{rrrrrrrr}
	\tablecolumns{8}
	\tablewidth{0pc}
	\tablecaption{Best-fitting parameters of the broad-band X-ray spectra of AT\,2018cow obtained combining \emph{Swift}-XRT, XMM, NuSTAR and INTEGRAL observations. We model the data with the combination of an absorbed power-law and cutoff power-law model (\texttt{ztbabs*(tbabs1*pow+tbabs2*cutoffpl)} within Xspec). The power-law and cutoff power-law are tied to have the same photon index $\Gamma$. We assume Galactic column density of neutral hydrogen $\NHsub{MW}=0.05\times 10^{22}\,\rm{cm^{-2}}$ \citep{Kalberla05}. Quantities without uncertainties have been frozen to the value reported. The absorbed cutoff power-law model is purely phenomenological. \label{Tab:specxray}}
	\tablehead{ \colhead{Epoch}
		& \colhead{Phase} &  \colhead{\NHsub{int}} & \colhead{$\Gamma$} & \colhead{\NHsub{hard}} & \colhead{$E_{\rm cutoff}$} & \colhead{$F_{\rm x,hard}$ (20-200 keV)} & \colhead{Unabsorbed $F_{\rm x,soft}$ (0.3-10 keV)}\\
        & (d) & ($10^{22}\rm{cm^{-2}}$) & & ($10^{22}\rm{cm^{-2}}$) & (keV) & ($10^{-11}\rm{erg\,cm^{-2}s^{-1}}$) & ($10^{-11}\rm{erg\,cm^{-2}s^{-1}}$)  }
	\startdata
1$^{a}$& 7.7 & $<0.06$ &$1.62\pm0.03$&$2100^{+150}_{-350}$&$72^{+25}_{-20}$&$2.29^{+0.72}_{-0.69}$& $1.04^{+0.09}_{-0.04}$\\
2$^{b}$& 10.1&$<0.07$ &$1.49\pm0.05$&$2100$&$31^{+21}_{-14}$			   &$3.00^{+0.94}_{-1.60}$&$1.04^{+0.07}_{-0.07}$	\\
3$^{c}$& 12.7&$<0.11$ &$1.51\pm0.06$&$2100$&$31$&$2.68^{+0.84}_{-1.20}$&$0.97^{+0.03}_{-0.06}$	\\
4$^{d}$& 16.5&$<0.04$ &$1.43\pm0.08$&$2100$&$31$&$<0.94$				&$0.79^{+0.04}_{-0.04}$	\\
5$^{e}$& 28.2&$<0.39$ &$1.58\pm0.04$&$2100$&$31$&$<0.40$				&$0.34^{+0.05}_{-0.05}$	\\
6$^{f}$& 36.5&$<0.03$ &$1.67\pm0.02$&$2100$&$31$&$<1.11$				&$0.21^{+0.01}_{-0.03}$	\\
	\enddata
\tablecomments{$^{a}$ \emph{Swift}-XRT data between $\delta t=7.3-8.0$ days; NuSTAR ID 90401327002 ($\delta t=6.8-8.3$ days); INTEGRAL orbit 1968 ($\delta t=6.3-8.5$ days). \\
$^{b}$\emph{Swift}-XRT data between $\delta t=9-11.2$ days; INTEGRAL orbit 1969 ($\delta t=8.9-11.2$ days).\\
$^{c}$\emph{Swift}-XRT data between $\delta t=11.6-13.8$ days; INTEGRAL orbit 1970 ($\delta t=11.6-13.8$ days).\\
$^{d}$\emph{Swift}-XRT data between $\delta t=16.1-16.9$ days; NuSTAR ID 90401327004 ($\delta t=16.1-16.9$ days).\\
$^{e}$\emph{Swift}-XRT data between $\delta t=27.8-28.1$ days; NuSTAR ID 90401327006 ($\delta t=27.8-28.5$ days).\\
$^{f}$\emph{Swift}-XRT data between $\delta t=36.1-37.0$ days; XMM ID 0822580401 ($\delta t=36.6-36.9$ days); NuSTAR ID 90401327008 ($\delta t=36.1-36.9$ days).\\} 
\end{deluxetable*}

\begin{deluxetable*}{rrrrrr}
	\tablecolumns{6}
	\tablewidth{0pc}
	\tablecaption{Radio flux-density measurements with the VLA and VLBI. For VLBI measurements the listed uncertainties include systematics. For the VLA we list statistical uncertainties only (systematic uncertainties are estimated to be at the level of $\sim$5\%). \label{Tab:Radio}}
	\tablehead{ \colhead{Start Date (UT)} & \colhead{Phase} &   \colhead{Frequency} &    \colhead{Bandwidth} &\colhead{Flux Density} & Instrument \\
        & (d) &  (GHz)  & (GHz)  & ($\rm{mJy}$)  & }
	\startdata
2018-07-08& 21.6 & 22.3 & 0.3 &$5.85 \pm 0.61$ &VLBI 	 \\
2018-09-06 & 82.51  & 5.0 & 1.0 & $5.04\pm0.04$ &VLA\\
2018-09-06 & 82.51  & 7.1 & 1.0& $7.76\pm0.09$&VLA\\			
2018-09-07 & 83.51  & 9.0 & 2.0 &	$9.10\pm0.30$&VLA\\	
2018-09-07 & 83.51  & 11.0 & 2.0 & $9.80\pm0.40$&VLA\\
2018-09-07 & 83.52  & 2.5 &1.0 & $2.25\pm0.11$&VLA \\
2018-09-07 & 83.52  & 3.5 &1.0& $3.24\pm0.06$&VLA	\\
2018-09-07 & 83.52  & 1.3 &0.5&$0.80\pm0.26$&VLA \\
2018-09-07 & 83.52  & 1.8 &0.5&$1.40\pm0.13$& VLA \\
2018-09-16 & 91.61  & 19. &2.0& $8.62\pm0.06$&VLA\\
2018-09-16 & 91.61 &  21 &2.0&$7.52 \pm 0.09$&VLA\\
2018-09-16 & 91.61 &  23 &2.0&$6.81\pm 0.11$&VLA\\
2018-09-16 & 91.61 &  25 &2.0&$6.10\pm0.07$&VLA \\
\enddata
\end{deluxetable*}

\begin{deluxetable*}{rrrrrr}
	\tablecolumns{6}
	\tablewidth{0pc}
	\tablecaption{Ground-based optical photometry (Vega magnitudes). Observed magnitudes.  \label{Tab:GroundPhotOpt}}
	\tablehead{ \colhead{Phase} & \colhead{V} &   \colhead{B} &    \colhead{R} &\colhead{I} & Telescope \\
     (d)   & (mag) &  (mag)  & (mag)  & (mag) & }
	\startdata
            5.63 &$          15.10 \pm           0.04$ &$          14.90 \pm           0.07$ &$          14.95 \pm           0.05$ &$          15.05 \pm           0.08$ &SMARTS+ANDICAM\\
           6.66 &$          15.30 \pm           0.14$ &$          15.24 \pm           0.04$ &$          15.21 \pm           0.07$ &$          15.22 \pm           0.16$ &SMARTS+ANDICAM\\
           7.69 &$          15.65 \pm           0.05$ &$          15.50 \pm           0.04$ &$          15.48 \pm           0.12$ &$          15.37 \pm           0.10$ &SMARTS+ANDICAM\\
           8.68 &$          15.72 \pm           0.16$ &$          15.46 \pm           0.14$ &$          15.54 \pm           0.17$ &$          15.34 \pm           0.17$ &SMARTS+ANDICAM\\
           9.61 &$          16.03 \pm           0.05$ &$          15.99 \pm           0.04$ &$          15.81 \pm           0.08$ &$          15.67 \pm           0.11$ &SMARTS+ANDICAM\\
          11.63 &$          16.27 \pm           0.07$ &$          16.27 \pm           0.06$ &$          16.18 \pm           0.08$ &$          16.08 \pm           0.10$ &SMARTS+ANDICAM\\
          12.63 &$          16.42 \pm           0.06$ &$          16.30 \pm           0.06$ &$          16.23 \pm           0.07$ &$          16.24 \pm           0.16$ &SMARTS+ANDICAM\\
          14.61 &$          16.47 \pm           0.16$ &$          16.53 \pm           0.21$ &$          16.45 \pm           0.22$ &$          16.26 \pm           0.30$ &SMARTS+ANDICAM\\
          15.64 &$          16.71 \pm           0.05$ &$          16.67 \pm           0.07$ &$          16.58 \pm           0.05$ &$          16.49 \pm           0.10$ &SMARTS+ANDICAM\\
          17.62 & -- &$          17.03 \pm           0.04$ &$          16.98 \pm           0.06$ &$          16.75 \pm           0.09$ &SMARTS+ANDICAM\\
          19.63 &$          17.09 \pm           0.06$ &$          17.17 \pm           0.05$ &$          17.10 \pm           0.08$ &$          16.86 \pm           0.08$ &SMARTS+ANDICAM\\
          21.61 & -- &$          17.36 \pm           0.05$ &$          17.25 \pm           0.06$ &$          17.05 \pm           0.09$ &SMARTS+ANDICAM\\
          22.64 &$          17.50 \pm           0.06$ &$          17.56 \pm           0.06$ &$          17.42 \pm           0.08$ &$          17.13 \pm           0.08$ &SMARTS+ANDICAM\\
          23.65 &$          17.63 \pm           0.07$ &$          17.61 \pm           0.09$ &$          17.46 \pm           0.08$ &$          17.18 \pm           0.09$ &SMARTS+ANDICAM\\
          24.62 &--&$          17.75 \pm           0.07$ &$          17.61 \pm           0.08$ &$          17.40 \pm           0.10$ &SMARTS+ANDICAM\\
          26.61 &$          18.05 \pm           0.08$ &$          18.07 \pm           0.08$ &$          17.93 \pm           0.08$ &$          17.48 \pm           0.10$ &SMARTS+ANDICAM\\
          34.61 &$          18.47 \pm           0.09$ &$          18.78 \pm           0.13$ &$          18.62 \pm           0.14$ &$          18.20 \pm           0.14$ &SMARTS+ANDICAM\\
          36.62 &$          18.94 \pm           0.14$ &$          18.94 \pm           0.15$ &$          18.79 \pm           0.20$ &$          18.20 \pm           0.17$ &SMARTS+ANDICAM\\
          37.59 &$          19.06 \pm           0.16$ &$          18.91 \pm           0.19$ &$          18.88 \pm           0.21$ &$          18.48 \pm           0.22$ &SMARTS+ANDICAM\\
          38.61 &$          19.22 \pm           0.12$ &$          19.01 \pm           0.11$ &$          18.69 \pm           0.10$ &$          18.51 \pm           0.22$ &SMARTS+ANDICAM\\
          44.61 &$          19.32 \pm           0.13$ &$          19.35 \pm           0.18$ &$          19.31 \pm           0.17$ &$          19.05 \pm           0.40$ &SMARTS+ANDICAM\\
          46.62 & -- &$          19.70 \pm           0.21$ &$          19.20 \pm           0.21$ &$          19.15 \pm           0.31$ &SMARTS+ANDICAM\\
          62.88 &$          21.20 \pm           0.07$ &$          20.97 \pm           0.08$ &$          20.68 \pm           0.06$ &$          20.34 \pm           0.07$ &Keck-II+DEIMOS\\
         111.82 &$          22.50 \pm           0.30$ &$          22.01 \pm           0.30$ &$          21.37 \pm           0.30$ &$          21.19 \pm           0.30$ &Keck-I+LRIS\\
	\enddata
\end{deluxetable*}

\begin{deluxetable*}{rrrrrrr}
	\tablecolumns{7}
	\tablewidth{0pc}
	\tablecaption{Ground-based NIR photometry (Vega magnitudes). Observed magnitudes. \label{Tab:GroundPhotNIR}}
	\tablehead{ \colhead{Phase (J)} & \colhead{J$^{a}$} &   \colhead{Phase (H)} &    \colhead{H} &\colhead{Phase (K)} & \colhead{K}  & Telescope\\
     (d)   & (mag) &  (d)  & (mag)  & (d)   & (mag)}
	\startdata
	9.86 & $15.83 \pm 0.06$ &  9.86 & $15.54 \pm 0.11$ & 9.86 & $15.17 \pm 0.22$&WIYN \\
            9.90&$          15.90\pm           0.04$&           9.91&$          15.24\pm           0.04$&           9.92&$          14.99\pm           0.05$ & UKIRT \\
          13.83&$          16.27\pm           0.05$&          13.84&$          15.72\pm           0.04$&          13.84&$          15.50\pm           0.05$ & UKIRT \\
          22.85&$          16.76\pm           0.10$&          22.86&$          15.75\pm           0.05$&          22.87&$          15.87\pm           0.08$ & UKIRT \\
          26.82&$          17.51\pm           0.20$&          26.83&$          15.84\pm           0.05$&          26.85&$          15.76\pm           0.07$ & UKIRT \\
          30.86&$          18.07\pm           0.37$&          30.88&$          16.11\pm           0.06$&          30.89&$          15.90\pm           0.07$ & UKIRT \\
          35.82&$          18.24\pm           0.40$&          34.93&$          17.18\pm           0.14$&          35.84&$          15.98\pm           0.08$ & UKIRT \\
          39.80&$          18.99\pm           0.87$&          35.83&$          15.81\pm           0.05$&          41.84&$          16.76\pm           0.12$ & UKIRT \\
          41.82&$          18.52\pm           0.51$&          39.82&$          15.85\pm           0.05$&           --&--& UKIRT \\
           --&--&          41.83&$          17.13\pm           0.07$&           --&-- & UKIRT \\
           75.70 & >20.06 &75.70  & >19.22 & 75.70 &  >19.03 & MMT\\
	\enddata
	\tablecomments{$^{a}$ Host-galaxy subtracted photometry.}
\end{deluxetable*}

\begin{deluxetable*}{cccccccccccc}
	\tablecolumns{6}
	\tablewidth{0pc}
	\tablecaption{\emph{Swift}-UVOT $v$, $b$ and $u$-band photometry. Observed magnitudes. \label{Tab:UVOT1}}
	\tablehead{ \colhead{Phase ($v$)} & \colhead{$v$} &   \colhead{Phase ($b$)} &    \colhead{$b$} &\colhead{Phase ($u$)} & \colhead{$u$}   \\
     (d)   & (mag) &  (d)  & (mag)  & (d)   & (mag)}
	\startdata
             3.06&$          13.95\pm           0.05$&           3.06&$          13.96\pm           0.05$&           3.06&$          12.53\pm           0.05$\\
           3.79&$          14.15\pm           0.05$&           3.79&$          14.28\pm           0.07$&           3.78&$          12.92\pm           0.07$\\
           5.26&$          14.83\pm           0.11$&           3.79&$          14.25\pm           0.05$&           3.79&$          12.89\pm           0.05$\\
           5.26&$          14.73\pm           0.05$&           5.25&$          15.02\pm           0.08$&           5.25&$          13.42\pm           0.07$\\
           6.26&$          15.03\pm           0.12$&           5.25&$          14.96\pm           0.05$&           5.25&$          13.46\pm           0.05$\\
           6.26&$          15.03\pm           0.05$&           6.25&$          15.23\pm           0.09$&           6.25&$          13.71\pm           0.08$\\
           6.84&$          15.18\pm           0.13$&           6.25&$          15.23\pm           0.05$&           6.25&$          13.72\pm           0.05$\\
           6.84&$          15.19\pm           0.06$&           6.84&$          15.35\pm           0.09$&           6.84&$          13.88\pm           0.08$\\
           9.18&$          15.73\pm           0.06$&           6.84&$          15.37\pm           0.05$&           6.84&$          13.85\pm           0.05$\\
          10.18&$          15.87\pm           0.08$&           9.17&$          15.88\pm           0.05$&           6.64&$          13.83\pm           0.05$\\
          11.24&$          15.96\pm           0.07$&          10.17&$          15.88\pm           0.06$&           9.17&$          14.49\pm           0.05$\\
          12.96&$          16.14\pm           0.07$&          11.24&$          16.09\pm           0.06$&          10.17&$          14.63\pm           0.06$\\
          14.29&$          16.14\pm           0.07$&          12.96&$          16.20\pm           0.05$&          11.10&$          14.80\pm           0.05$\\
          15.22&$          16.31\pm           0.08$&          14.29&$          16.30\pm           0.05$&          11.23&$          14.78\pm           0.05$\\
          16.36&$          16.51\pm           0.08$&          15.21&$          16.49\pm           0.06$&          11.23&$          14.82\pm           0.06$\\
          16.64&$          16.63\pm           0.09$&          16.35&$          16.64\pm           0.06$&          12.95&$          14.90\pm           0.05$\\
          18.35&$          16.78\pm           0.10$&          16.64&$          16.63\pm           0.06$&          14.28&$          15.07\pm           0.06$\\
          21.34&$          16.96\pm           0.11$&          18.34&$          16.88\pm           0.07$&          15.14&$          15.27\pm           0.05$\\
          21.73&$          16.87\pm           0.10$&          20.20&$          16.88\pm           0.07$&          15.21&$          15.29\pm           0.05$\\
          22.72&$          16.98\pm           0.10$&          21.33&$          17.11\pm           0.07$&          15.21&$          15.26\pm           0.06$\\
          23.79&$          17.02\pm           0.11$&          21.72&$          17.18\pm           0.07$&          16.35&$          15.49\pm           0.06$\\
          24.78&$          17.13\pm           0.12$&          22.72&$          17.19\pm           0.07$&          16.64&$          15.50\pm           0.06$\\
          25.72&$          17.21\pm           0.12$&          23.79&$          17.21\pm           0.07$&          18.34&$          15.71\pm           0.06$\\
          26.32&$          17.38\pm           0.14$&          24.78&$          17.36\pm           0.08$&          18.83&$          15.80\pm           0.05$\\
          27.01&$          17.30\pm           0.10$&          25.72&$          17.52\pm           0.08$&          20.20&$          15.86\pm           0.07$\\
          29.31&$          17.35\pm           0.15$&          26.32&$          17.53\pm           0.09$&          21.33&$          16.04\pm           0.07$\\
          31.09&$          17.34\pm           0.11$&          27.01&$          17.54\pm           0.07$&          21.72&$          16.05\pm           0.07$\\
          27.98&$          17.27\pm           0.09$&          29.31&$          17.68\pm           0.10$&          22.71&$          16.20\pm           0.07$\\
          33.50&$          17.25\pm           0.11$&          31.09&$          17.67\pm           0.08$&          23.79&$          16.35\pm           0.07$\\
          36.80&$          17.39\pm           0.11$&          27.97&$          17.55\pm           0.07$&          24.78&$          16.34\pm           0.07$\\
          34.81&$          17.57\pm           0.12$&          33.49&$          17.83\pm           0.08$&          25.71&$          16.54\pm           0.08$\\
          38.69&$          17.73\pm           0.11$&          36.80&$          17.80\pm           0.08$&          26.32&$          16.64\pm           0.08$\\
          41.72&$          17.57\pm           0.12$&          34.80&$          17.87\pm           0.08$&          27.01&$          16.64\pm           0.07$\\
          40.73&$          17.47\pm           0.13$&          38.82&$          18.02\pm           0.09$&          29.38&$          16.79\pm           0.08$\\
          39.93&$          17.52\pm           0.12$&          38.46&$          17.92\pm           0.10$&          31.09&$          16.89\pm           0.08$\\
          46.50&$          17.62\pm           0.12$&          41.71&$          18.12\pm           0.09$&          27.97&$          16.76\pm           0.07$\\
          45.44&$          17.69\pm           0.13$&          40.73&$          18.03\pm           0.10$&          33.49&$          17.14\pm           0.09$\\
          43.78&$          17.68\pm           0.13$&          39.92&$          18.27\pm           0.10$&          36.80&$          17.35\pm           0.09$\\
          47.96&$          17.80\pm           0.21$&          46.49&$          18.18\pm           0.10$&          39.19&$          17.47\pm           0.07$\\
          46.93&$          17.64\pm           0.13$&          45.43&$          18.04\pm           0.09$&          34.80&$          17.21\pm           0.08$\\
          60.14&$          17.72\pm           0.12$&          43.77&$          18.08\pm           0.09$&          40.73&$          17.44\pm           0.10$\\
          54.46&$          17.85\pm           0.27$&          47.96&$          18.18\pm           0.13$&          44.60&$          17.64\pm           0.08$\\
          52.27&$          17.90\pm           0.11$&          46.92&$          18.24\pm           0.11$&          41.71&$          17.45\pm           0.09$\\
          49.95&$          17.54\pm           0.10$&          54.46&$          18.24\pm           0.20$&          47.23&$          17.66\pm           0.08$\\
          57.71&$          17.67\pm           0.18$&          52.27&$          18.23\pm           0.08$&          49.95&$          17.76\pm           0.10$\\
          55.83&$          17.80\pm           0.13$&          49.95&$          18.06\pm           0.08$&          57.55&$          17.93\pm           0.16$\\
           --&       --    &60.14&$          18.44\pm           0.10$&          54.19&$          17.89\pm           0.08$ \\
           --&           --&          57.71&$          18.48\pm           0.18$&          60.14&$          17.93\pm           0.11$\\
           --&           --&          55.82&$          18.19\pm           0.09$&           --& --\\
\enddata
\end{deluxetable*}

\begin{deluxetable*}{cccccccccccc}
	\tablecolumns{6}
	\tablewidth{0pc}
	\tablecaption{\emph{Swift}-UVOT $w1$, $w2$ and $m2$-band photometry (Vega magnitudes). Observed magnitudes. \label{Tab:UVOT2}}
	\tablehead{ \colhead{Phase ($w1$)} & \colhead{$w1$} &   \colhead{Phase ($w2$)} &    \colhead{$w2$} &\colhead{Phase ($m2$)} & \colhead{$m2$}   \\
     (d)   & (mag) &  (d)  & (mag)  & (d)   & (mag)}
	\startdata
            3.06&$          11.84\pm           0.06$&           3.06&$          11.55\pm           0.07$&           3.07&$          11.68\pm           0.06$\\
           3.78&$          12.07\pm           0.06$&           3.79&$          11.82\pm           0.09$&           3.79&$          11.94\pm           0.06$\\
           5.25&$          12.77\pm           0.09$&           3.79&$          11.85\pm           0.07$&           5.05&$          12.63\pm           0.06$\\
           6.25&$          12.98\pm           0.09$&           5.06&$          12.47\pm           0.07$&           5.26&$          12.63\pm           0.07$\\
           6.84&$          13.33\pm           0.10$&           5.26&$          12.60\pm           0.09$&           6.26&$          12.95\pm           0.06$\\
           6.12&$          12.95\pm           0.06$&           5.26&$          12.59\pm           0.07$&           6.85&$          13.34\pm           0.06$\\
           6.12&$          12.95\pm           0.06$&           6.26&$          12.96\pm           0.10$&           9.18&$          14.03\pm           0.06$\\
           9.17&$          14.00\pm           0.06$&           6.65&$          12.98\pm           0.07$&           9.16&$          13.81\pm           0.06$\\
          10.11&$          14.06\pm           0.06$&           8.37&$          13.45\pm           0.07$&          10.18&$          14.00\pm           0.06$\\
          10.17&$          14.05\pm           0.06$&           9.17&$          14.08\pm           0.07$&          11.24&$          14.27\pm           0.06$\\
          11.23&$          14.35\pm           0.06$&           8.23&$          13.45\pm           0.07$&          12.75&$          14.51\pm           0.07$\\
          12.36&$          14.45\pm           0.13$&           9.11&$          13.85\pm           0.07$&          12.95&$          14.61\pm           0.07$\\
          12.37&$          14.46\pm           0.06$&          10.17&$          14.07\pm           0.07$&          12.96&$          14.60\pm           0.06$\\
          12.95&$          14.52\pm           0.06$&          11.24&$          14.42\pm           0.07$&          14.29&$          14.83\pm           0.07$\\
          14.08&$          14.73\pm           0.06$&          12.09&$          14.48\pm           0.08$&          15.22&$          15.03\pm           0.07$\\
          14.28&$          14.73\pm           0.06$&          12.11&$          14.59\pm           0.08$&          16.36&$          15.33\pm           0.07$\\
          14.28&$          14.79\pm           0.06$&          12.36&$          14.47\pm           0.08$&          16.94&$          15.43\pm           0.07$\\
          15.21&$          14.93\pm           0.07$&          12.36&$          14.44\pm           0.14$&          16.64&$          15.32\pm           0.07$\\
          16.35&$          15.25\pm           0.07$&          12.77&$          14.50\pm           0.08$&          18.35&$          15.72\pm           0.08$\\
          16.63&$          15.32\pm           0.07$&          12.96&$          14.58\pm           0.07$&          20.82&$          15.88\pm           0.09$\\
          17.77&$          15.48\pm           0.06$&          14.29&$          14.83\pm           0.07$&          21.34&$          16.11\pm           0.09$\\
          18.34&$          15.59\pm           0.07$&          15.16&$          15.09\pm           0.08$&          21.73&$          16.06\pm           0.08$\\
          20.20&$          15.68\pm           0.07$&          15.21&$          15.05\pm           0.07$&          22.72&$          16.17\pm           0.08$\\
          21.33&$          15.90\pm           0.08$&          16.21&$          15.28\pm           0.08$&          23.79&$          16.25\pm           0.09$\\
          21.72&$          16.00\pm           0.08$&          16.22&$          15.32\pm           0.10$&          24.78&$          16.36\pm           0.10$\\
          22.25&$          15.99\pm           0.08$&          16.35&$          15.32\pm           0.08$&          25.28&$          16.62\pm           0.10$\\
          22.32&$          16.05\pm           0.08$&          16.36&$          15.38\pm           0.08$&          25.72&$          16.61\pm           0.10$\\
          22.71&$          16.08\pm           0.08$&          16.96&$          15.49\pm           0.09$&          26.32&$          16.70\pm           0.10$\\
          23.79&$          16.03\pm           0.08$&          16.64&$          15.42\pm           0.08$&          27.02&$          16.68\pm           0.09$\\
          24.77&$          16.21\pm           0.08$&          17.82&$          15.77\pm           0.10$&          29.32&$          17.06\pm           0.14$\\
          25.79&$          16.54\pm           0.07$&          18.34&$          15.78\pm           0.08$&          31.10&$          17.12\pm           0.10$\\
          27.01&$          16.64\pm           0.08$&          18.84&$          15.79\pm           0.09$&          27.98&$          16.72\pm           0.08$\\
          30.22&$          16.80\pm           0.11$&          19.76&$          15.80\pm           0.08$&          34.82&$          17.29\pm           0.10$\\
          29.38&$          16.76\pm           0.09$&          19.77&$          15.82\pm           0.09$&          33.50&$          17.14\pm           0.10$\\
          31.08&$          16.90\pm           0.09$&          20.20&$          15.82\pm           0.08$&          38.82&$          17.46\pm           0.12$\\
          27.97&$          16.63\pm           0.08$&          20.84&$          16.03\pm           0.09$&          38.46&$          17.30\pm           0.13$\\
          34.16&$          17.11\pm           0.09$&          21.34&$          16.02\pm           0.08$&          36.81&$          17.22\pm           0.11$\\
          34.15&$          17.14\pm           0.07$&          21.72&$          16.13\pm           0.08$&          41.72&$          17.54\pm           0.12$\\
          33.48&$          16.98\pm           0.09$&          22.30&$          16.21\pm           0.09$&          40.73&$          17.71\pm           0.14$\\
          38.46&$          17.15\pm           0.12$&          23.37&$          16.25\pm           0.09$&          39.93&$          17.69\pm           0.12$\\
          36.79&$          17.15\pm           0.10$&          22.72&$          16.32\pm           0.08$&          45.44&$          17.72\pm           0.13$\\
          34.80&$          17.06\pm           0.09$&          23.79&$          16.26\pm           0.08$&          43.78&$          17.74\pm           0.12$\\
          40.73&$          17.44\pm           0.12$&          24.78&$          16.46\pm           0.09$&          47.96&$          17.69\pm           0.17$\\
          39.92&$          17.39\pm           0.11$&          26.02&$          16.73\pm           0.08$&          46.93&$          17.70\pm           0.12$\\
          38.82&$          17.20\pm           0.10$&          27.01&$          16.78\pm           0.08$&          46.50&$          17.59\pm           0.12$\\
          43.77&$          17.51\pm           0.11$&          29.31&$          16.92\pm           0.10$&          54.46&$          17.92\pm           0.21$\\
          42.80&$          17.60\pm           0.25$&          31.09&$          17.00\pm           0.09$&          52.28&$          17.88\pm           0.11$\\
          41.71&$          17.45\pm           0.11$&          27.97&$          16.73\pm           0.08$&          49.95&$          17.62\pm           0.12$\\
          47.22&$          17.73\pm           0.09$&          33.49&$          17.17\pm           0.09$&          60.14&$          17.98\pm           0.11$\\
          45.43&$          17.43\pm           0.11$&          38.69&$          17.39\pm           0.09$&          57.71&$          18.02\pm           0.16$\\
          49.94&$          17.66\pm           0.10$&          36.80&$          17.28\pm           0.09$&          55.83&$          18.04\pm           0.13$\\
          55.42&$          17.89\pm           0.11$&          34.81&$          17.23\pm           0.09$&           --&--\\
          55.82&$          17.82\pm           0.12$&          39.92&$          17.41\pm           0.10$&           --&--\\
          58.84&$          17.95\pm           0.10$&          43.78&$          17.58\pm           0.10$&           --&--\\
           --&--&          41.71&$          17.48\pm           0.10$&           --&--\\
           --&--&          40.73&$          17.44\pm           0.10$&           --&--\\
           --&--&          46.76&$          17.69\pm           0.09$&           --&--\\
           --&--&          45.43&$          17.58\pm           0.10$&           --&--\\
          --&--&          49.95&$          17.73\pm           0.09$&           --&--\\
           --&--&          47.96&$          17.62\pm           0.13$&           --&--\\
           --&--&          55.82&$          17.86\pm           0.11$&           --&--\\
           --&--&          54.46&$          17.68\pm           0.16$&           --&--\\
           --&--&          52.27&$          17.81\pm           0.09$&           --&--\\
           --&--&          60.14&$          18.00\pm           0.10$&           --&--\\
           --&--&          57.71&$          18.11\pm           0.15$&           --&--\\
\enddata
\end{deluxetable*}


\bibliographystyle{yahapj}
\bibliography{ms,AT2018cow,integral}

\end{document}